\input harvmac
\input epsf

\def\mp#1{{\overline{\cal M}}_{0,1}(\vec{#1};{\bf P})}
\def\mpp#1#2{{\overline{\cal M}}_{0,1}(\vec{#1} - \vec{#2};{\bf P})}
\def\m#1{{\overline{\cal M}}_{0,0}(\vec{#1};{\bf P})}
\def\mnp#1{{\overline{\cal M}}_{0,0}((1,\vec{#1});{\bf P}^1 \times {\bf P})}

\def\p{{{\bf P}^1}}
\def\pp{{{\bf P}^2}}
\def\proj{{\bf P}}\def\tbs{\phantom{-}}


\overfullrule=0pt

\def\IP{{\bf P}}

\def\bfone{\relax{\rm 1\kern-.35em 1}}
\def\inbar{\vrule height1.5ex width.4pt depth0pt}
\def\IC{\relax\,\hbox{$\inbar\kern-.3em{\mss C}$}}
\def\ID{\relax{\rm I\kern-.18em D}}
\def\IF{\relax{\rm I\kern-.18em F}}
\def\IH{\relax{\rm I\kern-.18em H}}
\def\II{\relax{\rm I\kern-.17em I}}
\def\IN{\relax{\rm I\kern-.18em N}}
\def\IQ{\relax\,\hbox{$\inbar\kern-.3em{\rm Q}$}}
\def\us#1{\underline{#1}}
\def\IR{\relax{\rm I\kern-.18em R}}
\font\cmss=cmss10 \font\cmsss=cmss10 at 7pt
\def\ZZ{\relax\ifmmode\mathchoice
{\hbox{\cmss Z\kern-.4em Z}}{\hbox{\cmss Z\kern-.4em Z}}
{\lower.9pt\hbox{\cmsss Z\kern-.4em Z}}
{\lower1.2pt\hbox{\cmsss Z\kern-.4em Z}}\else{\cmss Z\kern-.4em
Z}\fi}
\def\nup#1({Nucl.\ Phys.\ $\us {B#1}$\ (}
\def\plt#1({Phys.\ Lett.\ $\us  {B#1}$\ (}
\def\cmp#1({Comm.\ Math.\ Phys.\ $\us  {#1}$\ (}
\def\prp#1({Phys.\ Rep.\ $\us  {#1}$\ (}
\def\prl#1({Phys.\ Rev.\ Lett.\ $\us  {#1}$\ (}
\def\prv#1({Phys.\ Rev.\ $\us  {#1}$\ (}
\def\mpl#1({Mod.\ Phys.\ Let.\ $\us  {A#1}$\ (}
\def\ijmp#1({Int.\ J.\ Mod.\ Phys.\ $\us{A#1}$\ (}
\def\tit#1|{{\it #1},\ }

\def\Coe#1.#2.{{#1\over #2}}

\def\coe#1.#2.{\relax{\textstyle {#1 \over #2}}\displaystyle}

\def\del{\partial}

\def\t#1{{\theta_#1}}

\def\dd{\rm d}

\lref\morvaf{D. R. Morrison and C. Vafa, ``Compactifications
of F-Theory on Calabi-Yau threefolds -- II,'' 
Nucl. Phys. {\bf B476} (1996) 437-469; hep-th/9603161.}

\lref\vafaoog{H. Ooguri and C. Vafa, ``Summing up Dirichlet Instantons,''
Phys. Rev. Lett. {\bf 77} (1996) 3296-3298.}
  
\lref\cfkmII{P. Candelas, A. Font, S. Katz, and D. R. Morrison
    ``Mirror Symmetry for Two Parameter Models -- II,'' 
      Nucl. Phys. {\bf B429} (1994) 626-674.}

\lref\cdfkmI{P. Candelas, X. de la Ossa, A. Font, S. Katz, and D. R. Morrison
    ``Mirror Symmetry for Two Parameter Models -- I,'' 
      Nucl.Phys. {\bf B416} (1994) 481-538.}

\lref\grabp{T. Graber, private communication.}

\lref\katzcox{D. A. Cox, and S. Katz, ``{\it Mirror Symmetry and Algebraic
Geometry},'' AMS, Providence, RI USA, 1999, to appear.}

\lref\vain{I. Vainsencher, ``Enumeration of n-fold Tangent
Hyperplanes to a Surface,'' J. Algebraic Geom. {\bf 4},
1995, 503-526, math.AG/9312012.}

\lref\fultoni{W. Fulton, ``{\it Intersection Theory},'' Second edition,
Springer-Verlag (Berlin) 1998.}

\lref\faber{C. Faber, ``Algorithm for Computing Intersection Numbers
on Moduli Spaces of Curves, with an Application to the Class of
the Locus of the Jacobians,'' math.AG/9706006.}

\lref\penglu{P. Lu, ``Special Lagrangian Tori on a
Borcea-Voisin Threefold,'' math.DG/9902063.}

\lref\vafagop{C. Vafa and R. Gopakumar, ``M-theory and
Topological Strings-I \& II,'' hep-th/9809187, hep-th/9812127.}

\lref\litian{J. Li and G. Tian, ``Virtual Moduli Cycles
and Gromov-Witten Invariants of Algebraic Varieties,'' 
J. Amer. Math. Soc. {\bf 11} (1998), no. 1, 119-174.}

\lref\pandh{R. Pandharipande, ``Hodge Integrals and
Degenerate Contributions,'' math.AG/9811140.}

\lref\klemmhg{A. Klemm, ``Mirror Symmetry at Higher Loops: A Precision 
Test for M-theory/IIA Duality,'' to appear.}

\lref\behfan{K. Behrend and B. Fantechi, ``The Intrinsic
Normal Cone,'' math.AG/9601010.}

\lref\locvir{T. Graber and R. Pandharipande, ``Localization
of Virtual Classes,'' math.AG/9708001.}

\lref\kontloc{M. Kontsevich, ``Enumeration of Rational Curves
via Torus Actions,'' in {\sl The Moduli Space of Curves,}
Dijkgraaf et al eds., Progress in Mathematics {\bf 129},
Birkh\"auser (Boston) 1995.}

\lref\gasd{ C. Vafa, ``Gas of D-Branes and Hagedorn Density of BPS States,''
Nucl. Phys. {\bfB463} (1996) 415-419; hep-th/9511088.}


\lref\yz{ S.-T. Yau and E. Zaslow, ``BPS States, String Duality, and Nodal
Curves on K3,'' Nucl. Phys. {\bf B471} (1996) 503-512; hep-th/9512121.}

\lref\yztwo{ S.-T. Yau
and E. Zaslow, ``BPS States as Symplectic Invariants from String
Theory,''  in {\sl Geometry and Physics,} Proceedings of the
Special Session on Geometry and Physics, Aarhus, Denmark, 1996.}


\lref\mcl{R. McLean, ``Deformations of Calibrated Submanifolds,"
Duke preprint 96-01:  www.math.duke.edu/preprints/1996.html.}

\lref\hl{F. R. Harvey and H. B. Lawson, ``Calibrated Geometries,''
Acta Math. {\bf 148} (1982) 47;
F. R. Harvey, {\sl Spinors and Calibrations,} Academic Press,
New York, 1990.}

\lref\ooguri{ K. Becker, M. Becker, D. R. Morrison, H. Ooguri,
Y. Oz, and Z. Yin, ``Supersymmetric Cycles in Exceptional Holonomy
Manifolds and Calabi-Yau Fourfolds,'' Nucl. Phys. {\bf B480} (1996)
225-238.}

\lref\vafmir{ C. Vafa, ``Extending Mirror Conjecture to Calabi-Yau
with Bundles,'' hep-th/9804131.}

\lref\zaslow{ E. Zaslow, ``Solitons and Helices:  The Search for 
a Math-Physics Bridge,'' Commun. Math. Phys. {\bf 175} (1996) 337-375.}

\lref\kach{ S. Kachru, A. Klemm, W. Lerche, P. Mayr, C. Vafa,
``Nonperturbative Results on the Point Particle Limit of N=2 Heterotic String,''
Nucl. Phys. {\bf B459} (1996) 537.}

\lref\klmvw{A. Klemm, W. Lerche, P. Mayr, C. Vafa and N. Warner, 
``Selfdual Strings and N=2 Supersymmetric Field Theory,''
Nucl. Phys. {\bf B477} (1996) 746-766, hep-th/9604034} 

\lref\helix{  For several articles on the subject of helices,
see Rudakov, A. N. et al, {\sl Helices and Vector Bundles.}
Seminaire Rudakov, London Mathematical Society Lecture Note
Series {\bf 148,} Cambridge University Press (Cambridge), 1990.}

\lref\kkv{ S. Katz, A. Klemm, and C. Vafa,
``Geometric Engineering of Quantum Field Theories,''
Nucl. Phys. {\bf B497} (1997) 173-195.}

\lref\wittdgrav{E.Witten,
``Two-dimensional Gravity and Intersection Theory on Moduli Space,''
Surveys in Differential Geometry {\bf 1} (1991) 243-310.}

\lref\lly{ B. Lian, K. Liu, and S.-T. Yau, ``Mirror Principle I,''
Asian J. of Math. Vol. {\bf 1} No. 4 (1997) 729-763; math.AG/9712011.}

\lref\llyt{ B. Lian, K. Liu, and S.-T. Yau, ``Mirror Principle II,''
in preparation.}

\lref\givental{ A. Givental, ``A Mirror Theorem for Toric Complete 
Intersections,''
{\it Topological Field Theory, Primitive Forms and Related Topics (Kyoto,
1996)}, Prog. Math. {\bf 160}, 141-175; math.AG/9701016.}

\lref\syz{ A. Strominger,
S.-T. Yau, and E. Zaslow, ``Mirror Symmetry is T-Duality,''
Nuclear Physics {\bf B479} (1996) 243-259; hep-th/9606040.}

\lref\slag{ D. Morrison, ``The Geometry Underlying Mirror Symmetry,''
math.AG/9608006;
M. Gross and P. Wilson, ``Mirror Symmetry via 3-tori for a Class of Calabi-Yau
Threefolds,'' to appear in Math. Ann., math.AG/9608009;  B. Acharya, ``A
Mirror Pair of Calabi-Yau Fourfolds in Type II String Theory,'' Nucl. Phys.
{\bf B524} (1998) 283-294, hep-th/9703029;
N. C. Leung and C. Vafa, ``Branes and Toric Geometry,'' Adv. Theor.
Math.
Phys. {\bf 2} (1998) 91-118, hep-th/9711013;
N. Hitchin, ``The Moduli Space of Special Lagrangian Submanifolds,''
math.DG/9711002.}

\lref\kontsevich{ M. Kontsevich, ``Homological Algebra of Mirror Symmetry,''
Proceedings of the 1994 International Congress of Mathematicians {\bf I},
Birk\"auser, Z\"urich, 1995, p. 120;
math.AG/9411018.}

\lref\kmvt{ S. Katz, P. Mayr and C. Vafa, ``Mirror Symmetry and
Exact Solution of 4D N=2 Gauge Theories - I,'' Adv. Theor. 
Math. Phys. {\bf 1} (1997) 53-114.}

\lref\kkmv{ S. Katz, A. Klemm, and C. Vafa, ``Geometric Engineering of
Quantum Field Theories,'' Nucl. Phys. {\bf B497} (1997) 173-195, hep-th/9609239;
and A. Klemm, P. Mayr, and C. Vafa, ``BPS States of Exceptional Non-Critical 
Strings,''
hep-th/9607139.}

\lref\guzz{ D. Guzzetti, ``Stokes Matrices and Monodromy for the
Quantum Cohomology of Projective Spaces,'' preprint SISSA 87/98/FM.}

\lref\dub{ B. Dubrovin,
{\sl Geometry of 2D Topological Field Theories,}
Lecture Notes in Math {\bf 1620} (1996) 120-348.}

\lref\ttstar{ S. Cecotti and C. Vafa, ``Topological Anti-Topological
Fusion,'' Nucl. Phys. {\bf B367} (1991) 359-461.}

\lref\class{ S. Cecotti and C. Vafa, ``On Classification of
N=2 Supersymmetric Theories,'' Commun. Math. Phys. {\bf 158} (1993) 569-644.}

\lref\dubconj{ ``Geometry and Analytic Theory of Frobenius
Manifolds,'' math.AG/9807034.}

\lref\ad{ P. Aspinwall and R. Donagi,
``The Heterotic String, the Tangent Bundle,
and Derived Categories,'' hep-th/9806094.}

\lref\candelas{P.\  Candelas, X. C. De La Ossa, P. S. Green, L. Parkes, ``A Pair 
of Calabi-Yau Manifolds as an Exactly Soluble Superconformal Theory,'' Nucl.Phys. 
{\bf B359} (1991) 21-74.}

\lref\batyrev{V.\ Batyrev, ``Dual Polyhedra and Mirror Symmetry for Calabi-Yau 
Hypersurfaces 
in Toric Varieties,'' J. Algebraic Geom. {\bf 3} (1994) 493-535.}

\lref\koelman{R.\ J.\  Koelman, ``A criterion for the ideal of a projectively embedded toric 
surface to be generated by quadrics'', Beitr\"age  Algebra Geom. 34 (1993), no. 1, 57-62.}

\lref\batyrevII{V.\ Batyrev, ``Variations of the Mixed Hodge Structure 
of Affine Hypersurfaces in Algebraic Tori,''
Duke Math. Jour. {\bf 69}, 2 (1993) 349.}

\lref\hkty{ S.\  Hosono, A.\  Klemm, S.\ Theisen and S.T. Yau, 
``Mirror Symmetry, Mirror Map and Applications to Calabi-Yau Hypersurfaces,''
Commun. Math. Phys. {\bf 167} (1995) 301-350, hep-th/9308122;
and ``Mirror Symmetry, Mirror Map and Applications to Calabi-Yau Hypersurfaces,'' 
Nucl. Phys. {\bf B433} (1995) 
501-554, hep-th/9406055.} 

\lref\cox{D.\ Cox, ``The Homogeneous Coordinate Ring of a Toric Variety,''
J.\ Alg. Geom {\bf 4} (1995) 17; math.AG/9206011}

\lref\danilov{V. I. Danilov, {\sl The Geometry of Toric Varieties,} Russian Math. 
Surveys,{\bf  33} (1978) 97.}

\lref\oda{T.\ Oda,  {\sl Convex Bodies and Algebraic Geometry, An Introduction to 
the 
Theory of Toric Varieties,} Ergennisse der Mathematik und ihrer Grenzgebiete, 3. 
Folge, Bd. {\bf 15}, 
Springer-Verlag (Berlin) 1988.}

\lref\fulton{W.\ Fulton, {\sl Introduction to Toric Varieties,} Princeton Univ. 
Press {\bf 131} (Princeton) 1993.}   
 
\lref\batyrevborisov{V.\ Batyrev and L.\ Borisov, ``On Calabi-Yau Complete 
Intersections in 
Toric Varieties,'' in {\sl Higher-dimensional Complex Varieties,}
(Trento, 1994), 39-65, de Gruyter  (Berlin) 1996.}

\lref\secondaryfan{I. M. Gel'fand, M. Kapranov, A. Zelevinsky, {\sl 
Multidimensional Determinants,
Discriminants and Resultants}, Birkh\"auser (Boston) 1994; 
L. Billera, J. Filiman and B.\ Sturmfels, ``
Constructions and Complexity of Secondary Polytopes,'' Adv. Math. {\bf 83} 
(1990),  155-17.}

\lref\kmv{A. Klemm, P. Mayr, and C. Vafa, ``BPS States of Exceptional Non-Critical 
Strings,'' Nucl. Phys. {\bf B}
(Proc. Suppl.) 58 (1997) 177-194; hep-th/9607139.}

\lref\mnw{J. A. Minahan, D. Nemechansky and N. P. Warner, ``Partition function for 
the BPS States 
of the Non-Critical $E_8$ String,'' Adv. Theor. Math.Phys. {\bf 1} (1998) 167-183; 
hep-th/9707149 and
``Investigating the BPS spectrum of the Non-Critical $E_n$ strings,'' 
Nucl.Phys. {\bf B508} 
(1997) 64-106;  hep-th/9705237.} 

\lref\mnvw{J. A. Minahan, D. Nemechansky, C. Vafa and N. P. Warner,
``$E$-Strings and $N=4$ Topological Yang-Mills Theories,''
Nucl. Phys. {\bf B527} (1998) 581-623.}

\lref\lmw{W. Lerche, P. Mayr, and N. P. Warner,
``Non-Critical Strings, Del Pezzo Singularities
and Seiberg-Witten Curves,'' hep-th/9612085.}

\lref\morrison{D.\ Morrison, 
``Where is the large Radius Limit?'' Int. Conf. on Strings 93, 
Berkeley; hep-th/9311049}
\lref\griffith{P. Griffiths, ``On the Periods of certain
Rational Integrals,'' Ann. Math. {\bf 90} (1969) 460.}

\lref\klty{A. Klemm, W. Lerche, S. Theisen and S. Yankielowicz,        
      ``Simple Singularities and N=2 Supersymmetric Yang-Mills Theory,'' 
        Phys. Lett. {\bf B344} (1995) 169; A. Klemm, W. Lerche, S. Theisen,
       ``Nonperturbative Effective Actions of N=2 Supersymmetric Gauge Theories,'' 
       J.Mod.Phys. {\bf A11} (1996) 1929-1974.}

\lref\linearsigmamodel{E. Witten, ``Phases of $N=2$ Theories in Two 
Dimensions,'' Nucl. Phys. {\bf B403} (1993) 159.}

\lref\klry{A. Klemm, B. Lian, S.-S. Roan, S.-T. Yau. ``
         Calabi-Yau fourfolds for M- and F-Theory compactifications, ''
          Nucl.Phys. {\bf B518} (1998) 515-574.}

\lref\gmp{B. R. Greene, D. R. Morrison, M. R. Plesser, ``Mirror Manifolds in 
Higher Dimension,'' 
          Commun. Math. Phys. {\bf 173} (1995) 559-598.}

\lref\mayr{P. Mayr, ``N=1 Superpotentials and Tensionless Strings on Calabi-Yau 
Four-Folds,''
           Nucl.Phys. {\bf B494} (1997) 489-545.}

\lref\lv{ N. C. Leung and C. Vafa, ``Branes and Toric Geometry,'' Adv. Theor.Math.
Phys. {\bf 2} (1998) 91-118, hep-th/9711013.}

\lref\ln{A. Lawrence and N. Nekrasov, ``Instanton sums and five-dimensional 
theories,'' 
Nucl. Phys. {\bf B513} (1998) 93.}

\lref\bcov{M. Bershadsky, S. Cecotti, H. Ooguri, C. Vafa (with an appendix by S. Katz), 
      ``Holomorphic Anomalies in Topological Field Theories'',
        Nucl. Phys. {\bf B405} (1993) 279-304.}

\def\bP{{\bf P}}

\def\nec#1{\overline{NE}(#1)}
  
%
\Title{\vbox{\hbox{hep-th/9903053}
\hbox{IASSNS-HEP--99/26}}}
{\vbox{\centerline{Local Mirror Symmetry:}
\vskip 0.2in
\centerline{Calculations and Interpretations}}}
\vskip 0.2in
\centerline{T.-M. Chiang,\footnote{$^{\dagger}$}
{email:  chi@math.harvard.edu, yau@math.harvard.edu} A. Klemm,\footnote{$^{*}$}
{email:  klemm@ias.edu} S.-T. Yau,$^\dagger$ and E. Zaslow\footnote{$^{**}$}
{email:  zaslow@math.nwu.edu}}
\vskip 0.1in
\centerline{\it$^{\dagger}$Department of Mathematics, Harvard University,
Cambridge, MA 02138, USA}
\centerline{\it$^{*}$School of Natural Sciences, IAS, Olden Lane, Princeton, NJ 
08540, USA}
\centerline{\it$^{**}$Department of Mathematics, Northwestern University,
Evanston, IL 60208, USA}

\vskip 0.3 in

\centerline{\bf Abstract}
\vskip 0.1in

We describe local mirror symmetry from a mathematical
point of view and make several A-model calculations using
the mirror principle (localization).  Our results agree
with B-model computations from solutions
of Picard-Fuchs differential equations
constructed form the local
geometry near a Fano surface within a Calabi-Yau manifold.
We interpret the Gromov-Witten-type numbers from
an enumerative point of view.  We also describe the geometry
of singular surfaces and show how the local
invariants of singular surfaces
agree with the smooth cases when they occur as complete
intersections.
  
\Date{}

\centerline{{\bf Table of Contents}}
\nobreak
 $$\eqalign{
&\hbox{\bf 1.  Introduction}\cr
&\hbox{\bf 2.  Overview of the A-Model}\cr
&\qquad\hbox{\sl 2.1  ${\cal O}(5)\rightarrow {\bf P}^4$} \cr
&\qquad\hbox{\sl 2.2  ${\cal O}(-3)\rightarrow {\bf P}^2$}\cr
&\hbox{\bf 3. The Mirror Principle for General Toric Manifolds}\cr
&\qquad\hbox{\sl 3.1 Fixed points and a gluing identity}\cr
&\qquad\hbox{\sl 3.2 The spaces $M_{\vec d}$ and $N_{\vec d}$}\cr
&\qquad\hbox{\sl 3.3 Euler data}\cr
&\qquad\hbox{\sl 3.4 Linked Euler data}\cr
&\hbox{\bf 4.  Explicit Verification through Fixed-Point Methods}\cr
&\qquad\hbox{\sl 4.1 Some examples} \cr
&\qquad\hbox{\sl 4.2 General procedure for fixed-point computations} \cr
&\hbox{\bf 5.  Virtual Class and the Excess Intersection Formula}\cr
&\qquad\hbox{\sl 5.1  Rational curves on the quintic threefold}\cr
&\qquad\hbox{\sl 5.2  Calabi-Yau threefolds containing an algebraic
surface}\cr
&\qquad\hbox{\sl 5.3 Singular geometries}\cr
&\hbox{\bf 6.  Local Mirror Symmetry:  The B-Model}\cr
&\qquad\hbox{\sl 6.1  Periods
and differential equations for global mirror symmetry}\cr
&\qquad\hbox{\sl 6.2 The limit of the large elliptic fiber}\cr
&\qquad\hbox{\sl 6.3 Local mirror symmetry for the canonical line bundle of
a torically described surface}\cr
&\qquad\hbox{\sl 6.4 Fibered $A_n$ cases and more general toric grid
diagrams}\cr
&\qquad\hbox{\sl 6.5 Cases with constraints}\cr
&\hbox{\bf 7.  Discussion}\cr
&\hbox{\bf 8.  Appendix:  A-Model Examples}\cr
&\qquad\hbox{\sl 8.1  ${\cal O}(4)\rightarrow {\bf P}^3$}\cr
&\qquad\hbox{\sl 8.2  ${\cal O}(3)\rightarrow {\bf P}^2$}\cr
&\qquad\hbox{\sl 8.3  $K_{F_n}$}\cr
}$$
\vfill
\eject

\newsec{Introduction}
  
``Local mirror symmetry'' refers to a specialization of
mirror symmetry techniques to address the geometry
of Fano surfaces within Calabi-Yau manifolds.  The procedure
produces certain ``invariants'' associated to the surfaces.
This paper is concerned with the proper definition and
interpretation of these invariants.  The techniques
we develop are a synthesis of 
results of previous works (see \kach, \kmv, \kkv, \kmvt, \mnvw), with 
several new constructions. We have not found a cohesive explanation
of local mirror symmetry in the literature.  We offer
this description in the hope that it will add to
our understanding of the subject and perhaps help to advance
local mirror symmetry towards higher genus computations.

Mirror symmetry, or the calculation of Gromov-Witten
invariants in Calabi-Yau threefolds,\foot{We restrict the
term ``mirror symmetry'' to mean an equivalence of quantum
rings, rather than the more physical interpretation as an
isomorphism of conformal field theories.} can now be approached
in the traditional (``B-model'') way or by using localization techniques.
The traditional approach involves
solving the Picard-Fuchs equations
governing the behavior of period integrals
of a Calabi-Yau manifold under deformations
of complex structure, and converting the coefficients of the
solutions near a point of maximal monodromy into Gromov-Witten
invariants of the mirror manifold.  Localization techniques, first developed by
Kontsevich \kontloc\ and then improved by others
\givental \lly, offer a
proof -- without reference to a mirror manifold --
that the numbers one obtains in this way are
indeed the Gromov-Witten invariants as defined
via the moduli space of maps.

Likewise, local mirror symmetry has these two approaches. One finds
that the mirror geometry is a Riemann surface with a meromorphic differential. 
{}From this one is  able to derive differential equations which 
yield the appropriate numerical invariants.  Recall the geometry.
We wish to study a neighborhood of a surface $S$
in a Calabi-Yau threefold $X$, then take a limit where
this surface shrinks to zero size. 
In the first papers on the subject,
these equations were derived by first finding a
Calabi-Yau manifold containing the surface, then  
finding its mirror and ``specializing'' the
Picard-Fuchs equations by taking an appropriate
limit corresponding to the local geometry.
Learning from this work, one is now able to write
down the differential equations directly from
the geometry of the surface (if it is toric).
We use this method to perform our B-model calculations.

We employ a localization approach developed in \lly\
for computing the Gromov-Witten-type invariants 
directly (the ``A-model'').  Since the adjunction
formula and the Calabi-Yau condition of $X$
tell us that the normal bundle of the
surface is equal to the canonical bundle (in the
smooth case), the local geometry is intrinsic
to the surface.  We define the Gromov-Witten-type
invariants directly from $K_S,$ following \givental\ \lly.
We require $S$ to be Fano (this should be related to
the condition that
$S$ be able to vanish in $X$), which makes the bundle
$K_S$ ``concave,'' thus allowing us to construct
cohomology classes on moduli space of maps.
We consider the numbers constructed in this way
to be of Gromov-Witten type.

In section two,
we review the mirror principle and apply it
to the calculation of invariants for several surfaces.
In section three, we give the general procedure for toric varieties.
We then calculate the invariants ``by hand''
for a few cases, as a way
of checking and illucidating the procedure.
In section five, we describe the excess
intersection formula and show that the local
invariants simply account for the effective
contribution to the number of curves in a
Calabi-Yau manifold due to the
presence of a holomorphic surface.

In section six, we develop all the machinery
for performing B-model calculations without
resorting to a specialization of period equations from
a compact Calabi-Yau threefold containing the relevant local geometry.
Actually, a natural Weierstrass compactification exists
for toric Fano geometries, and its decompactification
(the limit of large elliptic fiber) produces expressions
intrinsic to the surface.  In this sense, the end result
makes no use of compact data.  The procedure closely
resembles the compact B-model technique of solving
differential equations and taking combinations of
solutions with different singular behaviors to produce
a prepotential containing enumerative invariants as
coefficients.  Many examples are included.

\vskip0.1in
$\underline{\qquad\qquad\qquad\qquad}$
\vskip0.15in

In order to accommodate readers with either mathematical
or physical backgrounds, we have tried to be reasonably
self-contained and have included several examples written
out in considerable detail.  Algebraic geometers may find
these sections tedious, and may content themselves with
the more general sections (e.g., $3, 4.2, 6.3$).  Physicists
wishing to get a feel for the mathematics of A-model computations
may choose to focus on the examples of section $4.1.$

\newsec{Overview of the A-model}

In this section, we review the techniques for calculating
invariants using localization.  We will derive the numbers
and speak loosely about their interpretation, leaving
more rigorous explanations and interpretations
for later sections.

For smooth hypersurfaces in toric varieties, we define the Gromov-Witten
invariants to be Chern classes of certain bundles over the moduli
space of maps, defined as follows.  Let $\m{d}$ be Kontsevich's
moduli space of stable maps of genus zero with no marked points.
A point in this space will be denoted $(C,f),$ where $f: C\rightarrow
{\bf P},$ ${\bf P}$ is some toric variety, and $[f(C)] = \vec{d} \in H_2({\bf
P}).$
Let $\mp{d}$ be the same but with one marked point.  Consider the diagram
$$\m{d} \longleftarrow \mp{d} \longrightarrow {\bf P},$$
where ${\bf P}$ is the toric variety in question,
$$ev : \mp{d} \longrightarrow {\bf P}$$
is the evaluation map sending $(C,f,*) \mapsto f(*),$
and
$$\rho : \mp{d} \longrightarrow \m{d}$$ is the 
forgetting map sending $(C,f,*) \mapsto (C,f).$
Let $Q$ be a Calabi-Yau defined as the zero locus of sections of a
convex\foot{``Convex'' means that $H^1(C,f^*V) = 0$ for $(C,f)\in \m{d}.$
For the simplest example, ${\bf P} = {\bf P}^4$ and $V = {\cal O}(5),$
as in the next subsection.}
bundle $V$ over ${\bf P}.$  Then $U_{\vec d}$ is the bundle over $\m{d}$
defined by
$$U_{\vec d} = \rho_* ev^* V.$$
The fibers of $U_{\vec d}$ over $(C,f)$ are $H^0(C,f^*V).$
We define the Kontsevich numbers $K_d$ by
$$K_d \equiv \int_{\m{d}} c(U_{\vec d}).$$
It is most desirable when $\dim \m{d} = {\rm rank} U_{\vec d},$ so that $K_d$
is the top Chern class.

The mirror principle is a procedure for evaluating the
numbers $K_d$ by a fancy version of localization.  The idea,
pursued in the next section, is as follows.  When all spaces and
bundles are torically described, the moduli spaces and the bundles
we construct over them inherit torus actions (e.g., by moving the image
curve).  Thus, the integrals we define can be localized to the fixed
point loci.
As we shall see in the next section, the multiplicativity of the
characteristic classes we compute implies relations among their
restrictions
to the fixed loci.  The reason for this is that the fixed loci of degree
$\gamma$ maps includes stable curves constructed by gluing degree
$\alpha$ and $\beta$ maps, with $\alpha + \beta = \gamma.$
One then constructs an equivariant map to a ``linear sigma model,''
which is an easily described toric space.  Indeed, the linear sigma model
is another compactification of the smooth stable maps, which
can be modeled as polynomial maps.  We then
push/pull our problem to this linear sigma model, where
the same gluing relations are found to hold.  The notion
of Euler data is any set of characteristic classes on the
linear sigma model obeying these relations.  They are not
strong enough to uniquely determine the classes, but
as the equivariant cohomology can be modeled as polynomials,
two sets of Euler data which agree upon restriction to enough
points may be thought of as equivalent (``linked Euler data'').
It is not difficult to construct Euler data linked to the
Euler data of the characteristic classes in which we are
interested.  Relating the linked Euler data, and therefore
solving the problem in terms of simply-constructed polynomial
classes, is done via a mirror transform, which involves
hypergeometric series familiar to B-model computations.
However, no B-model constructions are used.  These polynomial
classes can easily be integrated, the answers then
related to the numbers in which we are interested by the
mirror transform.  This procedure is used to evaluate
the examples in this section which follow, as well as all
other A-model calculations.

Using the techniques of the mirror principle, we are able to 
build Euler data from many bundles over toric varieties.
Typically, we have a direct sum of $\bigoplus_i{\cal O}(l_i)$
and $\bigoplus_j{\cal O}(-k_j)$ over ${\bf P}^n,$ with $l_i, k_j > 0.$
In such a case, if $\sum_i l_i + \sum_j k_j = n + 1$ then we 
can obtain linked Euler data for the bundle $U_{\vec d}$ 
whose fibers over a point $(f,C)$ in 
$\overline{{\cal M}_{0,0}}(d ; {\bf P}^n)$ is a direct sum of
$\bigoplus_i H^0(C,f^*{\cal O}(l_i))$ and
$\bigoplus_j H^1(C,f^*{\cal O}(k_j)).$
In this situation, the rank of the bundle
(which is $\sum_i (dl_i + 1) + \sum_j (dk_j - 1)$) may be greater than the
dimension $D$ of $\overline{{\cal M}_{0,0}}(d ; {\bf P}^n)$ (which is $D \equiv
(n+1)(d+1) - 4$).
In that case, we compute the integral over moduli
space of the Chern class $c_D(U_{\vec d}).$  The interpretations will
be discussed in the examples.

We begin with a convex bundle.

\subsec{${\cal O}(5) \rightarrow {\bf P}^4$}

Recall that this is the classic mirror symmetry calculation.
We compute this by using the Euler data $P_d = \prod_{j=1}^{5d} (5 H - m)$
As the rank of $U_d$ equals the dimension of moduli space,
we take the top Chern class of the bundle and call this
$K_d.$  This has the standard
interpretation:  given a generic section with isolated zeros,
the Chern class counts the number of zeros.  If we take as
a section the pull back of a quintic polynomial
(which is a global section of ${\cal O}(5)$), then its zeros
will be curves $(C,f)$ on which the section vanishes identically.
As the section vanishes along a Calabi-Yau quintic threefold, the
curve must be mapped (with degree $d$) into the quintic -- thus
we have the interpretation as ``number of rational curves.''  However,
the contribution of curves of degree $d/k,$ when $k$ divides $d,$
is also non-zero.  In this case, we can compose any $k-$fold cover of
the curve $C$ with a map $f$ of degree $d/k$ into the quintic.
This contribution is often called the ``excess intersection.''
To calculate the contribution to the Chern class, we must look at how
this space of $k-$fold covers of $C$ (which, as $C \cong {\bf P}^1$ in
the smooth case, is equal to $\overline{{\cal M}_{0,0}}(k,{\bf P}^1)$) sits in
the moduli space (i.e. look at its normal bundle).  This calculation
yields $1/k^3,$ and so if $n_d$ is the number of rational curves of
degree $d$ in the quintic, we actually count
\eqn\doubcov{K_d = \sum_{k\vert d}{n_{d/k}\over k^3}.}
This double cover formula will be discussed in detail
in section 5.1.

In the appendix, several examples of other bundles over projective
spaces are worked out.  In the cases where the rank of $U_d$ is
greater (by $n$) than the dimension of moduli space, we take the highest
Chern class that we can integate.  The resulting numbers count
the number of zeros of $s_0 \wedge ... \wedge s_n,$ i.e. the places
where $n + 1$ generic sections gain linear dependencies.  A zero of
$s_0 \wedge ... \wedge s_n$ represents a point $(C,f)$ in moduli
space where $f(C)$ vanishes somewhere in the $n$-dimensional
linear system
of $s_0, ..., s_n.$  See the appendix for details.
We turn now to the study of some concave bundles.

\subsec{${\cal O}(-3) \rightarrow {\bf P}^2$}

We think of ${\cal O}(-3)$ as $K_{{\bf P}^2},$ the canonical bundle. 
This case is relevant
to Calabi-Yau manifolds containing projective surfaces.
A tubular neighborhood of the surface is equivalent to the 
total space of the canonical bundle (by the adjunction formula and
the Calabi-Yau condition $c_1 = 0$).

In this case, the rank of $U_d$ (which is the
bundle whose fiber over $(C,f)$ is $H^1(C,f^*K_{{\bf P}^2})$)
is equal to the dimension of moduli
space, so we are computing the top Chern class
$K_d = \int_{\overline{{\cal M}_{0,0}}(d,{\bf P}^2)} c_{3d-1}(U_d).$ From
the $K_d$'s we arrive at the following $n_d$'s.

{\vbox{\ninepoint{
$$        
  \vbox{\offinterlineskip        
  \hrule        
  \halign{ &\vrule# & \strut\quad\hfil#\quad\cr     
  \noalign{\hrule}        
  height1pt&\omit&   &\omit&\cr        
  &$d$&& $n_d$ &\cr       
  \noalign{\hrule}        
  &1&& $3$ &\cr        
  &2&& $-6$ &\cr        
  &3&& $27$ &\cr        
  &4&& $-192$ &\cr        
  &5&& $1695$ &\cr        
  &6&& $-17064$ &\cr        
  &7&& $188454$ &\cr        
  &8&& $-2228160$ &\cr        
  &9&& $27748899$ &\cr        
 &10&& $-360012150$ &\cr        
  height1pt&\omit&   &\omit&\cr        
\noalign{\hrule}}}
$$
\vskip-7pt
\noindent
\centerline{{\bf Table 1}: Local invariants for $K_{{\bf P}^2}$}
\vskip7pt}}}

The interpretation for the $n_d$'s
is not as evident as for positive bundles, since no sections of $U_d$
can be pulled back from sections of the canonical bundle (which has no
sections).  Instead, we have the following interpretation.

Suppose the ${\bf P}^2$ exists within a Calabi-Yau manifold, and we are
trying to count the number of curves in the
same homology class as $d$ times the hyperplane
in ${\bf P}^2.$  The analysis for the Calabi-Yau would go along the
lines of the quintic above.  However, there would necessarily be
new families of zeros of your section
corresponding to the families of degree $d$
curves in the ${\bf P}^2$ within the Calabi-Yau.  These new
families would be isomorphic to $\overline{{\cal M}_{0,0}}(d,{\bf P}^2).$
On this space, we have to compute the contribution to the total
Chern class.  To do this, we would need to use the excess intersection
formula.  The result (see section 5.2) is precisely given
by the $K_d.$  Let us call once again
$n_d$ the integers derived from the $K_d$. 
Suppose now that we have two Calabi-Yau's, $X_0$ and $X_1,$
in the same family of
complex structures, one of which (say $X_1$ contains) a ${\bf P}^2.$  Then
the difference between $n_d(X_0)$ and $n_d(X_1)$ should be given by
the $n_d.$  

There are Calabi-Yau's, however, which generically contain a ${\bf P}^2$.
The simplest examples are the following elliptic fibrations over a $\bP^2$, 
A.) the degree 18 hypersurface in  denoted by 
${\bf P}_{6,9,1,1,1}[18]$, with 
$\chi=-540$, $h^{21}=272$, $h^{11}=2(0)$
B.) ${\bf P}_{3,6,1,1,1}[12]$,
$\chi=-324$, $h^{21}=165$, $h^{11}=3(1)$
C.) ${\bf P}_{3,3,1,1,1}[9]$,
$\chi=-216$, $h^{21}=112$, $h^{11}=4(2)$. $h^{11}$ contributions in brackets are 
non-toric divisors, in these cases they correspond to additional components
of the section, see below.

{\vbox{\ninepoint{
$$
\vbox{\offinterlineskip\tabskip=0pt
\halign{\strut\vrule#
&\hfil~$#$
&\vrule#&~
\hfil ~$#$~
&\hfil ~$#$~
&\hfil $#$~
&\hfil $#$~
&\vrule#\cr
\noalign{\hrule}
& && &  &    A&   & \cr
\noalign{\hrule}
& && d_F& 0 &  1  &  2 & \cr
\noalign{\hrule}
&d_B &&  & &        &       &      \cr
&0   && & &  540   &      540  &     \cr
&1   &&  &3& -1080       &    143370    &  \cr
&2   &&  &-6&  2700      &   -574560     &  \cr
&3   &&  &27 &  -17280     &   5051970     &  \cr
&4   &&  &-192& 154440   &    -57879900    &\cr
\noalign{\hrule}}
\hrule}
\vbox{\offinterlineskip\tabskip=0pt
\halign{
\strut\vrule#
&\hfil~$#$
&\hfil $#$
&\hfil $#$
&\vrule#\cr
\noalign{\hrule}
&     &       B&      &     \cr
\noalign{\hrule}
&    0&       1&      2 &      \cr
\noalign{\hrule}
&     &      &       &                     \cr	
&     &     216&     324&     \cr
&    6&    -432&   10260& \cr
&  -12&    1080&  -41688& \cr
&  54 &   -6912&  378756& \cr
&-384 &   61776&-4411260& \cr 
\noalign{\hrule}}
\hrule}
\vbox{\offinterlineskip\tabskip=0pt
\halign{
\strut\vrule#
&\hfil ~$#$~
&\hfil $#$
&\hfil $#$
&\vrule#\cr
\noalign{\hrule}
&       &  C  &   &        \cr
\noalign{\hrule}
& 0       &  1   &  2 &        \cr
\noalign{\hrule}
 &     &      &       &                     \cr
 &     &   162&    162&         \cr
 &    9&  -324&   3645&   \cr
 &  -18&   810& -14904&    \cr
 &  81 & -5184& 137781&   \cr
 & -576& 46332&-1617570&\cr 
\noalign{\hrule}}
\hrule}$$
\vskip-7pt
\noindent
{\bf Table 2}: Invariants of the three elliptic fibrations over ${\bf P}^2$  
($B$ and $F$ denote the class of
a section and the elliptic fiber, respectively.)
\vskip7pt}}

For such Calabi-Yau's the Gromov-Witten invariants of the homology
class of the base would be a multiple of the invariants
for $K_{base}$. In the above examples, we see from the first
column the different multiples which arise as we are counting curves in
the homology class of a curve which is dual to the hyperplane
class of the base ${\bf P}^2$. This homology class sits inside
a section of the elliptic fibration, and the multiplicities come from
the fact that the $A,B,C$  fibrations admit $1,2,3$ sections. For example, if 
we write the ambient toric variety for case C as $\bP({\cal O}_{\bP^2} \oplus
{\cal O}_{\bP^2} \oplus {{\cal O}_{\bP^2}(-3)})$, a section is given by 
one of the three components of  the vanishing locus of the coordinate 
on the fiber that transforms as ${{\cal O}(-3)}$ over $\bP^2$. 

Another interpretation of this number is as follows.  The space
$H^1(C,f^*K)$ represents obstructions to deformations of the curve $C.$
Therefore, the top Chern class of the bundle whose fibers are $H^1(C,f^*K)$
represents the number of
infinitesimal deformations in the family which represent finite
deformations.  Note this interpretation is equivalent to the one above.
The numbers represent the effective number of curves of
 degree $d$ in the Calabi-Yau ``coming from'' the ${\bf P}^2.$

This procedure can be performed for any Fano surface.  The Hirzebruch
(rational, ruled) surfaces are described in the Appendix A.  Next we
discuss the general toric case.

\newsec{The Mirror Principle for General Toric Manifolds}

	In this section, we review the mirror principle
for computations of Gromov-Witten invariants of
a toric variety. For a summary of what follows,
we refer the reader to the start of section 2. 
 Our treatment is somewhat more
general than that of \lly, as we consider general toric varieties,
though we omit some proofs which will be included in \llyt\ .
	
	Throughout this section, we take our target manifold
to be a smooth, toric and projective manifold $\bP$.
That is, we are interested in rational curves that map
into $\bP$. Let us write $\bP$ as a quotient
of an open affine variety:

$$ \bP = {{{\bf C}^{N_C} - \Delta} \over {G}},$$
where $G \cong ({\bf C}^*)^{N_C - M}$.
We can write the $i$th action of $G$ as

$$(x_1,...,x_{N_C}) \rightarrow (\nu^{q_{i,1}} x_1,...,\nu^{q_{i,N_C}} x_{N_C}),$$
where $\nu$ is an arbitrary element of ${\bf C}^*$.
There is a $T \equiv (S^1)^{N_C}$ action on
$\bP$ induced from its usual action on
${\bf C}^{N_C}$. This action has $N_C$ fixed points
which we denote by $p_1,...,p_{N_C}$. 
For example, for $\bP = {\bf P}^4$,
${\bf T}$ is $(S^1)^5$ and the fixed points
are the points with one coordinate nonvanishing.

	The ${\bf T}$-equivariant cohomology ring
can be obtained from the ordinary ring as follows.
Write the ordinary ring as a quotient

$$ {{\bf Q}[B_1, ..., B_{N_C}] \over I}, $$
where $B_i$ is the divisor class of ${x_i = 0}$ and $I$ is an ideal 
generated by elements homogeneous in the $B_i$'s. For example,
for ${\bf P}^4$ we have the ring

$$ {{\bf Q}[B_1,..., B_5] \over (B_1 - B_2, B_1 - B_3, B_1 - B_4,
B_1 - B_5, B_1 B_2 B_3 B_4 B_5)}. $$
Let $J_i, i = 1,...,M$ be the basis of nef divisors
in $H^2(\bP, {\bf Z})$. We can write the
$B_l$'s in terms of the $J_k$'s:

$$ B_i = \sum b_{ij} J_j. $$
The equivariant ring is then

$$ H_{\bf T}(\bP) = {{{\bf Q}[\kappa_1, ..., \kappa_{M},
\lambda_1,..., \lambda_{N_C}] \over I_{\bf T}}}, $$
where $I_{\bf T}$ is generated by $\sum q_{i,j} \lambda_j$ for
$i = 1,...,N_C$ and the nonlinear relations in $I$ with $B_i$ replaced by
$\sum b_{ij} \kappa_j - \lambda_i$. In the case of ${\bf P}^4$ this
is

$$ {{\bf Q}[\kappa, \lambda_1,..., \lambda_5] \over (\prod_{i=1}^{5} (\kappa - 
\lambda_i), \sum \lambda_i)}.$$
Clearly, setting $\lambda_i$ to zero in $H_{\bf T}(\bP)$ gives us
the ordinary ring in which $\kappa_j$ can be identified 
with $J_j$.

	Having described the base and the torus
action, we also need a bundle $V$ to define
the appropriate Gromov-Witten problem. For instance,
if we are interested in rational curves 
in a complete intersection of divisors in $\bP$, then
$V$ is a direct sum of the associated
line bundles. For local mirror symmetry, we can also
take a concave line bundle as a component of $V$.
More generally, we take $V = V^+ \oplus V^-$, with
$V^+$ convex and $V^-$ concave.

	Before proceeding to the next section,
we introduce some notation for later use.
Let $F_j$ be the associated divisors
of the line bundle summands of $V$, by
associating each line bundle to a divisor in the
usual way. We write $F_j$ as greater or less than zero,
depending on whether it is convex or concave.
Homology classes of curves in $\bP$ 
will be written in the basis $H_j$ Poincar\'e dual to $J_j$.
For instance, $\m{d}$ is the moduli space of stable
maps with image homology class $\sum d_i H_i$.
Finally, $x$ denotes a formal variable
for the total Chern class.

\subsec{Fixed points and a Gluing Identity}

	The pull-back of $V$ to $\mp{d}$
by the evaluation map gives a bundle of the form
$ev^*(V^+) \oplus ev^*(V^-)$. Then, in terms of the
forgetful map from $\mp{d}$ to $\m{d}$, we obtain a bundle on $\m{d}$
$\rho_{*} ev^*(V^+) \oplus {\cal R}^1 \rho_{*}
ev^*(V^-)$. The latter is the obstruction bundle $U_{\vec d}$.

	On $\m{d}$, there is a torus action induced by
the action on $\bP$, i.e. by moving the image curve 
under the torus action.  A typical fixed point of
this action is $(f, {\bf P}^1)$\foot{We apologize
for reversing notation from the previous section, and writing
$(f,C)$ instead of $(C,f)$. This is to agree with \lly\ which 
we closely follow in this section.}, where $f({\bf P}^1)$ is a 
${\bf P}^1$ joining two ${\bf T}$-fixed points in $\bP$.

	Another type of fixed point we consider
is obtained by gluing. Let
 $(f_1, C_1, x_1) \in \mp{r}$ and
$(f_2,C_2,x_2) \in \mpp{d}{r}$ be two fixed points. Then
$f_1(x_1)$ is a fixed point of $\bP$, i.e.
one of the $p_i$'s, say $p_k$. If $f_2(x_2)$
is also $p_k$, let us glue them at the marked points to obtain
$(f, C_1 \cup C_2) \in \m{d}$,
where $f|_{C_1} = f_1$, $f|_{C_2} = f_2$ and $f(x_1 
~ x_2) = p_k$. Clearly, $(f, C_1 \cup C_2)$ is a fixed point
as $(f_1,C_1,x_1)$ and $(f_2, C_2,x_2)$ are fixed points.
 Let us denote the loci of fixed
points obtained by gluing as above $FL(p_k, \vec r, \vec{d} -\vec{r})$.
Over $C_1 \cup C_2$, there is an exact sequence
for $V$:

\eqn\convexactseq{0 \rightarrow f^*{V} \rightarrow 
f_1^*{V} \oplus f_2^*{V} \rightarrow 
{V}|_{f_1(x_1) = f_2(x_2)} \rightarrow 0.}
The long exact cohomology sequence then gives
us a gluing identity

\eqn\eclassident{ \Omega_{\bf T}^{V} c_{\bf T}(U_{\vec d}) = c_{\bf T}(U_{\vec r}) 
c_{\bf T}(U_{\vec{d} - \vec{r}}),}
where $\Omega_{\bf T}^{V} = c_{\bf T}(V^+)/c_{\bf T}(V^-)$ is
the ${\bf T}$-equivariant Chern class of $V$.		

This relation will generate one on the linear sigma
model to which we now turn.

\subsec{The spaces $M_{\vec d}$ and $N_{\vec d}$}
	
Because $\m{d}$ is a rather unwieldy space, 
the gluing identity we found in the last section seems 
not to be useful. However, we will find, using
the gluing identity, a similar identity on  
a toric manifold $N_{\vec d}$.
We devote this section mainly to describing $N_{\vec d}$
and its relation to $\m{d}$.

First we consider $M_{\vec d} \equiv \mnp{d}$.
We will call $\pi_1$ and $\pi_2$ the projections
to the first and second factors of ${\bf P}^1 \times \bP$ respectively.
Since $\pi_2$ maps to $\bP$, one might
consider a map from $M_{\vec d}$ to $\m{d}$ 
sending $(f,C)$ to $(\pi_2 \circ f, C)$.
However, this is not necessarily a stable map. 
If it is unstable, $\pi_2 \circ f$ maps some components of $C$
to points, so if we let $C'$ be the curve 
obtained by deleting these components,
there is a map $\pi : M_{\vec d} \rightarrow \m{d}$ 
which sends $(f,C)$ to $(\pi_2 \circ f, C')$.

Let us now recall some facts about maps from ${\bf P}^1$.
A regular map to $\bP$ is equivalently
a choice of generic sections of
${\cal O}_{{\bf P}^1}(f^*B_i \cdot H_{{\bf P}^1})$, $i = 1,...,N_C$.
For example, a map of degree $d$ from ${\bf P}^1$
to ${\bf P}^4$ gives five generic sections 
of ${\cal O}_{{\bf P}^1}(d)$, i.e., five degree $d$ polynomials. 
If one takes five arbitrary sections, constrained only
by being not all identically zero, one gets a rational 
map instead.  Generalizing this, arbitrary sections
of ${\cal O}_{{\bf P}^1}(f^*B_i \cdot H_{{\bf P}^1})$ which are
not in $\Delta$ give rational maps to $\bP$.
The space $N_{\vec d}$ is the space of all such maps with
$f^*(J_i) = d_i J_{{\bf P}^1}$, where $J_{{\bf P}^1} \cdot H_{{\bf P}^1} = 1$. 
Explicitly, we can write it as a quotient space. Defining $D = \sum d_j H_j$,
we have

\eqn\nddef{N_{\vec{d}} = {{\oplus_{i} H^0({\bf P}^1, {\cal O}(B_i \cdot D))
- \Delta} \over G}.}

	There is a map $\psi : M_{\vec d} \rightarrow N_{\vec d}$, which
we now describe. Take $(f,C) \in M_{\vec d}$ and decompose
$C$ as a union $C_0 \cup C_1 \cup ... \cup C_N$ of not 
neccesarily irreducible curves, 
so that $C_j$ for $j > 0$ meets $C_0$ at a point, 
and $C_0$ is isomorphic to ${\bf P}^1$ under
$\pi_1 \circ f$. Since $C_0 \cong {\bf P}^1$, 
$\pi_2 \circ f|_{C_0}$ can be regarded as a point in 
$N_{[\pi_2 \circ f(C_0)]}$, where $[\mu]$ denotes the homology class
of $\mu$. We can also represent $\pi_2 \circ f|_{C_j}$ for
$j > 0$ by elements in $N_{[\pi_2 \circ f(C_j)]}$,
except that since the maps are to have domain $C_0$,
in this case we take the rational map from $C_0$ that
vanishes only at $x_j = C_j \cap C_0$ and belongs to
$N_{[\pi_2 \circ f(C_j)]}$.

	Having now $N+1$ representatives, compose them via the map
$$ N_{\vec{r}_1} \otimes N_{\vec{r}_2} \rightarrow N_{\vec{r}_1 + \vec{r}_2}$$
given by multiplying sections of ${\cal O}(B_i)$.
The result, since $\sum_{i=0}^{N} [\pi_2 \circ f(C_i)] = D$, is a point
in $N_{\vec{d}}$. Thus we have obtained a map 
from $M_{\vec d}$ to $N_{\vec d}$.

To illustrate, let us take the case of ${\bf P}^4$ again.
Here $(f,C)$ is a degree $(1,d)$ map. Let us decompose
$C$ as before into $C_0 \cup ... \cup C_N$, with $x_i = C_i \cap C_0$
and $\pi_1 \circ f|_{C_0}$ an isomorphism.
The image of $\psi$ is a rational morphism
given as a map by $\pi_2 \circ f|_{C_0}$ except at
the points $x_i$. At $x_i$, a generic hyperplane
of ${\bf P}^4$ pulled back vanishes to
the order given by the multiplicity of $[\pi_2 \circ f(C_i)]$
in terms of a generator.

So far we have discussed the spaces and the maps
between them. We now briefly describe 
the torus actions they admit. Clearly, $M_{\vec d}$ has 
an $S^1 \times {\bf T}$ action induced
from an action on ${\bf P}^1 \times \bP$.
In suitable coordinates, the $S^1$ action
is $[w_0,w_1] \rightarrow [e^\alpha w_0, w_1]$.

Since sections of ${\cal O}_{{\bf P}^1}(1)$
is also a one-dimensional projective space, there
is an $S^1$ action on $H^0({\bf P}^1, {\cal O}_{{\bf P}^1}(1))$. 
This induces an action on sections of ${\cal O}_{{\bf P}^1}(d)$.
$N_{\vec d}$ is defined by the latter,
so it admits an $S^1$ action. In addition,
it has a ${\bf T}$-action induced
from the action on ${\cal O}(B_i)$.

The map $\pi$ is obviously
${\bf T}$-equivariant,
since the ${\bf T}$-actions are induced from 
$\bP$. It is shown in \llyt\ that $\psi$ is 
$(S^1 \times {\bf T})$-equivariant. Summarizing,
we have the following maps:
$$N_{\vec{d}}\matrix{{}_\pi \cr \longleftarrow\cr {}} M_{\vec{d}}
\matrix{ {}_\psi \cr \longrightarrow \cr {}} \overline{{\cal M}}_{0,0}
(\vec{d};{\bf P})\matrix{ {}_\rho \cr \longleftarrow \cr {}} 
\overline{{\cal M}}_{0,1}(\vec{d};{\bf P})
\matrix{ {}_{ev} \cr \longrightarrow \cr {}} {\bf P}.$$
Pushing and pulling our problem to $N_{\vec d}$,
we define 

$$Q_{\vec d} = \psi_{!} \pi^{*} c_{\bf T}(U_{\vec d}).$$

\subsec{Euler data}

	In this section we will derive from the gluing identity 
a simpler identity on $N_{\vec d}$. Recall that the gluing identity
holds over fixed loci 
$FL(p_i, \vec r, \vec {d}-\vec{r}) \in \m{d}$. 
Therefore, an identity holds over 
 $\pi^{-1}(FL(p_i, \vec r, \vec{d} - \vec{r}))$ in $M_{\vec d}$
under pull-back by $\pi$. We next turn to describing a sublocus
of $\pi^{-1}(FL(p_i, \vec r, \vec{d} - \vec{r}))$ which,
as we will see later is mapped by $\psi$ to
a fixed point in $N_{\vec d}$.

	Let $F_{p_i, \vec r}$ denote the fixed point loci in $\mp{r}$
with the marked point mapped to $p_i$.
Let $(f_1,C_1,x_1) \in F_{p_i, \vec r}$ and 
$(f_2, C_2, x_2) \in F_{p_i,\vec {d}-\vec{r}}$
be two points. We define a point $(f,C)$
in $M_{\vec d}$ as follows. For $C$ we take $(C_0 \equiv {\bf P}^1) \cup C_1
\cup C_2$, with $C_0 \cap C_1 = x_1$ and $C_0 \cap C_2 = x_2$.
For the map $f$, we define it by giving the projections
$\pi_1 \circ f$ and $\pi_2 \circ f$. We require
$\pi_1 \circ f(C_1) = 0$, $\pi_1 \circ f(C_2) = \infty$ and
$\pi_1 \circ f|_{C_0}$ be an isomorphism. This ``fixes'' $\pi_1 \circ f$,
since any other choice is related by an automorphism
of the domain curve preserving $x_1$ and $x_2$, which
is irrelevant in $M_{\vec d}$. We require $\pi_2 \circ f$
to map $C_1$ as $f_1$, $C_2$ as $f_2$ and $C_0$ to $f_1(x_1)$.
Clearly, $\pi$ maps $(f,C)$ to a point in $FL(p_i, \vec r, \vec {d}-\vec{r})$.
Let us denote the loci of such $(f,C)$'s by
$MFL(p_i, \vec r, \vec {d} - \vec{r})$. By construction, it is
isomorphic to $F_{p_i,\vec r} \times F_{p_i,\vec{d}-\vec{r}}$.

	We verify $(f,C)$ is a fixed point.
$f_i(C_i)$ and $p_i$ are fixed in
$\bP$, so $f$ is ${\bf T}$-fixed.
The $S^1$-action fixes only $0$ and $\infty$ on 
the first factor of ${\bf P}^1 \times \bP$.
Nevertheless, the point $(f, C_0 \cup C_1 \cup C_2)$ remains 
fixed under the $S^1$ action, as we need to
divide out by automorphisms of $C_0$
preserving $x_1$ and $x_2$.

	We now compare the maps just constructed
with the fixed points of $N_{\vec d}$. It will
be most convenient to do so by describing the latter
in terms of rational morphisms. So take a point
in $N_{\vec d}$, viewed as a rational morphism from 
$C_0 \equiv {\bf P}^1$.
Let $x_1,...,x_N$ be the points where it is undefined.
At $x_i$, the chosen sections of
${\cal O}(B_j \cdot D)$ vanish 
to certain orders, including possibly zero. A generic section
of ${\cal O}(J_j \cdot D) = {\cal O}(d_j)$ vanishes to
order, say $r_j$ at $x_1$. Any 
section of a line bundle ${\cal O}(L)$ pulled back 
by the map then vanishes at $x_1$ at least to order $L \cdot \sum r_j H_j$.

	Therefore the rational morphism is equivalent to
the data of a regular map from $C_0$ and a curve class
for each bad point. The classes for the bad points 
indicate the multiplicity of vanishing of a 
generic section of a pulled-back line bundle.
Altogether, the class of the image of the regular map
and the curve classes we associate to the bad points
 sum to $D$, since a generic
section of ${\cal O}(L)$ must have exactly $L \cdot D$
zeroes.

	Now we can deduce the fixed points of $N_{\vec d}$.
The ${\bf T}$-action moves the 
image of a rational morphism,
whereas the $S^1$ action rotates the domain ${\bf P}^1$ about 
an axis joining $0$ and $\infty$. So a fixed point
is a rational map, undefined at $0$ and 
$\infty$, whose image is a fixed point of $\bP$.
Let us denote them by $p_{i,\vec{r}}$,
where $p_i$ denotes a fixed point of $\bP$, and $\sum r_i H_i$
determines orders of vanishing of pulled-back line bundles
at the point $0 \in {\bf P}^1$.
Clearly, $\psi$ maps the fixed points in $M_{\vec d}$ 
discussed earlier to the fixed point $p_{i,\vec{r}}$. 

	We now use the Atiyah-Bott formula for localization
to relate restrictions of $Q_{\vec{d}}$ to
$p_{i,\vec{r}}$ (which we denote by $Q_{\vec{d}}(p_{i,\vec{r}})$) to 
$c_{\bf T}(\pi^*U_{\vec{d}})$. Explicitly, 

$$ Q_{\vec{d}}(p_{i, \vec{r}})
= \int_{N_{\vec{d}}} \phi_{p_{i,\vec{r}}} Q_{\vec{d}}
= \int_{M_{\vec{d}}} \psi^*(\phi_{p_{i,\vec{r}}}) 
c_{\bf T} (\pi^*(U_{\vec{d}})).$$
where $\phi_{p_{i,\vec{r}}}$ is the equivariant Thom class
of the normal bundle of $p_{i,\vec{r}}$ in $N_{\vec d}$.
To evaluate the last integral,
we need the equivariant euler class of
the normal bundle of $MFL(p_i, \vec{r}, \vec{d} - \vec{r})$ in $M_{\vec d}$.

	Since $MFL(p_i, \vec{r},\vec{d} - \vec{r}) \cong
F_{p_i, \vec r} \times F_{p_i,\vec{d} - \vec{r}}$,
we have contributions from the normal bundles of 
$F_{p_i, \vec{r}} \in \mp{r}$ and 
$F_{p_i, \vec{d} - \vec{r}} \in \mpp{d}{r}$.
They are respectively $e(N(F_{p_i,\vec{r}}
/\mp{r}))$ and 
$e(N(F_{p_i,\vec{d} - \vec{r}}/\mpp{d}{r}))$.
Points in $MFL(p_i, \vec{r}, \vec{d} - \vec{r})$ have
domain of the form $C_0 \cup C_1 \cup C_2$,
where $C_1 \cap C_0 = x_1$, and $C_2 \cap C_0 = x_2$.
Now let $L_{\vec r}$ denote the line bundle on $\mp{r}$ whose
fiber at $(f_1, C_1, x_1)$ is the tangent line at $x_1$.
Then we can write the contributions from deforming
$x_1$ and $x_2$ as $e(L_{\vec{r}} \otimes T_{x_1} C_0) = 
\alpha + c_1(L_{\vec{r}})$ and 
$\alpha + c_1(L_{\vec{d} - \vec{r}})$, respectively.
In addition, automorphisms of $C_0$ which
do not fix $x_1$ and $x_2$ need to be included. 
They can be shown to give weights of 
$T_{x_1} C_0$ and $T_{x_2}C_0$, so there is an
extra factor of $(\alpha)(- \alpha)$. Finally,  
normal directions which move the image of 
the marked point from $p_i$ have to be excluded, so we 
divide by the weights of $T_{p_i} \bP$.

	This yields, after using \eclassident,

$$\eqalign{\Omega^{V}(p_i) Q_{\vec{d}}(p_{i,\vec{r}}) = 
&{-1 \over \alpha^2}
e(T_{p_i}\bP) e(p_{i,\vec{r}}/N_{\vec{d}})
\sum_{F_{p_i,\vec{r}}} \int_{F_{p_i,\vec{r}}} 
{\rho^*c_{\bf T}(U_{\vec{r}}) \over e(N(F_{\vec{r}}))(\alpha 
+ c_1(L_{\vec{r}}))} \cr
&\sum_{F_{p_i,\vec{d} - \vec{r}}} \int_{F_{p_i,\vec{d} - \vec{r}}} 
{\rho^*c_{\bf T}(U_{\vec{d} - \vec{r}}) \over e(N(F_{p_i,\vec{d} - 
\vec{r}}))(\alpha 
+ c_1(L_{\vec{d} - \vec{r}}))}}. $$

	We introduce some
more notation. 
Let $\kappa_{j,\vec{d}}$ be the  
 member of the $S^1 \times {\bf T}$
equivariant cohomology ring of $N_{\vec d}$
whose weight at the fixed point $p_{i,\vec r}$
is $\kappa_j(p_i) + r_j \alpha$.
Clearly, $\kappa_{j, \vec{0}} \equiv \kappa_{j}$.
The identity $e (T_{p_i} \bP)
e (p_{i,\vec{r}}/N_{\vec{d}}) = e (p_{i, \vec{r}}/N_{\vec{r}})
e(p_{i,\vec{0}}/N_{\vec{d} - \vec{r}})$ then  implies

\eqn\eeulerdat{ \Omega^V(p_i) Q_{\vec{d}} (p_{i,\vec{r}}) 
= \overline{Q_{\vec{r}}(p_{i,\vec{0}})} Q_{\vec{d} - \vec{r}}(p_{i,\vec{0}}). }
Here the overbar $\overline{\;}$ is an automorphism of the 
$S^1 \times {\bf T}$
equivariant cohomology ring 
with $\overline{\alpha} = -\alpha$ and $\overline{\kappa_{j,\vec{d}}}
= \kappa_{j,\vec{d}}$. A sequence of equivariant cohomology classes
satisfying \eeulerdat\ is called in \lly\ a set of 
$\Omega^{V}$-Euler data.

\subsec{Linked Euler data}

	If we knew the values of $Q_{\vec{d}}$ at
all fixed points, we would also know $Q_{\vec{d}}$
as a class. Since we do not know this,
we will use equivariance
to compute $Q_{\vec{d}}$ at certain points, for example,
those that correspond to the ${\bf T}$-invariant ${\bf P}^1$'s
in $\bP$. It turns out that this is also sufficient,
as we will find Euler data which agree
with $Q_{\vec{d}}$ at those points, and
a suitable comparison between the two 
gives us the rest.

	Before we begin the computation, we first describe
a ${\bf T}$-equivariant map from
$N_{\vec{0}} = \bP$ to $N_{\vec{d}}$. 
Sections of ${\cal O}(B_i \cdot D)$
over ${\bf P}^1$ are polynomials in $w_0$ and $w_1$, where
$w_0$ and $w_1$ are as before coordinates
so that the $S^1$ action takes the form
$[w_0, w_1] \rightarrow [e^\alpha w_0, w_1]$.
Each polynomial contains 
a unique monomial invariant under the $S^1$ action. By
sending the coordinates of a point
to the coefficients of the invariant monomials,
we hence obtain a map $I_{\vec d}$
from $N_{\vec{0}}$ to $N_{\vec d}$.

	We begin with the case of a convex line bundle ${\cal O}(L)$,
where $L$ denotes the associated divisor.	
Let $(f, {\bf P}^1)$ be a point in $\m{d}$ with
 $f({\bf P}^1)$ being  a
 multiple of the ${\bf T}$-invariant
${\bf P}^1$ joining $p_i$ and $p_j$ in $\bP$.
The fiber of the obstruction bundle at $(f,{\bf P}^1)$ is
 $H^0({\cal O}(L \cdot D))$, which
is spanned in appropriate coordinates for the ${\bf P}^1$
by $u_0^k u_1^{L \cdot D - k}$, $k = 0, ..., L \cdot D$.

	Choose a basis so that $u_1 = 0$
is mapped to $p_i$, and $u_0 = 0$ mapped to $p_j$.
Since the section $u_0^{L \cdot D}$ does not vanish
at $u_1 = 0$, its weight is equal to the weight
of $L$ at $p_i$, which we denote by $L(p_i)$. 
Similarly, $u_1^{L \cdot D}$
has weight $L(p_j)$. Hence the induced weight on $u_1/u_0$
is $(L(p_j) - L(p_i))/(L \cdot D)$, giving us the
weights of all sections. 
 
	Since $(f, {\bf P}^1)$ is 
fixed by ${\bf T}$,  the 
corresponding loci $\psi(\pi^{-1}((f, {\bf P}^1)))$ 
in $N_{\vec{d}}$ is, by equivariance,
fixed by a ${\rm dim} {\bf T}$ subgroup of $S^1 \times {\bf T}$. 
The points in $\psi(\pi^{-1}((f, {\bf P}^1)))$ 
represent a regular map from a ${\bf P}^1$ to
the ${\bf T}$-invariant ${\bf P}^1$ joining $p_i$ and $p_j$.
Therefore any two points in $\psi(\pi^{-1}((f, {\bf P}^1)))$ 
differ by an automorphism of the domain.
Explicitly, we can consider coordinates $[w_0,w_1]$ as before.
Let $\eta$ denote the point which sends $[w_0 = 1,w_1 = 0]$ to
$p_i$ and $[w_0 = 0, w_1 = 1]$ to $p_j$. Any other point,
thought of as a map, factors via $\eta$ by an automorphism
$[w_0,w_1] \rightarrow [a w_0 + b w_1, c w_0 + d w_1]$.
We can therefore find the relevant subgroup of
$S^1 \times {\bf T}$ by choosing the element of 
$S^1$ to cancel the induced weight of ${\bf T}$ on $w_1/w_0$.
Explicitly, we have $\alpha = (L(p_i) - L(p_j))/(L \cdot D)$,
so that the value of $Q_{\vec d}$ is
		
$$ \prod_{k} (x + L(p_j) + k \alpha). $$
The weight of $\kappa_{i,\vec{d}}$ under 
$(S^1 \times {\bf T})/(\alpha - (L(p_i) - L(p_j))/(L \cdot D))$
at any point in $\psi(\pi^{-1}((f, {\bf P}^1)))$ is the same as its
weight under ${\bf T}$ at the corresponding point obtained by 
setting $w_0$ to zero.
For $\eta$, setting $w_0$ to zero gives a point in $N_{\vec d}$
which can be thought of as a rational map to $p_j$. Since
$I_{\vec d}$ is equivariant, the weight
of $\kappa_{i,\vec d}$ at this point is the same as the weight
of $\kappa_i$ at $p_j$. Thus, as $\sum l_i \kappa_i(p_j) = L(p_j)$,
the value of $Q_{\vec d}$ is the same as the value of $P_{\vec{d}}$,
where $P_{\vec{d}}$ is given by

$$ \prod_{k} (x + \sum l_i \kappa_{i, \vec{d}} + k \alpha). $$

	To simplify our computations, we 
take a partial nonequivariant limit by replacing
$\kappa_{i}$ with $J_i$. Then
$P_{\vec{d}}$ reduces to
$\tilde{\Gamma}(x, L, L \cdot D + 1, 0)$, where 
$\tilde{\Gamma}$ is defined as follows:

\eqn\gammaeqn{{\tilde{\Gamma}(y, K, i, j)} = \cases{
\prod_{k = j}^{i - 1} (y + K +  (k - 1)\alpha )& if $ i > j$ \cr\cr
	1 & if $i = j$ \cr\cr
\displaystyle{{1 \over \prod_{k=i}^{j-1} (y + K +  k\alpha)}} & if  $i < j$}.}
Similarly, the case of a concave line bundle gives
$\tilde{\Gamma}(x, L, -L \cdot C, 1)$.
For a direct sum, since the Chern
class is multiplicative and the product of the
cases just considered is a valid Euler
data, the product gives us the 
appropriate value of $P_{\vec{d}}$.

 	Let us form a series:
\eqn\hypserthr{HGA[Q]({\vec T}) = e^{-\sum T_i J_i/\alpha} 
{(\sum_{D \in {\nec{\bP} - {\vec{0}}}} I_{D}^{*}(Q_{\vec d}) 
\prod_i \tilde{\Gamma}(0, B_i, -B_i \cdot D, 0)
e^{\sum T_i J_i \cdot D} + \Omega_{\bf T}^{V})},}
where by $Q_{\vec d}$ we mean its partially nonequivariant limit,
as described above, and $\nec{\bP}$ is the set of curve classes in $\bP$
which have nonnegative intersection with the effective divisors of $\bP$.

	A similar series can be constructed
from the partial nonequivariant limit of 
$P_{\vec d}$ as follows:
\eqn\hypser{\eqalign{HGB[\vec{t}] = & e^{-\sum t_i J_i/\alpha}
(\sum_{D \in \nec{\bP}
- {\vec{0}}} \prod_{i} {
\tilde{\Gamma}(0, B_i, -B_i \cdot D, 0)}
\prod_{i: F_i < 0} {
{\tilde{\Gamma}(x, F_i, - F_i \cdot D, 1)}} \times. \cr
& \prod_{i: F_i > 0} {
{\tilde{\Gamma}(x, F_i, F_i \cdot D + 1, 0)}}
e^{J \cdot D} + \Omega_{\bf T}^{V})}.}

	
	If we take the example of the quintic in ${\bf P}^4$,
${\rm dim} H_2(\bP, {\bf Z}) = 1$, so we have

\eqn\hypserq{HGB[t] = e^{-J/\alpha}({\sum_{d > 0} {{\prod_{m = 1}^{5d} (x + 
5H - m\alpha)} \over {\prod_{m=1}^{d} (x + H - m\alpha)}^5} + 
5 H}).}

For local mirror symmetry, we can take $V = K_{{\bf P}^2}$, giving
\eqn\hypserl{HGB[t] = e^{-J/\alpha} ({\sum_{d > 0} {{\prod_{m=1}^{3d - 1} (x
-3H + m\alpha)} \over {\prod_{m=1}^{d} (x + H - m\alpha)}^3} 
- {1 \over 3H}}).}

To compare $HGB[\vec{t}]$ and $HGA[\vec{T}]$, let us
expand $HGB[\vec t]$ at large $\alpha$, keeping terms 
to order $1/\alpha$. It is shown in \llyt\
that $HGA[\vec{T}]$ has the form
$\Omega_{\bf T}^{V} (1 - (\sum T_i J_i)/\alpha)$. 
Equating the two expressions gives us
$\vec{T}$ in terms of $\vec{t}$. Then
setting $HGA[\vec{T}(\vec{t})] = HGB[\vec{t}]$ gives 
us $Q_{D}$ as a function of $J_i$ and $\alpha$, from
which we may obtain $K_{D}$ \llyt:

\eqn\Kdeqn{{{(2 - \sum d_i t_i) K_{D}} \over \alpha^3} = 
\int_{\bP} e^{-\sum t_i J_i/\alpha} {Q_{\vec d}
\prod_j \tilde{\Gamma}(0, B_j, -B_j \cdot D, 0)}.}
	
\newsec{\bf Explicit Verification Through Fixed-Point Methods}

\subsec{Some examples}

Though the techniques of section one are extremely powerful,
it is often satisfying -- and a good check of one's methods --
to do some computations by hand.  In this section, we outline
fixed point techniques for doing so, and walk through several
examples.  In this way, we have verified many of the
results in the appendices for low degrees ($d = 1,2,3$).
Readers familiar with such
exercises
may wish to skip to the next subsection.

All the bundles described in section two
are equivariant with respect to the torus $T$-action
which acts naturally on the toric manifold ${\bf P}.$  Let $t \in T$ be
a group
element acting on ${\bf P}.$  Then if $(C,f,*) \in \mp{d}$ and $(C,f)
\in \m{d}$
the induced torus action sends theses points to $(C, t\circ f, *)$ and
$(C, t\circ f),$ respectively.  The bundle actions are induced by the
natural $T$-action on the canonical bundle, $K.$

Now that we understand the torus action, what are the fixed point
theorems?
First of all, we work in the realm of equivariant characteristic
classes, which live in the equivariant cohomology ring
$H_T^*(M)$ of a manifold, $M.$  Let $\phi \in H_T^*(M)$ be an
equivariant cohomology
class.  The integration formula of Atiyah and Bott is
$$\int_M \phi = \sum_{P} \int_P \left( i_P^*\phi \over e(\nu_P)
\right),$$
where the sum is over fixed point sets $P,$ $i_P$ is the embedding in
$M,$
and $e(\nu_P)$ is the Euler class of the normal bundle $\nu_P$ along
$P.$
For $P$ consisting of isolated points and $\phi$ the Chern class
(determinant),
we get the ratio of the product of the weights of the $T$-action on the
fibers at $P$ (numerator) over the product of the weights of the
$T$-action
on the tangent bundle to $M$ at $P$ (denominator).  One needs only
to determine
the fiber and tangent bundle at $P$ and figure out the weights.

Let's start with degree one ($K_1$) for the quintic
(${\cal O}(5)\rightarrow {\bf P}^4$).  Let
$X_i \mapsto \alpha_i^{\lambda_i}X_i,$ $i=1,...,5,$ be the $({\bf
C}^*)^5$
action on ${\bf C}^5$ (${\bf P}^4 = {\bf P}({\bf C}^5)$), where
$\vec{\alpha}\in ({\bf C}^*)^5$ and the $\lambda_i$ are the weights.
The fixed curves are $P_{ij},$
where $i,j$ run from $1$ to $5:$  $P_{ij} = \{X_k=0, k\neq i,j\}.$
Since $f$ is a degree one map, we may equate
$C \cong f(C)$ and the pull-back of ${\cal O}(5)$ is therefore
equal to ${\cal O}(5)$ on $C_e.$  Recall the bundle ${\cal O}(5)$ on
$\p.$  Its global sections are degree five polynomials in the
homogeneous
coordinates $[X,Y],$ so a convenient basis is $\{ X^a Y^{5-a}, a =
0,...,5\}$
(or $u^a$ in a local coordinate $u = X/Y$).
The weights of these sections are $a \mu + (5-a)\nu,$ if $\mu$ and
$\nu$ are the weights of the torus
$({\bf C}^*)^2$ action on ${\bf C}^2.$
The map $f: C_e \rightarrow {\bf P}^4$ looks like
$[X,Y] \mapsto [...,X,...,Y,...]$ with non-zero entries only in
the $i$th and $j$th positions.  Therefore, the weights of
$U_1$ at the fixed point $(C,f)$ are $a\lambda_i + (5-a)\lambda_j,$
$a = 0...5.$  To take the top Chern class we take the product of these
six weights.

We have to divide this product by the product of the weights of the
normal bundle, which in this case (the image $f(C)$ is smooth) are
the weights of $H^0(C,f^*N),$ where $N$ is the normal bundle to $f(C).$
More generally, we take sections of the pull-back
of $T{\bf P}^4$ and remove sections of $T_C.$
The normal bundle of $P_{ij}$ is equal to ${\cal O}(1)\oplus{\cal
O}(1)\oplus
{\cal O}(1),$ each corresponding to a direction normal to $f(C)$ and
each of which has two sections.  Let $w = X_j/X_i$ be a coordinate
along $C \cong P_{ij}$ on the patch $X_i \neq 0.$
If $z_k = X_k/X_i$ is a local coordinate of ${\bf P}^4,$ then
$del_k \equiv {\del\over \del z_k}$ is a normal vector field 
on $C$ with weight $\lambda_k-
\lambda_i.$  $w\del_k$ is the other normal vector
field corresponding to the direction $k,$ and has weight $\lambda_k
-\lambda_j.$  Summing over the $\pmatrix{5\cr 2}=20$
choices of image curve $P_{ij}$
gives us
$$K_1 = \sum_{(ij)} {\prod_{a=0}^{5} [a\lambda_i + (5-a)\lambda_j]
\over \prod_{k\neq i,j} (\lambda_k - \lambda_i)(\lambda_k -
\lambda_j)} = 2875,$$
the familiar result.

Let's try degree two $(K_2)$ for $K_{\pp}.$  The dimension of
${\cal M}_{0,0}(d;\pp)$ is
$3d-1 = 5$ for $d = 2.$  What are the fixed points?  Well, the image of
$(C,f)$
must an invariant curve, so there are two choices for degree
two.  Either the image is a smooth $\p$ or the union of two $\p$'s.
There
are three fixed points on $\pp$ and
therefore $\pmatrix{3\cr 2} = 3$ invariant $P_{ij}$'s.
If the image is a smooth $\p,$ the domain curve may either be a smooth
$\p,$ in which case the map is a double cover (let's call this
case 1a), or it may have two
component $\p$'s joined at a node (let's call this case 1b).
If the image has
two components, the domain must as well.  Let's call this case 2.

For case 1a, the situation is similar to the quintic case above.
The tangent space to moduli space consists of sections of $H^0(C,f^*N),$
where $N$ is the normal bundle to the image curve.  That is, we take
$H^0(C,f^*T\pp)$ and delete those sections from $H^0(TC).$
Since $T\pp\vert_{f(C)} \cong {\cal O}(2)\oplus {\cal O}(1)$ and $f$ is
a degree 2 map, we have $f^*T\pp \cong {\cal O}(4)\oplus {\cal O}(2),$
which has 5 + 3 = 8 sections.  $TC \cong {\cal O}(2)$ has three
sections,
leaving us with five total.  If $f(C) = P_{ij},$ then the ${\cal O}(2)$
sections
are $\del_k, w\del_k, w^2\del_k,$ where $k \neq i,j,$ $w$ is the
coordinate on $C,$ and $z_m = X_m/X_i$ are inhomogeneous
coordinates on $\pp.$  The degree two map is, in these coordinates,
$w \mapsto (z_j = w^2, z_k = 0).$
Note that $\del_k$ is the only non-vanishing section
at $w=0,$ and the others are obtained by successive multiplications
by $w.$  Notice that $w$ inherits the
weight $(\lambda_j - \lambda_i)/2$ by requiring equivariance.
The weights are, so far, $\lambda_i - \lambda_k, (\lambda_i
+ \lambda_j)/2 - \lambda_k, \lambda_j - \lambda_k.$  For the
${\cal O}(4)$ sections, the procedure is similar, only we must
remove the weights $0, \pm (\lambda_i - \lambda_j)/2,$ as
these correspond to the tangent vectors $\del_w, w\del_w, w^2\del_w.$
We are left with $\pm (\lambda_j - \lambda_i)$ giving a total of five.

The weights of $H^1(C,f^*K_{\pp})$ are easily calculated for the curve
$C$
by using Serre duality.  That is, if one thinks
(naively) of sections of a vector bundle $E$
as elements of $\overline{\del}$ cohomology, and recalling that the
canonical bundle $K$ is the bundle of holomorphic top forms, then
$H^k(E)$
pairs with $H^{n-k}(E^*\otimes K)$ by wedging and contracting $E$ with
its dual $E^*,$ then integrating.
Thus, $H^k(E) \cong H^{n-k}(E^*\otimes K)^*.$
For a curve, $C,$ we have $H^1(f^*K_{\pp}) \cong H^0(f^*K^{-1}_{\pp}
\otimes K_C)^*.$  Let's compute.  $K^{-1}_{\pp} \cong {\cal O}(-3)$
as a bundle, and $K_C \cong {\cal O}_{\p}(-2),$ so
$f^*K^{-1}_{\pp}\otimes
K_C \cong {\cal O}(2\cdot(+3) - 2) = {\cal O}(4),$ and the five sections
can be obtained in the usual way once we have a non-vanishing one at
$w=0.$  Such a section is ${\del\over \del z_j}\wedge{\del\over \del
z_k}\otimes
dw,$ and has weight $2\lambda_i - \lambda_j - \lambda_k + (\lambda_j-
\lambda_i)/2.$  In all, then the weights of $H^0(f^*K^{-1}_{\pp}
\otimes K_C)$ are $2\lambda_i - \lambda_j - \lambda_k + m(\lambda_j-
\lambda_i)/2,$ $m = 1,...,5.$  For the dual space $H^1(f^*K_{\pp})$
we must take the negatives of these weights.  So much for case 1a.

Cases 1b and 2 the domain curves have two components
(say, $C_1$ and $C_2$), so we must
understand what is meant, for example, by $f^*T{\bf P}^2$ and
$TC$ in order to calculate the normal bundle.  $TC$ is locally
free (like a vector bundle) everywhere except at the singularity.
There, we require tangent vectors to vanish.  The canonical bundle,
$K_C,$ is defined as a line bundle of holomorphic differentials,
with the following construction at the singularity.  Let $f(w)dw$
be a differential along $C_1$, and let $g(z)dz$ be a
differential along $C_2$ where the singularity is taken
to be $(z = 0)\sim (w = 0).$  To define a differential along the
total space, we allow $f$ and $g$ to have up to simple poles
at the origin, with the requirement that the total residue vanish:
${\rm res}_wf + {\rm res}_zg = 0.$  What this does is serve as an
identification, at the singularity,
of the fibers of the canonical bundles of the two components.
In this way, we arrive at a line "bundle."
This canonical bundle, when restricted to a component $C_i$, looks
like $K_{C_i}(p),$ where the ``$(p)$'' indicates twisting by the
point, i.e. allowing poles.

At this point, we can proceed with the calculation.  Consider
case 2, for which the map $f$ is a bijection.  The sections
of $H^1(f^*K_{\pp})\cong H^0(f^*(K_{\pp})$ can be looked at
on each component, where $K_{C}\vert_{C_i}$ is as above.
Hence on $C_i$ we have
$${1\over w} {\del\over\del z_1}\wedge {\del\over \del z_2}
\otimes dw,
\qquad 1 { \del\over\del z_1}\wedge {\del\over
\del z_2} \otimes dw,
\qquad w{ \del\over\del z_1}\wedge {\del\over
\del z_2} \otimes dw.$$
On the the other component, we have three analogous sections,
but two with poles need to be identified, since they are related
by the requirement of no total residue.  Indeed, this identification
is compatible with equivariance, since ${1\over w}dw$
has zero weight.  All in all, we have weights
(recalling duality) $\lambda_j + \lambda_k - 2\lambda_i,
\lambda_k - \lambda_i, \lambda_k - \lambda_j,
\lambda_j - \lambda_i, \lambda_j - \lambda_k.$

The normal bundle to moduli space consists of sections
of the pull-back of tangent vectors on ${\bf P}^2,$
less global tangent vectors on $C.$  In addition, we
include $T_pC_1\otimes T_pC_2,$ which is a factor
corresponding to a normal direction in which the node
is resolved \kontloc\ .  Since the maps from components
are degree one for these cases, we can take as sections
of the normal bundles (two each) $\del_w, z\del_w$ and
$\del_z, w \del_z.$  Here we have identified the
coordinate of the other component with the coordinate
of $\pp$ normal to the component.  The $T_pC_1\otimes T_pC_2$
piece gives $\del_w \otimes \del_z.$  In total,
the weights are
$\lambda_i-\lambda_j, \lambda_k-\lambda_j,
\lambda_i-\lambda_k, \lambda_j - \lambda_k, 2\lambda_i
-\lambda_j-\lambda_k.$

One checks that the product of the numerator weights divided
by the denominator weights is equal to $-1.$  Since there are
three graphs of this type, the total contribution to $K_2$
is $-3.$  Graphs whose image is a single fixed $\p$ contribute
$-21/8,$\foot{Because we are considering integrals in the
sense of orbifolds, we must divide out the contribution
of each graph by the
order of the automorphism group of the map.  Automorphisms
are maps $\gamma: C\rightarrow C$ such that $f\circ \gamma
= f.$  Cases 1a and 1b have ${\bf Z}_2$
automorphism groups.} giving $K_2 = -45/8.$

In physics, local mirror symmetry is all that is needed to
describe the effective quantum field theory from compactification
on a Calabi-Yau manifold which contains a holomorphic surface, if we
take an appropriate limit.
In this limit, the global structure of the Calabi-Yau manifold becomes
irrelevant (hence the term ``local''), and we can learn about
the field theory by studying the local geometry of the surface --
its canonical bundle.  We can therefore construct
appropriate surfaces to study aspects of
four-dimensional gauge theories of our choosing \kkv\ .  
The growth of the Gromov-Witten
invariants (or their local construction) in a specified
degree over the base $\p$ is related to the Seiberg-Witten
coefficient at that degree in the instanton expansion of
the holomorphic prepotential of the gauge theory.
For example, the holomorphic
vanishing cycles of an $A_n$ singularity fibered over
a $\p$ give $SU(n+1)$ gauge theory (the
McKay correspondence, essentially), and one can construct
a Calabi-Yau manifold containing this geometry to check this \kkv\ \kmv.
In this case, the local surface is singular, as it is
several intersecting $\p$'s fibered over a $\p$ (for $A_1$
we can take two Hirzebruch surfaces intersecting in a common
section.  For this reason, it is important to understand the
case where the surface is singular, as well.  We will
have more to say about this in section six.

\subsec{General procedure for fixed-point computations}

Following \kontloc\ and \locvir, we can compute the
weights
of our bundles explicitly.  Each
connected component of the fixed point set is
described by a graph, $\Gamma,$
which is a collection of vertices, edges, and flags.
The graph contains the data of the fixed map, which includes the image
${\bf P}^{1}$'s, the degrees of the maps to the fixed curves, and the
way they
are glued together.

Let us fix some notation.
To each connected component of $f^{-1}(p),$ where $p$ is a fixed point
of
${\bf P},$ we have a vertex, $v.$  We call $C_v = f^{-1}(p)$ the
pre-image
of $p,$ and if
$p = p_j,$ we say that $i(v) = j$ (so $i$ is a map from $\{vertices\}$
to $\{ 1...n+1 \}$).  Let $val(v)$ be the number of special
(marked or nodal) points on $C_v$ (for us equal to the
number of edges with $v$ as their vertex).
The connected components of the pre-image of a
fixed line $P_{ij}$ are denoted $C_e.$  An edge consists of $C_e$
together with the data $i(e), j(e) \in \{1,...,n+1\}$ encoding the image $f(C) =
P_{i(e),j(e)},$ and $d_e$ the degree of the map
$f\vert_{C_e}.$  If there is no confusion, we will write
$i$ and $j$ for $i(e)$ and $j(e).$
Note that in the case of higher genus maps,
the genus $g >0$ components of the domain curve must map to fixed points
$C_v$ as there are no invariant curves of higher genus.  In particular,
the $C_e$ are all of genus zero.
We call a pair $(v,e)$ where $C_v$ and $C_e$ intersect non-trivially
a ``flag,'' $F.$  For $F = (v,e),$ we define $i(F) = i(v).$

The fixed point set corresponding to a graph
$\Gamma$ is then equal to a product
over vertices of
the moduli space of genus $g(v)$ curves with $val(v)$ marked points:
$M_{\Gamma} = \prod_{v}\overline{M}_{g(v),val(v)}.$

The calculation of the weights along the fixed point sets
follows from a simple, general observation.  Given two varieties, $Y_1$
and $Y_2,$ $X = Y_1 \cup Y_2$ may be singular, but
we can construct the maps
$Y_1\cap Y_2 \matrix{{}_\rightarrow\cr {}^\rightarrow} Y_1 \coprod Y_2
\rightarrow X,$
from which we construct maps of sheaves of holomorphic functions:
$${\cal O}_X \rightarrow {\cal O}_{Y_1}\oplus{\cal O}_{Y_2}
\rightarrow {\cal O}_{Y_1 \cap Y_2}.$$
All maps are obtained from inclusions except
the last map, which sends $(f_1,f_2)$ to $f_1 - f_2,$
so this sequence is exact.

For a graph with domain curve $C$ which is
equal to the union of all its components,
things are simple because there are at most pairwise
non-trivial intersections, those being points.  Thus
we have the sequence
\eqn\exseq{0 \rightarrow {\cal O}_C
\rightarrow\bigoplus_{v} {\cal O}_{C_v} \oplus
\bigoplus_{e} {\cal O}_{C_e} \rightarrow \bigoplus_{F}
{\cal O}_{x_F} \rightarrow 0,}
where $x_F = C_v \cap C_e$ if $F = (v,e),$ and the second
map sends $(g\vert_{C_V},h\vert_{C_e})$ to $g-h$ on the
point of intersection (if it exists).

We will use the long exact sequence associated to this
short exact sequence in two ways.  The fixed point
formula tells us we need to compute the weights of
our bundle $U_d$ (whose fibers are $H^1(C,f^*K_{{\bf P}^2})$).
When $C$ is singular, we need to use the above sequence
twisted by (or tensored by) $f^*K_{{\bf P}^2}.$  Then using
concavity of the canonical bundle, which states that
$f^*K_{{\bf P}^2}$ has no global sections on $C_e,$ the
long exact sequence reads
\eqn\canseq{\eqalign{0&\rightarrow H^0(C,f^*K_{\pp})\rightarrow
\bigoplus_{v}H^0(C_v,f^*K_{\pp})\oplus
\bigoplus_{e}H^0(C_e,f^*K_{\pp})\rightarrow
\bigoplus_{F}K_{\pp}\vert_{f(x_F)}\rightarrow \cr
&\rightarrow
H^1(C,f^*K_{\pp})\rightarrow \bigoplus_{v}H^1(C_v,f^*K_{\pp})
\oplus\bigoplus_{e}H^1(C_e,f^*K_{\pp})\rightarrow 0.}}
The last term in the first line follows since $x_F$ is a point, which
is why the last term in the second line is zero.  Note that $f^*K_{\pp}$
is trivial as a bundle on $C_v,$ since $C_v$ is mapped to a point.
However, this trivial line bundle has non-trivial weight
equal to
$\Lambda_i \equiv -3\lambda_i
+ \lambda_i + \lambda_j + \lambda_k,$ where $i = i(F).$
This
will affect equivariant Chern classes nontrivially.
For example, $H^0(C_v,f^*K_{\pp})$ is one-dimensional
(constant section) with
the same weight -- let's call it ${\bf C}_{\Lambda_i}.$
$H^1(C_v,f^*K_{\pp})$ is thus equal to
$H^1(C_v,{\cal O})\otimes {\bf C}_{\Lambda_i}.$
Also, since $H^1(C_v,{\cal O})$ are global holomorphic
differentials, which may be integrated against cycles, we see that
$H^1(C_v,{\cal O}),$ as a bundle over the fixed point component
$\overline{M}_{g(v),val(v)}$ in moduli space,
is equal to the dual $E^*$ of the rank $g(v)$ Hodge bundle, $E.$
We are interested in $c_g(E^*\otimes {\bf C}_{\Lambda_i}).$
The Chern character (not class) is well-behaved under tensor
product, from which we can conclude \locvir\
$$c_g(E^*\otimes {\bf C}_{\Lambda_i}) =
P_g(\Lambda_i,E^*) \equiv \sum_{r=0}^{g}\Lambda_i^r c_{g-r}(E^*),$$
where we have defined the polynomial $P_g(\Lambda_i,E^*)$

We know more about \canseq.  $H^1(C_e,f^*K_{\pp})$ can
be computed exactly as in case 1a from the previous section,
giving weights $\Lambda_i + m(\lambda_i-\lambda)j),$
$m=1,...,3d_e-1.$  Also, $H^0(C_e,f^*K_{\pp}) = 0$ by convexity,
which tells us as well that $H^0(C,f^*K_{\pp}) = 0$ (obvious if
you think of the map).  Therefore the map to flags on the first
line of \canseq is $1-1,$ which is also obvious as it is restriction of
constant sections (zero at a point iff the section is identically
zero).  Thus the weights from the top line which map into
$H^1(C,f^*K_{\pp})$ are $\prod_F\Lambda_{i(F)}/\prod_v\Lambda_{i(v)}.$
Noting that there are $val(v)$ flags with $v$ as their vertex,
and combining with the weights from the middle term on the second
line of \canseq, we have
\eqn\numwts{
\prod_{v}\Lambda_{i(v)}^{val(v)-1}P_{g(v)}(\Lambda_{i(v)},E^*)
\prod_{e}\left[\prod_{m=1}^{3d_e - 1}\Lambda_i + m(\lambda_i-\lambda_j)
\right].}  For the genus zero case, the polynomials involving the
Hodge bundle disappear.

If we twist the sequence \exseq\ by $f^*T\pp$ we can deduce
the information we need to compute $H^0(C,f^*T\pp)-H^1(C,f^*T\pp),$
which is most of what is needed to compute the virtual normal
bundle to the fixed point locus.\foot{When $g\neq 0,$
the moduli space of maps is not smooth (convexity/concavity is
no longer valid), and one has to take care to define integration
of forms in the expected (``top'') dimension, as the moduli
space will contain components other dimensions.  To do so,
one must define a cycle of the expected dimension -- the
virtual fundamental class (\behfan\ , \litian\ ).
\locvir\ proved that with these definitions,
the Atiyah-Bott localization formulas continue to hold,
with the normal bundle replaced by an
appropriately defined virtual normal bundle.}
However, a complete exposition for higher genus, where
concavity or convexity is not enough to guarantee a
smooth moduli space, is beyond
the scope of this paper, and we refer the reader to the
discussion in section four of
\locvir\ , with whose notation
this paper is largely compatible.  The genus zero case has
been worked out in full by \kontloc\ (see the formula
at the end of section 3.3.4).

The upshot is that we can determine all the weights
and classes of the
bundles restricted to the fixed point loci systematically.
After dividing numerator (Chern class)
by denominator (Euler class of normal bundle), one has
polynomial class of degree equal to the dimension of the
fixed locus.
What's left is to integrate these classes over the moduli
spaces of curves (not maps) $\overline{M}_{g(v),val(v)}$
at each vertex.  These integrals obey
famous recursion relations \wittdgrav, which entirely
determine them.  A program for doing just this has been
written by \faber\ .  With this, and an algorithm for
summing over graphs (with appropriate symmetry factors),
one can completely automate the calculation of higher
genus Gromov-Witten invariants.  Subtleties remain,
however, regarding multicovers \pandh, \klemmhg, \vafagop\ .

\newsec{Virtual Classes and the Excess Intersection Formula}

One of the foundations of the theory of the moduli space of
maps has been the construction of the virtual class \litian\ \behfan\ .  
A given
space of maps ${\overline{\cal M}(\beta,X)}$ may be of the wrong
dimension, and the virtual class provides a way to correct for this.
We consider the ``correct dimension'' to be one imposed either by physical 
theory, or by the requirement
that the essential behavior of the moduli space be invariant under deformations of 
$X$ 
(this may include topological deformations of $X$, or just deformations of 
symplectic or almost
complex structures).  The virtual class is a class in the cohomology (or Chow) 
ring of
${\overline{\cal M}(\beta,X)}$; its principal properties are that it is a class in 
the 
cohomology ring of the expected
dimension, and that numbers calculated by integrating over the virtual class are
invariant under deformations of $X$.  See \litian\ \behfan\ 
for more exact and accurate statements.  One main
theme of this paper is to use the invariance under deformation to either calculate 
these numbers,
or to explain the significance of a calculation.

The idea of a cohomology class (or cohomology calculation) which corrects 
for ``improper behavior'' has been around for a long time in intersection
theory. One example is the excess intersection formula.  If we attempt
to intersect various classes in a cohomology ring, and 
 if we choose representatives of
those classes which fail to intersect transversely, the resulting dimension
of the intersection may be too large. The excess intersection formula allows
us to perform a further calculation on this locus to determine the actual class
of the intersection.  The purpose of this section is to describe the excess 
intersection formula for degeneracy loci of vector bundles, and to use this
formula to evaluate or explain some mirror symmetry computations.  In the 
cases we examine, the moduli space of maps to our space $X$ can be given
as the degeneracy locus of a vector bundle on a larger space of maps.
In each case that these degeneracy loci are of dimension larger than expected,
the virtual class will turn out to be the same as the construction
given by the excess intersection formula.  The virtual class and the excess
intersection formula are both aspects of the single idea mentioned above, and
have as common element in their construction the notion of the refined 
intersection class \fultoni.

To a vector bundle $E$ of rank $r$ on a smooth algebraic
variety $X$ of dimension $n$, we associate the Chern classes, 
$c_j(E)$,  $j=0,\ldots,r$, and also the total Chern class
$c(E) = 1 + c_1(E) + c_2(E) + \cdots + c_r(E)$.  These classes are
elements of the cohomology ring of $X$.
The class $c_j(E)$ represents a class of codimension $j$ in
$X$, and in particular the class $c_n(E)$ is a class in
codimension $n$, and can be associated with a number. 
For any class $\alpha$ in the ring,  the symbol 
$\int_X \alpha$ means to throw away all parts of $\alpha$ except those parts
in degree $n$, and evaluate the number associated to those parts.

For a vector bundle $E$ whose rank is greater or equal to the dimension $n$
of $X$, we are often interested in calculating the number associated to $c_n(E)$,
or in the previous notation, $\int_X c(E)$.  One way to compute Chern classes is 
to
realize them as degeneracy loci of linear combinations of sections.  If $E$ is of 
rank
$r \ge n$, and we take $r-n+1$ generic global sections 
$\sigma_1,\ldots,\sigma_{r-n+1}$,
the locus of points where $\sigma_1,\ldots, \sigma_{r-n+1}$ fail to be linearly 
independent
represents the class $c_n(E)$. Often this way of interpreting the Chern classes is 
the one
which has the most geometric meaning.  The statement ``generic'' above means that 
if we carry out this
procedure and find out that the degeneracy locus is of the correct dimension (that 
is: points),
then the sections were generic enough.

Sometimes the sections we can get our hands on to try and calculate $\int_X c(E)$ 
with are not
generic in this sense, and the degeneracy locus consists of some components which 
are positive
dimensional.  In this situation, the excess intersection formula tells us how to 
associate to
each positive dimensional connected component of the degeneracy locus a number, 
called the ``excess
intersection contribution''. This number is the number of points which the 
component ``morally'' accounts
for.  Part of the excess intersection theorem is the assertion that the sum of the 
excess intersection contributions over all the connected
components of the degeneracy locus, and the sum of the remaining isolated points 
add up to $\int_{X} c(E)$.
This corresponds to the invariance of numbers computed using the virtual class 
under deformations of the target manifold.

Let $Y$ be one of the connected components described above.
Let's assume for simplicity that
$Y$ is actually a submanifold of $M$.  In this situation the
excess intersection formula says that
the excess intersection contribution of $Y$ is
\eqn\excess{\int_{Y} { c(E) \over c(N_{Y/M})}.}
Here $N_{Y/M}$ is the normal bundle of $Y$ in $M$, and
the expression after the integral sign makes sense, since
$c(N_{Y/M})$ is an element of
a graded ring whose degree zero part is $1$, and so
$c(N_{Y/M})$ may be inverted in that
ring.

\subsec{Rational curves on the quintic threefold}

As an example of an application of the excess intersection formula to explain the 
significance
of a calculation, let us review the count of the rational curves on a quintic 
threefold as
explained by Kontsevich \kontloc\ .
Let $M_d={\cal M}_{00}(d,\bP^4)$ be the moduli space of maps of genus zero curves 
of degree $d$ to
$\bP^4$.  $M_d$ is of dimension $5d+1$.  Let $U_d$ be the vector bundle on $M_d$ 
whose fiber at any
stable map $(C,f)$ is $H^0(C,f^{*}{\cal O}_{\bP^4}(5))$; this is a bundle of rank 
$5d+1$.  The numbers
$K_d=\int_{M_d} c(U_d)$ have been 
computed by mirror symmetry, and the first few are $K_1=2875$, $K_2=4876875/8$, 
and $K_3=8564575000/27$.
To try and find a geometric interpretation of these numbers, we compute 
$\int_{M_d} c(U_d)$ by finding a global section of $U_d$ and examining its 
degeneracy locus. Let $F$ be
a generic section of ${\cal O}_{\bP^4}(5)$ on $\bP^4$ which cuts out a smooth 
quintic threefold $X$.  
We pull $F$ back to give us a global section of $U_d$, which
we call $\sigma_d$.  The degeneracy locus of $\sigma_d$ in
$M_d$ consists of those maps $(C,f)$ with $f(C)$ contained in this quintic 
threefold. This observation allows us
to use the $K_d$ to compute the number of rational curves of degree $d$ on the 
quintic threefold $X$.

In degree $1$ the degeneracy locus consists of one point for every
line mapping into $X$, and so we see that 
$K_1 = 2875$ is the number of lines in a quintic threefold. In degree two, the 
degeneracy locus of
$\sigma_2$ consists of an isolated point for every degree two rational curve in 
$X$, and $2875$ positive
dimensional loci, each one consisting of maps which map two to one onto a line in 
$X$.  To calculate
the actual number of degree two rational curves in $X$, we compute the excess 
intersection 
contribution of each of these positive dimensional components, and subtract from 
the previously computed total
of $4876875/8$.  We now compute this excess intersection contribution.

For each line $l$ in $X$, let $Y_l$ be the submanifold of $M_2$ parameterizing two 
to
one covers of $l$.  The normal bundle of $l$ in $\bP^4$ is 
$N_{l/\bP^4}= {\cal O}_{l}(1) \oplus {\cal O}_{l}(1) \oplus {\cal O}_{l}(1)$. A 
calculation on the tangent
space of $M_2$ shows that the normal bundle $N_{Y_l/M_2}$ is (at a map $(C,f)$) 
equal to
$H^0(C,f^{*}N_{l/\bP^4})$. 

Since the line $l$ is sitting in the quintic threefold $X$, its normal bundle maps 
naturally to the
normal bundle of $X$ in $\bP^4$, with kernel the normal bundle of $l$ in $X$.  
This gives us an exact
sequence:

$$ 0\longrightarrow {\cal O}_{l}(-1) \oplus {\cal O}_{l}(-1) \longrightarrow 
N_{l/\bP^4} \longrightarrow
{\cal O}_{l}(5) \longrightarrow 0.$$

Let $V_2$ be the bundle on $Y_l$ whose fiber at a map $(C,f)$ is $H^1(C,f^{*}{\cal 
O}_{l}(-1))$.  The above
short exact sequence on $l$ gives us the sequence
$$ 0 \longrightarrow N_{Y_l/M_2} \longrightarrow E_2 \longrightarrow V_2 \oplus 
V_2 \longrightarrow 0$$
of bundles on $M_2$.  The multiplicative properties of Chern classes in short 
exact sequences shows us that
the excess contribution of $Y_l$ is:
$$\int_{Y_l} {c(E_2) \over c(N_{Y_l/M_2})} = \int_{Y_l} c(V_2)c(V_2).$$

This last number is the Aspinwall-Morrison computation of $1/d^3$, or in this 
case, $1/8$. 
This gives the number of actual degree two rational curves on a quintic threefold 
as
$4876875/8-2875/8= 609250.$

Under the assumption that each rational curve in $X$ is isolated and smooth,
then similar computations
give the famous formula \candelas\ 
\eqn\multcov{K_d = \sum_{k\vert d}{n_{d/k}\over k^3},}
where $n_d$ is the number of rational curves in degree $d$.
A caveat:  It has been shown that
this assumption is false in at least one instance -- 
in degree five, some of the rational curves are
plane curves with six nodes \vain.  This doesn't affect the
computation until you try to calculate
multiple cover contributions from these curves.  For
example,\foot{We thank N. C. Leung for describing this
example to us.}
in degree ten we have double covers of these nodal curves.
The moduli space of double covers of a (once) nodal rational
curve has two components:  one being degree two maps to the
normalization of the nodal curve, which contributes
$1/8$ as for the smooth case; the other being a single
point representing two disconnected copies of the normalization
mapping down to the singular curve.  If $P, Q$ represent the
points on the normalization which are to be identified for the
nodal curve, there is a uniqe map from a domain curve with
two components and one node, where the node is mapped to $P$
on one copy and $Q$ on the other (these points are identified).
This double cover does not factor through the normalization.
If we have $n$ such curves, their double covers contribute
$n/8 + n.$  The integers $n_d$ obtained from the formula \multcov\
need to be shifted to have the proper enumerative interpretation
(``experimentally,'' this shift is integral, though this has not
been proven \katzcox\ ).

\subsec{Calabi-Yau threefolds containing an algebraic surface}

Let us consider the situation where we have a Calabi-Yau threefold, $X,$
in a toric variety, $\proj,$ and a smooth
algebraic surface, $B,$ contained within $X:$
\eqn\seqbxp{B \subset X \subset \proj.} 
We assume as well that
$B$ is a Fano surface so that $X$ may be deformed so that $B$ shrinks \morvaf.  
This is the scenario of interest to us in this paper.
Now since there are holomorphic maps of many degrees into $B,$ which therefore all
lie within $X,$ we will have an enormous degeneracy locus.  If $X$ is cut from 
some
section $s,$ then at degree $\beta$ the whole space $\overline{\cal 
M}_{0,0}(\beta; B)$
will be a zero set of the pull-back section $\tilde{s}.$\foot{Actually,
$\beta$ labels a class in $X$ which may be the image of a number of classes
in $B.$  In such a case, our invariants are only sensitive to the image
class, and represent a sum of invariants
indexed by classes in $B.$}  Therefore, we will need
to use the excess intersection formula to calculate the contribution of the
surface to the Gromov-Witten invariants for $X.$  From this, we will extract
integers which account for the effective number of curves due to $B.$

Mapping tangent vectors, we have from \seqbxp\
the following exact sequence:
$0\rightarrow N_{B/X}\rightarrow N_{B/\proj}
\rightarrow N_{X/\proj}\rightarrow 0.$
Note that $N_{B/X} = K_B,$ by triviality of $\Lambda^3 TX$ and the exact sequence
$TB\rightarrow TX\rightarrow N_{B/X}.$  Therefore, we have
$$0\longrightarrow K_B\longrightarrow N_{B/\proj}\longrightarrow
N_{X/\proj}\longrightarrow 0.$$
Given $(C,f) \in \overline{\cal M}_{0,0}(\beta; B),$ we can pull back
these bundles and form the long exact sequence of cohomology:
\eqn\longseq{\eqalign{0&\longrightarrow H^0(C,f^*K_B)
\longrightarrow H^0(C,f^*N_{B/\proj})\longrightarrow H^0(C,f^*N_{X/\proj})
\longrightarrow \cr &\longrightarrow H^1(C,f^*K_B)\longrightarrow
H^1(C,f^*N_{B/\proj})\longrightarrow ....}}
Now $H^0(C,f^*K_B) = 0$ since $B$ is Fano (its canonical bundle
is negative), and $H^1(C,f^*N_{B/\proj}) = 0$ when $B$ is a 
complete intersection of (of nef divisors), which we assume.
As a result, \longseq\ becomes a short exact sequence of the bundles
over $\overline{\cal M}_{0,0}(\beta; B) \subset
\overline{\cal M}_{0,0}(\beta; \proj)$ whose fibers are the corresponding
cohomology groups.  The bundle (call it $U_\beta)$
with fiber $H^1(C,f^*K_B)$ is the one
we use to define the local invariants.  The bundle with fiber
$H^0(C,f^*N_{B/\proj})$ is $N_{{\cal M}(B)/{\cal M}(\proj)}$ (abbreviating
the notation a bit).  That with fiber $H^0(C,f^*N_{X/\proj})$ is the
one used to define the (global) Gromov-Witten invariants for
$X$ -- call it $E_\beta.$  Therefore, we have
$$0\longrightarrow U_\beta \longrightarrow N_{{\cal M}(B)/{\cal M}(\proj)}
\longrightarrow E_\beta \longrightarrow 0.$$

Now using \excess\ with
$E = E_\beta;$
$M = \overline{\cal M}_{0,0}(\beta; \proj);$ and
$Y = \overline{\cal M}_{0,0}(\beta; B);$
the multiplicativity of the Chern class gives the contribution 
to the Gromov-Witten invariant of the threefold $X$ from
a surface $B\subset X$ is
\eqn\locinv{K_\beta = \int_{\overline{\cal M}_{0,0}(\beta; B)} c(U_\beta),}
which is what we have been computing.

Typically, the presence of a surface $B\subset X$ may not be
generic, so that $X$ can be deformed to a threefold $X'$ not
containing such a holomorphic surface.
Let $K^X_\beta$
be the Gromov-Witten invariant of $X,$ and let $K^{X'}_\beta$
be the Gromow-Witten of $X'.$  These are equal, as the Gromov-Witten
invariant is an intersection independent of deformation: 
$K^X_\beta = K^{X'}_\beta.$  For $X'$
we have an enumerative interpretation\foot{Singular rational
curves notwithstanding.} of $K^{X'}_\beta$ in terms of
$n'_\beta,$ the numbers of rational curves on $X'.$  Let
$n_\beta$ be the numbers of rational curves on $X,$ and let
$K_\beta$ be the integral in \locinv.
For simplicity, let us assume that
${\rm dim}H_2(X') = 1,$ so that degree is labeled by an integer: 
$\beta = d.$
Then combining the enumerative interpretation with the interpretation
of the excess intersection above, we find
$$K^{X'}_d = \sum_{k\vert d} n'_{d/k}/k^3 =
\sum_{k\vert d} n_{d/k}/k^3 + K_d.$$
Subtracting, we find
$$K_d = \sum_{k\vert d} \delta n_{d/k}/k^3.$$
Here $\delta n = n' - n$ represents the effective number of curves
coming from $B.$  In the text, we typically write $n$ for
$\delta n.$

We therefore have an enumerative interpretation of the local invariants.
After performing the $1/d^3$ reduction we get an integer representing
an effective number of curves (modulo multiple covers of singular
curves, which should shift these integers).  We should note that
one might ask about rational curves in the Calabi-Yau manifold which intersect our
Fano surface.  Such a situation would make for a more complicated
degeneracy locus, but it turns out this situation does not arise.
Indeed, if $C'\subset X$ is a holomorphic curve in $X$
meeting $B$ transversely, then $C'\cdot B > 0$ (strictly greater).
However, for $C \subset B,$ we have
$$C \cdot B = \int_D c_1(N_{B/X}) = \int_D c_1(K_B) < 0,$$
by the Fano condition.  Therefore, $C'$ cannot lie in the
image of $H_2(B)$ in $H_2(X)$ -- the only classes in which
we are interested -- and so
our understanding of the numbers $n_d$ is
therefore complete.

\subsec{Singular geometries}

For physical applications, we will often want the surface $B$
to be singular.  For example, in order to geometrically engineer
$SU(n+1)$ supersymmetric
gauge theories in four dimensions, we consider the 
local geometry of an $A_n$ singularity fibered over a $\p.$
In fact, we take a resolution along each $A_n$ fiber, so that
the exceptional divisor over a point is a set of $\p$'s intersecting
according to the Dynkin diagram of $SU(n+1).$  The total geometry
of these exceptional divisors forms a singular surface, which
is a set of $\p$ bundles over $\p$ (Hirzebruch surfaces) intersecting
along sections.  In \kkv\ it is shown
how the local invariants we calculate can be used to derive the
instanton contributions to the gauge couplings.  Roughly speaking,
the number of wrappings of the $\p$ base determines the instanton
number, while the growth with fiber degree
of the number of curves with a fixed
wrapping along the base determines the corresponding invariant.

It is clearly of interest, then, to be able to handle singular
geometries.  Actually, we will be able to do so without
too much effort.  Let us consider an illustrative example.
Define the singular surface $B'$ to be two $\pp$'s
intersecting in a $\p.$ 
This can be thought of as a singular
quadric surface, since it can be represented as the zero locus of the
reducible degree two polynomial
$$XY = 0$$
in $\proj^3$ with homogeneous coordinates
$[X,Y,Z].$  The generic smooth quadric is a surface $B=\p\times\p.$
If we express $B$ as a hypersurface in $\proj^3,$
we can define the local invariants (indexed only by the generator of
$H_2(\proj^3))$ of $B$ as an intersection calculation in
$\overline{\cal M}_{0,0}(d;\proj^3)$ as follows.
Define the bundle
$$E \equiv {\cal O}_{\proj^3}(2)\oplus {\cal O}_{\proj^3}(-2).$$
Then let $s_E = (s,0)$ be a
global section of $E,$ where $s$ is a 
quadric and $0$ is the only global section of ${\cal O}_{\proj^3}(-2).$
Note that, by design,
$E$ restricted to the zero set of $s_E$ is equal to $K_B.$
We now define a bundle over $\overline{\cal M}_{0,0}(d;\proj^3)$
whose fibers over a point $(C,f)$ are
$H^0(C,f^*{\cal O}_{\proj^3}(2))\oplus H^1(C,f^*{\cal O}_{\proj^3}(-2)).$
We then compute the top Chern class of the bundle, which can be
calculated as in the previous subsections
in terms of the zero locus of $\tilde{s}_E,$ which
picks out maps into $B \cong \p\times\p.$  The calculation
gives the usual local invariant
for $\p\times\p,$ counting
curves by their total degree $d = d_1 + d_2,$ where $d_i$
is the degree in $\proj^1_i,$ $i = 1,2.$  The reason is that
${\cal O}(2)\vert_B = N_{B/\proj^3},$ so the contribution from this
part to the total Chern class cancels with the normal bundle
to the map. (The local invariants are listed in the first column of Table 7.) 

Now note that this intersection
calculation is independent of the section we use to compute it.
In fact, if we use a reducible quadric whose zero locus is $B',$
the calculation will reduce to one on
the singular space $\overline{\cal M}_{0,0}
(d; B').$  The excess intersection formula tells us exactly
which class to integrate over this (singular) space.  In fact,
integration over the singular space is only defined via the
virtual fundamental class -- which is constructed to yield the
same answer.  In degree one, this can all be checked explicitly
in this example \grabp .  The upshot is that as our calculations
are independent of deformations, we can deform our singular
geometries to do local calulations in a simpler setting.
In fact, this makes intuitive physical sense:  the A-model
should be independent of deformations.

Another phenomenon that we note in examples is that the calculation
of the mirror principle can be performed without reference to a
specific bundle.  In other words, the toric data defining any
non-compact Calabi-Yau threefold works as input data.  As a result, we can
consider Calabi-Yau threefolds containing singular divisors and
perform the calculation.  For the example of $A_2$ fibered over
a sphere, we get the numbers in Table 4.  
Though this technique has not yet been proven to work,
it is tantalizing to guess that the whole machinery makes sense
for any non-compact threefold, with intersections taking place
in the Chow ring and with an appropriately defined prepotential.

In the next section, we will use the B-model to
define differential equations whose solutions determine
the local contributions we have been discussing.

\def\bP{{\bf P}}

\def\nec#1{\overline{NE}(#1)}

\newsec{Local Mirror Symmetry:  The B-Model}

In this section we describe the mirror symmetry calculation of the Gromov-Witten
invariants for a $(n-1)$-dimensional manifold $B$ with $c_1(B)>0.$ 
We first approach this by using mirror symmetry  
for a compact, elliptically fibered Calabi-Yau $n$-fold\foot{We will state 
formulas for $n$-folds, when possible.} $\hat X$ which contains  
$B$ as a section, and taking then the volume of the fiber to infinity.  
If $B$ is a Fano manifold or comes from a $(n-1)$-dimensional reflexive 
polyhedron a smooth Weierstrass Calabi-Yau manifold $\hat X$ with $B$ as a section exists. 
Moreover the geometry of $\hat X$ depends only on $B$ and therefore the 
limit can be described intrinsically from the geometry of $B$.
This is an intermediate step. Later, we will define the objects relevant 
for the B-model calculation for $B$ intrinsically, without referring to 
any embedding. Such an embedding, in fact, is in general not possible.       

\subsec{Periods and differential equations for global mirror symmetry}
 
We briefly review the global case in the framework 
of toric geometry, following the ideas and notations 
of \batyrev\hkty. According to \batyrev, a mirror pair $(X,\hat X)$ with the 
property 
$h_{p,q}(X)=h_{n-p,q}(\hat X)$ can be represented 
as the zero locus of the Newton polynomials\foot{The generalization to complete 
intersections in toric ambient spaces is worked out in \hkty\batyrevborisov.} 
$(P=0,\hat P=0)$ associated to a dual pair of reflexive $(n+1)$-dimensional 
polyhedra  $(\Delta,\hat \Delta).$  $X$ is defined as hypersurface  
by the zero locus of
\eqn\defP{
P=\sum_{\nu^{(i)}} a_i \prod_{j=1}^{n+1} X_j^{\nu^{(i)}_j}}
in the toric ambient space ${\bf P}_{\Sigma(\hat \Delta)}$, constructed by the 
complete
fan $\Sigma(\hat \Delta)$ associated to $\hat \Delta$. The sum \defP~runs over $r$ 
``relevant'' points $\{\nu^{(i)}\}\in \Delta$, which do not lie on 
codimension one faces and with 
$\nu^{(0)}$ we denote the unique interior point in $\Delta$. 
The $a_i$ parametrize the complex structure deformations of $X$
redundantly because of the induced $({\bf C}^*)^{n+2}$ 
actions on the $a_i$, which compensate $X_i\rightarrow \lambda_i X_i$, 
$P\rightarrow \lambda_0 P,$ such that $P=0$ is invariant.  Invariant 
complex structure coordinates are combinations 
\eqn\complexcoord{z_i=(-1)^{l_0^{(j)}}\prod_{j=0}^{r-1} a_j^{l^{(i)}_j},}
where the $l^{(j)},$ $j=1,\ldots,k=r-(n+1)$ are an integral basis of 
linear relations among the extended ``relevant'' points 
$\bar \nu^{(i)}=(1,\nu^{(i)})$ with $\nu^{(i)}\in {\rm rel}(\Delta)$, i.e. 
\eqn\linearrelations{\sum_{i=0}^{r-1} 
l_i^{(j)}\bar \nu^{(i)}=\vec 0.}

The $l^{(j)}$ have in the gauged linear sigma model \linearsigmamodel~the r\^ole
of charge vectors of the fields with respect to $U(1)^k$.
Moreover, if the $l^{(j)}$ span a cone in the 
secondary fan of $\Delta$, which correspond to a
complete regular triangulation of $\Delta$ 
\secondaryfan, then $l^{(j)}$ span the
dual cone (Mori cone) to the K\"ahler cone of 
${\bf P}_{\Sigma(\hat \Delta)}$, which is
always contained in a K\"ahler cone of $\hat X$ 
and $z_i=0$ corresponds to a point of maximal unipotent monodromy, which by the 
mirror
map \candelas\hkty~corresponds to the large K\"ahler structure limit of $\hat X$.
 
The period integrals of $X$ contain the information 
about the Gromov-Witten invariants of $\hat X$ (and vice versa). 
They are defined as integrals of the unique holomorphic $(n,0)$-form
over the $b_n$ ($=2(h_{2,1}(X)+1)$ when $n=3$) cycles $\Gamma_i$ in
the middle cohomology of $X$. 
The $(n,0)$-form  is given by a generalization of Griffiths 
residue expressions \griffith        
\eqn\nform{
\Omega={1\over (2 \pi i)^{n+1}}\int_{\gamma_0} {a_0 \omega \over P}, 
\qquad {\rm with} \ \ 
\omega= {\dd X_1\over X_1} \wedge \ldots \wedge {\dd X_{n+1} \over X_{n+1}},}
and $\gamma_0$ is a contour around $P=0$. General periods are then  
$\Pi_i(z_i)= \int_{\Gamma_i} \Omega$,
and for a particular cycle\foot{Which, when $n=3$ is dual to the 
$S^3$ which shrinks to zero at the generic
point in the discriminant.} this leads to the following simple integral  
\eqn\simpleperiod{\Pi_0(z_i)= {1\over (2 \pi i)^{n+1}}\int_{|X_i|=1}
{a_0 \omega \over P}.}

Because of the linear relations among the points \linearrelations, the 
expression $\hat \Pi(a_i)={1\over a_0} \Pi(z_i)$ fulfills the differential 
identities 
\eqn\diffidentg{\prod_{l^{(k)}_i> 0} \left({\partial\over
\partial a_i}\right)^{l^{(k)}_i} \hat \Pi= 
\prod_{l^{(k)}_i< 0} \left({\partial\over \partial a_i}\right)^{-l^{(k)}_i} \hat 
\Pi .}
The fact that the $\bar \nu^{(i)}$ lie on a hyperplane, together
with \linearrelations\ imply
the same
numbers of derivatives on both sides of \diffidentg, assuring equality.
Unlike the $\Pi(z_i)$ the  $\hat \Pi(a_i)$ are however not well defined under the 
${\bf C}^*$ action $P\rightarrow \lambda_0P$ defined above. To obtain differential 
operators ${\cal L}_k(\theta_i,z_i)$ annihilating $\Pi_i(z_i)$  one uses 
\diffidentg, 
$[\theta_{a_i},a_i^r]= r a_i^r$,  and the fact that $\Pi_i$
depends on the $a_i$ only through 
the invariant combinations $z_i$.
Here we defined logarithmic derivatives $\theta_{a_i}=a_i 
{\partial \over \partial a_i}$, $\theta_i=z_i{\partial \over \partial z_i}$.

{\sl Example}: $\hat X$ is the degree 18 hypersurface
in ${\bf  P}(1,1,1,6,9)$, with Euler number $-540$ 
and $h_{1,1}(\hat X)=2$ and $h_{2,1}(\hat X)=272$. The toric data are 
\eqn\toricdata{\eqalign{
a.)& \ \ \ \ \ \hat\Delta={\rm 
conv}\{[-6,-6,1,1],[-6,12,1,1],[0,0,-2,1],[0,0,1,-1],[12,-6,1,1]\}\cr
b.)& \ \ \ {\rm 
rel}(\Delta)=\{[0,0,0,0];[1,0,2,3],[0,1,2,3],[-1,-1,2,3],[0,0,2,3],[0,0,-1,0],[0,0
,0,-1]\}\cr 
c.)& \ {\rm triang}=\{[0,1, 2, 4, 5], [0,1, 3, 4, 5], [0,1, 3, 4, 6], [0,2, 3, 4, 
5], [0,1, 2, 4, 6], [0,2, 3, 5, 6],\cr 
            &\qquad \qquad [0,1, 2, 5, 6], [0,1, 3, 5, 6], [0,2, 3, 4, 6] \}\cr
d.)& \ \ \  {\cal SRI}=\{x_1=x_2=x_3=0, x_4=x_5=x_6=0\}\cr
e.)& \ \ \ \ \ l^{(1)}=(-6;0,0,0,1,2,3),\qquad l^{(2)}=(0;1,1,1,-3,0,0)\cr
f.)& \ \ \ \ \ J_1^3=9 \quad J_1^2 J_2=3 \quad  J_1 J_2^2=1,\quad  c_2
J_1=102,\quad  c_2 J_2=36 .}}
Here $\rm triang$ is a regular star triangulation of $\Delta$, where the 4d
simplices are specified by the indices of the points in ${\rm
rel}(\Delta)$. ${\cal SRI}$ denotes the Stanley Reisner Ideal. $J_iJ_kJ_l$ and 
$c_2J_i$
are the triple intersection numbers and the evaluation of the second chern
class on the forms $J_i$, i.e. $\int c_2 J_i$.     
Then by \defP~$X$ is given by
$$P=a_0+X_3^2 X_4^3(a_1 X_1 +a_2 X_2+{a_3\over X_1 X_2}+a_4)+{a_5\over 
X_3}+{a_6\over X_4}=a_0+\Xi
$$    
and the period \simpleperiod~ is easily integrated in the 
variables\complexcoord~$z_1={a_4 a_5^2 a_6^3\over a_0^6}$, $z_2={a_1 a_2 a_3\over 
a_4^3}$ 
$$\eqalign{
\Pi_0&={1\over (2 \pi i)^{4}}\int{1\over {1+{1\over a_0}\Xi}}\omega=
\left[\sum_{n=0}^\infty \left(-{\Xi\over a_0}\right)^n\right]_{{\rm
term\;constant\;in\;}X_i}\cr
&=\left[\sum_{n=0}^\infty (-a_0)^{-n}\sum_{\nu_1+\ldots+\nu_6=n}
\left(n!\over \nu_1!\ldots \nu_6!\right)(a_1 X_1 X_3^2 X_4^3)^{\nu_1}\cdots 
\left(a_6\over X_4\right)^{\nu_6}\right]_{{\rm
term\;constant\;in\;}X_i}\cr
&=\sum_{r_1=0,r_2=0}^\infty 
{\Gamma(6 r_1+1)\over {\Gamma(r_2+1)^3\Gamma(r_1- 3 r_2+1)\Gamma(3 r_1+1)\Gamma(2 
r_1+1)}} 
z_1^{r_1} z_2^{r_2}.}$$
Likewise, it is easy to see that \diffidentg~leads to   
\eqn\diffpii{\eqalign{
{\cal L}_1&=\t1(\t1 - 3\t2) - 12(6\t1-5) (6\t1-1)z_1\cr
{\cal L}_2 &= \t2^3 - (1 + \t1 - 3\t2)(2 + \t1 - 3\t2)(3 + \t1 - 3\t2)z_2\ ,}}
where we factored from the first operator a degree four differential operator. 
This is equivalent to discarding four solutions which have incompatible 
behavior at the boundary of the moduli space to be periods, 
while the remaining $(2h_{2,1}(X)+2)$ solutions
can be identified with period integrals for $X$.

Note that in general at the point of maximal unipotent
monodromy \candelas\morrison\hkty\batyrevII\  
$z_i=0,$ $\Pi_0=1+O(z)$ is the only holomorphic solution to the Picard-Fuchs 
system. 
Let us now\foot{Some aspects for the case of arbitrary $n$ are discusses in 
\gmp\klry\mayr .} set $n=3$.  
In general, there will be $h_{2,1}(X)$ logarithmic solutions of the form
$\Pi_i={1\over 2 \pi i}\log(z_i) \Pi_0+holom.,$ and
\eqn\mirrormap{t_i={\Pi_i(z)\over \Pi_0(z)}}
defines affine complex structure parameters of $X$, which at $z_i=0$ can be 
identified with the 
complexified K\"ahler parameters $t_i=i Vol(C_i) +B(C_i)$ of $\hat X$, following 
\candelas . 
The relation \mirrormap\ is called the ``mirror map'' and in particular, 
in the limit $Vol(C_i)\rightarrow \infty$ one has 
\eqn\limit{\log(z_i)\sim -Vol(C_i).}
$h_{2,1}$ further solutions are quadratic, and one is cubic in the logarithm. 
These solutions are related to each other and to the quantum corrected triple 
intersection 
$c_{i,j,k}$ by special geometry, basically Griffith transversality\foot{If $n$ is 
even we get in general 
algebraic relations between the solutions {\sl and} differential relations. The  
algebraic relations in the $K3$ 
case are well known, in the $3$-fold case we have special geometry, for $4$-folds 
the algebraic 
and differential relations can be found in \gmp\mayr\klry.}    
$\int (\partial_i \Omega) \wedge \Omega=\int (\partial_i\partial_j \Omega) 
\wedge \Omega=0$. As a consequence, these quantities derive 
from a prepotential, which has the general form \candelas \hkty (${\rm 
Li}_3(x)=\sum_{k=1}^\infty {x^k\over k^3}$)
\eqn\prepotential{
{\cal F}={J_i\cdot J_k \cdot J_l\over 6}t_it_jt_k+{1\over 24} c_2\cdot J_i t_i-
i {\zeta(3)\over 2(2 \pi)^3} c_3+\sum_{d_1,\ldots,d_{h_{1,1}(\hat X)}}N_{\vec 
n}{\rm Li}_3(q^{d_1}
\cdots q^{d_{h_{1,1}(X)}})}
The relations are 
\eqn\allsolution{\eqalign{\vec \Pi&=
\Pi_0(1,t_i,\partial_{t_i} {\cal F},2{\cal F}-\sum_i t_i\partial_{t_i}{\cal 
F}),\cr 
c_{i,j,k}&=\partial_{t_i} \partial_{t_j} \partial_{t_k} {\cal F}=J_i \cdot 
J_j\cdot J_k+
\sum_{\{d_{i}\}} d_i d_j d_k N_{\vec d}
{{\vec{q}}^{\vec{d}}\over 1-{\vec{q}}^{\vec{d}}},}}
where ${\vec{q}}^{\vec{d}} \equiv \prod_{i}q_i^{d_i}$ and
the $N_{\vec{n}}$ are then
the Gromov-Witten invariants of the mirror $\hat{X}.$

\subsec{The limit of large elliptic fiber}
We now identify the classes of the curves $C_i$ and define the limit of 
large fiber volume.  Then we will use \mirrormap~to translate that into a limit 
in the complex structure deformations parameters $z_i$ of $X$. 
Note that in Batyrev's construction the points $\nu^{(i)}\in {\rm rel}(\Delta)$ 
correspond to monomials in $P$ as well as to divisors $D_i$ in 
${\bf P}_{\Sigma(\Delta)}$ (in the example, ${\bf P}_{\Sigma(\Delta)}=
{\bf P}(1,1,1,6,9)$) 
which intersect $\hat X $. Each $l^{(i)}$ defines a wall
in the K\"ahler cone of $\hat X$ at which curves in the class $[C_i]$ vanish.  
Moreover, the entries $l^{(j)}_i$, $i=1,\ldots,r$
are the intersection of these curves $C_j$ 
with the restriction $\tilde D_i$ of the  divisors $D_i$, $i=1,\ldots,r$ to $\hat 
X$. 
{}From this information one can identify the classes $[C_i]$ in $\hat X$. It is 
convenient to use the 
Cox coordinate ring representation \cox~  
${\bf P}_{\Sigma(\Delta)}=\{C[x_1,\ldots x_{r}]\setminus {\cal SRI}\}/({\bf 
C}^*)^k$, 
where the ${\bf C}^*$-actions are given by $x_i\rightarrow x_i 
(\lambda^{(j)})^{l^{(j)}_i}$ 
and ${\cal SRI}$ denotes the Stanley Reisner Ideal. 
In these coordinates $D_i$ is simply given by $x_i=0$ and the polynomial 
reads 
\eqn\batcoxp{\hat P=\sum_{i=0}^{\hat r-1}a_i\prod_{j=0}^{r-1} x_j^{\langle 
\nu^{(j)},\hat \nu^{(i)}\rangle+1}\ .} 

In the example, 
\eqn\batcoxpoly{\hat P=x_0(x_4^6 g_{18}(x_1,x_2,x_3)+x_4^4 x_5 f_{12}(x_1,x_2,x_3) 
+x_5^3+x_6^2)} 
has a smooth Weierstrass form.  We notice, taking into account \toricdata
parts d and e, 
that $D_4$ meets $\hat X$ in a ${\bf P}^2$, the section of the Weierstrass form. 
As $[C_2]\cdot D_4=-3$, $C_2$ must be contained in this $\IP^2,$ and from 
$C_2\cdot D_i=1,$ $i=1,2,3,$ it follows that $C_2$ lies in that $\IP^2$
with degree\foot{As a consistency check, 
note that $J_2$ which is the dual divisor to $C_2$ must then have a component 
in the base $P^2$ and one in the fiber direction,
hence on dimensional grounds it can at most 
intersect quadratically comp \toricdata, part f.} $1$. 
$[C_1]\cdot \tilde D_4=1$, hence $C_1$ meets the section once and must be a curve 
in the fiber direction, whose volume goes to infinity in the large fiber limit.
By this association $l^{(1)}=l^{(F)}$ and as $\log (z_F)\sim - Vol(C_F)$, 
$z_F={a_4 a_6^2 a_6^3\over a_0^6}\rightarrow 0$ is
the correct complex structure limit. Next
we pick the periods which stay finite in this limit -- that is,
whose cycle has still compact support.  From \prepotential,\allsolution,
and \toricdata, part f, we see that the finite solutions are 
$\Pi_0$, $\Pi_0 t_2$ and $\Pi_0(\partial_1-3\partial_2){\cal F}$. Moreover,
as they do not 
contain $\log(z_1)$ terms they satisfy in
this limit (and are in fact determined by) the 
specialization of \diffpii\ as $z_1 \rightarrow 0:$ 
\eqn\localdiffpii{{\cal L}= \theta^3 + 3 z \theta (3\theta+2)( 3\theta+1),}
where we have put $z \equiv z_2$ and $\theta \equiv \theta_2.$

This differential equation comes from the relation
of the points in the two dimensional face
$\Delta_B={\rm conv}\{\nu_1,\ldots, \nu_4\}$
in $\Delta.$  We call the Newton polynomial for this set of points
$P_B$. We want to define a special limit of the finite $\Pi_i(z)$. 
The limit is $a_5,a_6\rightarrow 0$ in $P$, which is compatible with $z_F=0$. 
We define $W=X_3^2X_4^3$ and $V=(X_3X_4)^{-2}.$  Then $Jac\sim X_3,$ 
$\omega'= d W\wedge {d V\over V}\wedge {dX_1 \over X_1}\wedge 
{dX_2 \over X_2},$ and $\omega''=d W\wedge {dX_1 \over X_1}\wedge 
{dX_2 \over X_2}$. As $1/W$ is the non-compact direction, perpendicular to
the compact plane, we remove the compactification point in the
$1/W$-loop, which becomes open. Hence   
$$\eqalign{&\Pi_0(z)=  {c\over (2 \pi)^4} 
\int_{|X_i|=1\atop |V|=1}\int_{\epsilon}^{1} 
{a_0\over {W (a_0+W P_B)}} \omega'+O(a_5,a_6)\cr 
&\approx
{c\over (2 \pi i)^3}\int_{|X_i|=1}\int_{\epsilon}^{1} 
{1\over {W (1+{W P_B\over a_0})}} \omega''= {c\over (2 \pi 
i)^3}\int_{|X_i|=1}\int_{\epsilon}^{1} 
W^{-1}\sum_{i=0}^\infty (-)^i\left({{W P_B\over a_0}}\right)^i\omega''\cr 
&=\!-\!\log(\epsilon)\!-\! {c\over (2 \pi i)^3}\int_{|X_i|=1} \! \! \! \! \! \! 
\log(1+{P_B\over a_0}) 
{d X_1\over X_1}\wedge{d X_2\over X_2}= C\!-\! {c\over (2 \pi i)^3} 
\int_{|X_i|=1} \! \! \! \! \! \! \log(P_B'){d X_1\over X_1}\wedge{d X_2\over X_2}  
 ,}
$$
where $P'_B$ is $P_B$ with rescaled $a_i$.
 
\subsec{Local mirror symmetry for the canonical line bundle of a torically 
described 
surface}

We will generalize the example of the previous section to the situation where one 
has 
as smooth Weierstrass form for the threefold over some base $B$. 
The known list of bases 
which lead to smooth Weierstrass forms are
Fano varieties and torically described 
bases whose fans are constructed from the polyhedra $\Delta_B$, which we
display in figure 1. 
For these bases,
we can demonstrate the property of admitting a smooth Weierstrass form by
explicitly showing that total space constructed from 
$\Delta^{fibration}={\rm conv}\{(\nu_i^{B},\nu^{E}),
(0,0,-1,0),(0,0,0,-1)\}$ is smooth. 
Here $\nu_i^{B}$ runs over the 2-tuple of the coordinates of points in $\Delta_B$ 
and for $\nu^{E_k}$ 
one has the choice $(2,3)$, $(1,2),$ and $(1,1)$ for the $E_8,E_7,E_6$
respective fiber types described below.

For the smooth fibrations, all topological
data of $\hat X$ are  expressible from the base topology
and we have a surjective map $i^*:H^{1,1}(X)\rightarrow H^{1,1}(B)$. 
Using the adjunction formula\foot{This calculation arises in the F-theory context 
\ref\FMW{R. Friedman, J. Morgan, E. Witten, ``Vector Bundles and F Theory,''
Commun. Math. Phys. {\bf 187} (1997) 679-743, Section 7;
hep-th 9701162.}.} 
one finds (here we understand that on the left side we integrate over $\hat X$ and 
on the right 
side over $B$) 
\eqn\top{\eqalign{
c_3(\hat X)&=-2 h c_1(B)^2 \cr
c_2(\hat X) J_E&= k  c_2(B)+k({12\over k}-1) c_1(B)^2, \quad 
c_2(\hat X) J_i= 12 k c_1(B) J_i\cr
J_E^3=& k  c_1^2(B), \quad  
J_E^2 J_i= k c_1(B) J_i, 
\quad J_E J_i J_k=k  J_i J_k,}}
Here $J_E$ is a cohomology element supported on the elliptic fiber; its
dual homology element is the base.
The $J_i$ are cohomology elements supported on curves in $B$,
with homology dual curves in $B$ together with their fibers. 
$k$ is the ``number'' of sections for the various Weierstrass forms, i.e. 1 for 
the 
$E_8$ form $X_6(1,2,3)$, 2 for the $E_7$ form $X_{4}(1,1,2)$, 3 for the $E_6$ 
form $X_3(1,1,1)$ and  4 for the $D_5$ form $X_{2,2}(1,1,1,1)$; $h$ is the dual 
Coxeter number associated with the groups:  $h=30,18,12,8,$ respectively.       

{}From \prepotential~and \allsolution~we get that (were $c_1=c_1(B)$)
$$\eqalign{
\partial_{t_E}{\cal F}&=t_E^2 c_1^2+t_E \sum_i t_i (c_1 J_i)+t_i t_j(J_j 
J_i)+O(q)\cr 
\partial_{t_i}{\cal F}&=t_E^2 (c_1 J_i)+t_E\sum_j t_j (J_j J_i)+O(q)}$$
and hence the unique finite combination in the $t_E\rightarrow i \infty$ limit 
is given by
\eqn\finite{\Pi_{fin}=(\partial_{t_E}-\sum_i x_i \partial_{t_i}){\cal F}, \ {\rm 
with}\  \sum_{i} x_i (J_iJ_j) =c_1 J_j}
We define new variables in the Mori cone $t_E=S$ and $t_i=\tilde t_i-x_i  S$ such
that $\Pi_{fin}=\partial_S {\cal F}|_{t_E\rightarrow i \infty},$ and by
\allsolution\ we get a general form
of the instanton expansion for the curves which live in the base 
\eqn\instbase{g_{i,k}:=C_{S,i,k}|_{q_E=0}=J_i J_k + {1\over 2}(c_1 J_i+c_1 J_k)+ 
\sum_{\{d_l\}} (\sum_{i=1}^{h^{1,1}(B)} -x_i d_i) d_j d_k
N_{\vec{d}} { {\vec{q}}^{\vec{d}}\over 1- {\vec{q}}^{\vec{d}} }\ ,}
where the $d_l$ run only over degrees of classes in the base. 
Note that ${\cal F}_{local}=\partial_S{\cal F}$ is a potential for the 
metric $g_{i,j},$. For the polyhedra $2-14$  ${\rm Im} (g_{i,j})$  becomes 
in a suitable limit the exact  
gauge coupling for an $N=2$ theory in 4 dimensions 
(Seiberg-Witten theory) \kkv .   
Furthermore, all intersections in \instbase\ are intersections in the 
two-dimensional base manifold and in this sense we have achieved the goal of 
formulating the mirror symmetry conjecture intrinsically from the
geometric data of the base. It remains to construct a local Picard-Fuchs system 
which has    
${\cal F}_{local}= \sum_{i,j=1}^{h^{1,1}(B)} (J_i J_j) 
\log z_i \log z_j + S_i \log z_i+S_0$ as the unique solution
quadratic in the logarithms. 

As a generalization of the situation discussed in the last section, 
we propose the following data for local mirror symmetry: 
A convex $n-1$ dimensional polyhedron  $\Delta_B$, $P_B$ its 
Newton polynomial,  
\eqn\lnform{
\eqalign{&\Pi_i(z_i)= \int_{\Gamma_i} \Omega,\cr 
\Omega&=\int_{\gamma_0} {\log (P_B) \omega }:  \ \ \gamma_0: \ {\rm contour\  
around }\ P_B=0, \quad 
\omega= {{\dd} X_1\over X_1} \ldots {{\dd} X_{n-1} \over X_{n-1}}\cr
z_i&=\prod_{j=1}^{\# \nu} a_j^{l^{(i)}_j}: \ \
({\bf C}^*)^{n+2}-{\rm invariant \ complex\ 
structure\ variables, }}}
$\{l^{(i)}\}$ a basis  of linear relations among the 
$\bar \nu^{(i)}\in\bar  \Delta_B=(1,\Delta_B)$, i.e. $\sum_{i=1}^{\# \nu} 
l^{(j)}_i 
\bar \nu^{(i)}=\vec 0$ spanning the Mori cone of $P_{\Delta_B}$. 

The non-compact local geometry $T_{\Sigma(\Delta_B)}$ is 
the canonical bundle over ${\bf P}_{\Delta_B}$. It is described by the 
incomplete fan $\Sigma(\Delta_B)$, which is spanned in three dimensions ($n=3$) 
by ``extended'' vectors $\bar \nu^{(i)}=(1,\nu^{(i)}),$ with 
$\nu^{(i)}\in \Delta_B$ generate the lattice.  

The local mirror geometry is in these cases given by an elliptic curve 
which is defined in coordinates as \batcoxp, w.r.t. to
$(\Delta_B,\hat\Delta_B)$
and a meromorphic two-form $\Omega$ with nonvanishing residue. 
The number of independent cycles increases for the polyhedra $1-15$ with 
the number of nonvanishing residue, which in physics play the r\^ole of
scale or mass parameters. This is in contrast to the situation in 
the next section, where the genus of the Riemann surface 
will increase and with it the number of double logarithmic solutions.

In particular, for the example discussed before we 
get from \batcoxp~that the mirror geometry is the elliptic curve given by the 
standard cubic in ${\bf P}^2$ 
$$P_B=a_1 x_1^3+a_1 x_2^3+a_3 x_3^3+a_0x_1x_2x_3.$$
An important intermediate step in the derivation of this form \batyrev, 
which be useful later, is that the polynomial
can also be represented by coordinates $Y_i,$
\eqn\intstep{P_B=a_1 Y_1+a_1 Y_2+a_3 Y_3+a_0 Y_0\ ,}
with $r$ relations $\prod_{i=1}^{\# \nu-1} Y_i^{l^{(k)}_i}=Y_0^{-l^{(k)}_0}$, 
$k=1,\ldots,r$.  
The $x_i$ are then introduced by an suitable \'etale map, here
$Y_1=x_1^3$, $Y_2=x_2^3$, $Y_3=x_3^3$ and 
$Y_0=x_1x_2x_3$, which satisfy
identically the relation(s).

The $\Pi_i(z)$ are now well-defined under ${\bf C}^*$-actions, up to a shift,
and they satisfy directly 
\eqn\diffident{\prod_{l^{(k)}_i> 0} 
\left({\partial\over \partial a_i}\right)^{l^{(k)}_i} \Pi= \prod_{l^{(k)}_i< 0} 
\left({\partial\over \partial a_i}\right)^{-l^{(k)}_i}  \Pi .}
In particular $\Pi_0=1$ is always a solution. 
By the same procedure as indicated below \diffidentg,
we now directly get the differential equation \localdiffpii. 
One can easily show that this
differential equation has besides the constant solution
a logarithmic and a double-logarithmic 
solution. The explicit form of the solutions can be given  in general using the 
$l^{(i)}$, $i=1,\ldots m$ in
specialized versions of the formulas which appeared in \hkty:
\eqn\expsol{\eqalign{
\Pi_0(z)&=\sum_{\vec n} c(\vec n,\vec \rho)
z_{\vec n}^{\vec n}|_{\vec \rho=0},\quad \quad 
c(\vec n,\vec \rho)=
{1\over \prod_i\Gamma(\sum_\alpha l^{(\alpha)}_i(n_\alpha+\rho_\alpha)+1)},\cr
\Pi_i(z)&=\partial_{\rho_i}\Pi_0|_{\vec \rho=0},\qquad \Pi_{m+1}=\partial_S{\cal 
F}=
\sum_{i,j} (J_i J_k) \partial_{\rho_i}\partial_{\rho_j}\Pi_0|_{\vec \rho=0}.}
}
The predictions for the local mirror symmetry 
are then obtained using \mirrormap~and \instbase.

{\baselineskip=12pt \sl
\goodbreak\midinsert
\centerline{\epsfxsize 5.5 truein\epsfbox{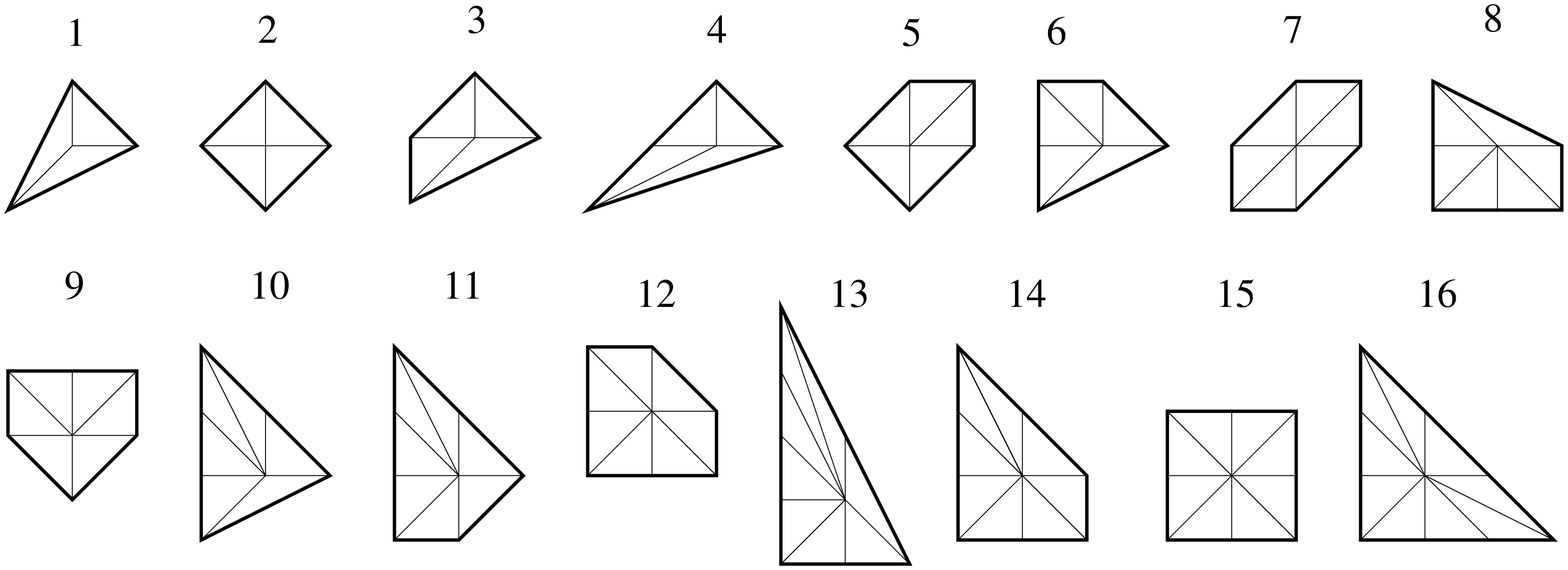}}
\leftskip 1pc\rightskip 1pc \vskip0.3cm
\noindent{\ninepoint  \baselineskip=8pt  
{{\bf Fig. 1:} 
Reflexive Polyhedra ($\Delta_B$) in two dimensions. 
$\hat \Delta_p=\Delta_{17-p}$ for $p=1,\ldots,6$. $\Delta_{7,8,9,10}$ are self-dual 
\koelman\batyrev$^{17}$. 
Case one is the polyhedron representing the ${\bf P}^2;$
two, three and four are the Hirzebruch surfaces 
$F_0={\bf P}^1\times {\bf P}^1,F_1$ and $F_2$
and the others are various blow-ups of these cases. 
Note that $c_2=\# {\rm 2\  simpl.}$ and $c_1^2=12-\# {\rm 2\  simpl}$. 
The labeling starts with
$0$ for the inner point. 
The point to its right is point $1$ and the labels of the others increase 
counterclockwise. 
}}
\endinsert}

Below we give further data for local
mirror symmetry calculation for some\foot{In the cases we do not treat explicitly, 
the Mori 
cone is non-simplicial. This means that there are several coordinate choices for 
the 
large complex structure variables, which correspond to the simplical cones in a 
simplicial 
decomposition of the Mori cone. This is merely a technical complication. We 
checked that 
for the simplicial subcones we get  consistent instanton expansions 
from \instbase.}  from  of the polyhedra\foot{The polyhedra appeared only in the preprint version of \batyrev~and are therefore 
reproduced here. We thank J. Stienstra for pointing out an omission in an earlier version.} in Fig. 1:

$$\eqalign{
1. \quad & l^{(1)}=(-3,1,1,1),\quad {\cal C}_1= 3 J_1,\quad {\cal R}=J_1^2\cr
2. \quad & l^{(1)}=(-2, 1, 0, 1, 0),\ \ l^{(2)}=(-2,0, 1, 0, 1),\quad {\cal 
C}_1=2J_1+2J_2,\ \ {\cal R}=J_1J_2 \cr
3.  \quad & l^{(1)}=(-2, 1, 0, 1, 0), \ \ l^{(2)}=(-1,0, 1, -1, 1),\quad {\cal 
C}_1=3J_1+2J_2\ \ {\cal R}=J_1J_2 + J_1^2\cr
4.  \quad & l^{(1)}=(-2, 1, 0, 1, 0), \ \ l^{(2)}=(0 ,0, 1, -2, 1),\quad {\cal 
C}_1=4J_1+2J_2\ \ 
{\cal R}=J_1J_2 + 2 J_1^2\cr}$$
$$\eqalign{
5. \quad & l^{(1)}=(-1, 1, -1, 1, 0, 0), \ \ l^{(2)}=(-1,-1, 1, 0, 0, 1),\ \ 
l^{(3)}=(-1, 0, 1, -1, 1, 0)\cr
\ \ \ \quad & {\cal C}_1=3J_1+2J_2+2J_3\quad {\cal R}=J_1^2+J_2J_1+J_1J_3+J_2J_3 
\cr
6.\quad  & l^{(1)}=(-1, 1,-1,  1, 0, 0), \ \ l^{(2)}=(0, 0, 0, 1, -2, 1),\ \ 
l^{(3)}=(-1,0,  1,-1,1, 0)\cr
\ \ \ \quad & {\cal C}_1=4J_3+2J_2+3J_1 \quad {\cal R}= J_2J_3 + 2J_3^2 + J_2J_1 + 
2J_3J_1 + J_1^2\cr
11.\quad  & l^{(1)}=(-1, 0,-1,  0, 0, 1,1), \ \ l^{(2)}=(0,0, 1, 0, 1, -2, 0),\ \ 
\cr 
 \ \ \ \quad & l^{(3)}=(0,1,0, 0,-2,1, 0),\ \ 
l^{(4)}=(0,0,1,1,0,0,-2), \quad {\cal C}_1=6J_1+4J_2+2J_3+3J_4 \cr
\ \ \ \quad &{\cal R}= 6J_1^2+4J_1J_2+2J_2^2+2J_1 J_3+J_2J_3+
3J_1J_4+2J_2J_4+J_3J_4J_4^2}$$
Here we use the short-hand notation ${\cal R}=\sum J_i J_k  \int_B J_i J_k$ and 
${\cal C}_1=\sum_{i} J_i \int c_1(B)J_i$. Below are further Picard-Fuchs 
systems derived using \diffident.

$$\eqalign{
{\cal L}^{F_0}_1 &= \t1^2 - 2 z_1 (\t1 + \t2)(1 + 2\t1 + 2\t2)\cr
{\cal L}^{F_0}_2 &= \t2^2 - 2 z_2 (\t1 + \t2)(1 + 2\t1 + 2\t2)}$$ 

$$\eqalign{
{\cal L}^{F_1}_1&= \t2^2 - z_1 (\t2 -\t1)(2\t1 + \t2)\cr
{\cal L}^{F_1}_2&= \t1(\t1 - \t2) - z_2(2\t1 + \t2) (1 + 2\t1 + \t2)\cr
}$$

$$\eqalign{
{\cal L}^{F_2}_1&= \t1(\t1+\t2) - z_1 2 \t1(2\t1 +1)\cr
{\cal L}^{F_2}_2&= \t1(\t1+\t2) - z_2 2\t2 ( 2 \t2+1)\cr
}$$
In general, linear combinations of the $l^{(i)}$ may lead to to independent 
differential 
operators. 
For example, a complete system for the
blown up $F_2$ (polyhedron 6) is obtained using
in addition to the $l^{(i)}$, $i=1,2,3,$ the linear relations $l^{(2)}+l^{(3)}$ 
and 
$l^{(1)}+l^{(3)}:$
$$
\eqalign{{\cal L}_1&= \t2(\t2 - \t3 + \t1) - (-2 + 2\t2 - \t3)(-1 + 2\t2 - \t3) 
z_1,\cr
         {\cal L}_2&= (2\t2 - \t3)(\t3 - \t1) -
(1 + \t2 - \t3 + \t1)(-1 + \t3 + \t1)z_2, \cr
         {\cal L}_3&= (-\t2 + \t3 - \t1)\t1 - (1 + \t3 - \t1)(-1 + \t3 + 
\t1)z_3,\cr
         {\cal L}_4&= \t2(\t3 - \t1) - (-1 + 2\t2 - \t3)(-1 + \t3 + \t1)z_1 z_2, 
\cr
         {\cal L}_5&= -((2\t2 - \t3)\t1) - (-2 + \t3 + \t1)(-1 + \t3 + \t1) z_2 
z_3.}$$

Using similar arguments as in section four of the second
reference in \hkty\ and the calculation of toric
intersections as described in \oda,\fulton,\danilov, one can show that \diffident\
and \expsol\ implies the appearance of the intersection numbers in \instbase. 

Concrete instanton numbers for $\IP^2$ appear in Table 1;
for $K_{F_0}$, $K_{F_1}$ and $K_{F_2}$ in the appendix; 
and for the canonical bundles over the geometry defined by the 
polyhedra $P_5$ and $P_6$ in table 3 below. 
\vskip 0.5cm
{\vbox{\ninepoint{
$$
\vbox{\offinterlineskip\tabskip=0pt
\halign{\strut\vrule#
&\hfil~$#$
&\vrule#&~
\hfil ~$#$~
&\hfil ~$#$~
&\hfil $#$~
&\vrule#\cr
\noalign{\hrule}
&d_1=0 && d_3    & 0   &  1  &  \cr
\noalign{\hrule}
&d_2 &&          &     &     & \cr
&0   &&          &     &   1 & \cr
&1   &&          &  1  &   0 & \cr
\noalign{\hrule}}}
\ \
\vbox{\offinterlineskip\tabskip=0pt
\halign{\strut\vrule#
&\hfil~$#$
&\vrule#&~
\hfil ~$#$~
&\hfil ~$#$~
&\hfil $#$~
&\vrule#\cr
\noalign{\hrule}
&d_1=1 && d_3    & 0   &  1  &  \cr
\noalign{\hrule}
&d_2 &&          &     &     & \cr
&0   &&          &  1  &  -2 & \cr
&1   &&          &  -2 &   3 & \cr
\noalign{\hrule}}}
\ \
\vbox{\offinterlineskip\tabskip=0pt
\halign{\strut\vrule#
&\hfil~$#$
&\vrule#&~
\hfil ~$#$~
&\hfil ~$#$~
&\hfil ~$#$~
&\hfil $#$~
&\vrule#\cr
\noalign{\hrule}
&d_1=2 && d_3    & 0   &  1  & 2 & \cr
\noalign{\hrule}
&d_2 &&          &     &     &    & \cr
&0   &&          &     &     &     & \cr
&1   &&          &     &  -4 & 5  & \cr
&2   &&          &     &  5  & -6 & \cr
\noalign{\hrule}}}
$$

$$
\vbox{\offinterlineskip\tabskip=0pt
\halign{\strut\vrule#
&\hfil~$#$
&\vrule#&~
\hfil ~$#$~
&\hfil ~$#$~
&\hfil ~$#$~
&\hfil ~$#$~
&\hfil $#$~
&\vrule#\cr
\noalign{\hrule}
&d_1=3 && d_3    & 0   &  1  & 2   &  3& \cr
\noalign{\hrule}
&d_2 &&          &     &     &     &    &\cr
&0   &&          &     &     &     &    &\cr
&1   &&          &     &     & -6  & 7  &\cr
&2   &&          &     & -6  &  35 & -32&\cr
&3   &&          &     &  7  & -32 & 27 &\cr
\noalign{\hrule}}}\ \
\vbox{\offinterlineskip\tabskip=0pt
\halign{\strut\vrule#
&\hfil~$#$
&\vrule#&~
\hfil ~$#$~
&\hfil ~$#$~
&\hfil ~$#$~
&\hfil ~$#$~
&\hfil ~$#$~
&\hfil $#$~
&\vrule#\cr
\noalign{\hrule}
&d_1=4 && d_3    & 0   &  1  & 2   &  3&     4&\cr
\noalign{\hrule}
&d_2 &&          &     &     &     &    &     &\cr
&0   &&          &     &     &     &    &     &\cr
&1   &&          &     &     &     & -8 &   9 &\cr
&2   &&          &     &     & -32 & 135& -110&\cr
&3   &&          &     &  -8 & 135 &-400&286  &\cr
&4   &&          &     &   9 &-110 & 286&-192 &\cr
\noalign{\hrule}}}
$$

$$
\vbox{\offinterlineskip\tabskip=0pt
\halign{\strut\vrule#
&\hfil~$#$
&\vrule#&~
\hfil ~$#$~
&\hfil ~$#$~
&\hfil ~$#$~
&\hfil ~$#$~
&\hfil ~$#$~
&\hfil $#$~
&\hfil $#$~
&\vrule#\cr
\noalign{\hrule}
&d_1=5 && d_3    & 0   &  1  & 2   &  3&     4&   5&\cr
\noalign{\hrule}
&d_2 &&          &     &     &     &     &       &       &\cr
&0   &&          &     &     &     &     &       &       &\cr
&1   &&          &     &     &     &     &  -10  &  11   &\cr
&2   &&          &     &     &     &-110 & 385   &-288   &\cr
&3   &&          &     &     & -110&1100 &-2592  & 1651  &\cr
&4   &&          &     &  -10& 385 &-2592& 5187  & -3038 &\cr
&5   &&          &     &   11&-288 & 1651& -3038 &  1695 &\cr
\noalign{\hrule}}}
$$
\vskip-7pt
\noindent
{\bf Table 3}: Invariants of $dP_2$(polyhedron 5). The  invariants
for $d_1=0$ sum to 2 and for all other degrees $d_1$  to zero. Note also that
the invariants for the blow up of $F_2$ (polyhedon 6) are related 
to the above by $n^{(6)}_{k,i,i+j}=n^{(5)}_{k,i,j}$. 
\vskip7pt}}
The $K_{F_0}$ and $K_{F_1}$ geometry\foot{This is true also for 
$K_{F_2}$, which can be seen as specialization in the complex structure moduli space of the $K_{F_0}$ case.}  
describes in the double scaling limit $N=2$ $SU(2)$ Super-Yang-Mills theory \kkv. Similarly in that
limit the geometry of the canonical bundle over ${\bf P}_{\Delta_5}$ and ${\bf P}_{\Delta_6}$ 
describes $N=2$ $SU(2)$ Super-Yang-Mills theory with one matter multiplet in the fundamental 
representation of $SU(2)$ \kkv.

\subsec{Fibered $A_n$ cases and more general toric grid diagrams}
The fibered $A_n$ geometry we will discuss here is motivated 
from physics \kkv. Note that the complexified K\"ahler 
moduli of this geometry yield the vector moduli space for the Type II-A compactification and 
the electrically/magnetically charged BPS states come from even dimensional D-branes wrapping 
holomorphic curves/four-cycles. Mirror symmetry on the fiber relates it 
to the geometry considered in \klmvw\ for which the vector moduli space of a type 
II-B compactification emerges from its complex deformations.
 
The type II-A geometry arises when inside a Calabi-Yau space 
an $A_n$ sphere tree is fibered over a ${\bf P}^1$. 
Again we consider the limit in which all 
other K\"ahler parameters of the threefold which do not control
the sizes of the mentioned ${\bf P}^1$'s 
become large.  In this case \instbase\ becomes the exact gauge coupling of 
$SU(n+1)$ $N=2$ Seiberg-Witten theory when one takes a double-scaling limit in 
which 
the size of the fiber ${\bf P}^1$  and the one of base ${\bf P}^1$ are taken small 
in a ratio
described in \kkv. 
We have already discussed
the simplest cases:  the Hirzebruch surfaces $F_2$, 
$F_0$ and $F_1$, which give rise to $A_1$ theory.  Next we consider 
the $A_n$ generalization of the $F_2$ case. We can describe it by 
$T_{\Sigma(\Delta_B)}$ 
as before (below \lnform), but it clearly
does not have the structure of a canonical bundle over a space.

{\baselineskip=12pt \sl
\goodbreak\midinsert
\centerline{\epsfxsize 2.5 truein\epsfbox{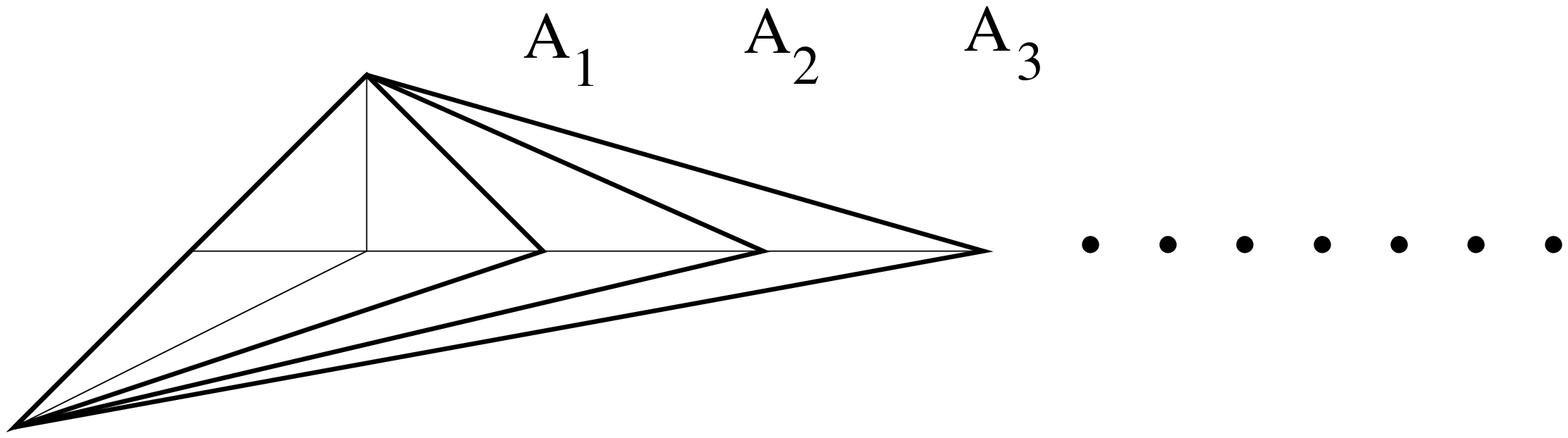}}
\leftskip 1pc\rightskip 1pc \vskip0.3cm
\noindent{\ninepoint  \baselineskip=8pt  
{{\bf Fig. 2:} 
Toric diagrams ($\Delta_B)$ for the $A_n$ singularity fibered over ${\bf P}^1$. 
Note that these diagrams have unique triangulations.  
}}
\endinsert}

To obtain the local situation as a limit of a compact case, we can consider a 
polyhedron $\Delta$
defined analogous to $\Delta^{fibration}$ described at the beginning of  
subsection 6.3. It turns out
that for $A_2$ ($A_3$) $\Delta$ becomes reflexive only after adding the point(s) 
$[1, 0, 1, 1]$ 
($[1, 0, 0, 1], [2, 0, 1, 2]$), giving us $2$ ($3$) line bundles normal to the  
to the compact part of the local geometry. 
This is true more generally, i.e. one has $n$ line bundles for 
$A_n$, but for large $n$ it is not possible to embed the $A_n$ singularity in a 
compact Calabi-Yau space.
The $n$ normal directions in the K\"ahler moduli space mean
that the limit \finite\ leads to an $n$-dimensional linear space of finite 
double-logarithmic solutions whose coefficients reflect 
the directions in the normal K\"ahler moduli space in which the limit can be taken 
and for 
combinatorial reasons their number corresponds to the number of `inner' points in 
the
toric diagram. We denote a basis for these directions $S_i$, $i=1,\ldots,n$. 
The $n$ double-logarithmic 
solutions come as we will see, given the meromorphic differential \lnform, from a 
genus $n$ Riemann 
surface for the local mirror. The Gromov-Witten invariants, on the other hand,
should not depend 
on the direction of the limit we take to obtain $\Pi_{fin}$ before matching it to 
the 
$N_{\vec d}$ in \instbase.  They therefore
become rather non-trivial invariants of the 
differential system \diffident.  In the following, we explicitly
describe for all $n$ the solutions of 
this system corresponding to a preferred period basis, up to a choice 
of the double-logarithmic solutions.

The generators of linear relations are 	
\eqn\lingen{\eqalign{
l^{(b)}&=(1,1,-2,\tbs 0,\tbs 0,\tbs 0,\tbs 0,\ldots,\tbs 0,\tbs 0,\tbs 0,\tbs 
0),\cr
l^{(1)}&=(0,0,\tbs 1, -2 ,\tbs 1,\tbs 0,\tbs 0,\ldots,\tbs 0,\tbs 0,\tbs 0,\tbs 
0),\cr
l^{(2)}&=(0,0,\tbs 0,\tbs 1, -2,\tbs 1,\tbs 0,\ldots,\tbs 0,\tbs 0,\tbs 0,\tbs 
0),\cr
&\quad \vdots\quad \quad \quad \quad \vdots\cr
l^{(n-1)}&=(0,0,\tbs 0,\tbs 0, \tbs 0,\tbs 0,\tbs 0,\ldots,\tbs 1,-2,\tbs 1,\tbs 
0),\cr
l^{(n)}&=(0,0,\tbs 0,\tbs 0, \tbs 0,\tbs 0,\tbs 0,\ldots,\tbs 0,\tbs 1, -2,\tbs
1).}}
Using the \'etale map $Y_0=z s$,
$Y_1={s\over z}$, $Y_2=s$ and $Y_k=s t^{k-2}$, $k=3,\ldots, n+3,$   
solving $Y_0 Y_1=Y_2^2$, $Y_iY_{i+2}=Y^2_{i+1}$, $i=2,\ldots,n+1$ on the 
polynomial 
$P=\sum_{i=0}^{n+3} a_i Y_i$ one gets \kkv
\eqn\localm{P=s z+{s\over z}+a_{n+1} s + a_{n} st+\ldots+b_0 st^{n+1} \ .}
This can be indentified, upon going to an affine patch $s=1$ after 
trivial redefinitions and the \kkv~limit, with the genus $n$ 
$SU(n+1)$ curves for $N=2$ Super-Yang-Mills theory \klty . 

To obtain the complete solutions of the period system to the local mirror 
geometry \localm, we have to specify the classical intersection terms in 
the $n$ double-logarithmic solutions $\partial_{S_i}{\cal F}$. 
Using \lingen,\diffident, the relation between the ideal of the 
principal part of the differential operator at the maximal unipotent point $z_i=0$ 
and the logarithmic solutions of GKZ-systems as in section 4 of \hkty,
as well as some algebra, we arrive at the following structure of the general 
double-logarithmic 
solution for arbitrary $n$  ($J_i\sim \log(z_i)$):
\eqn\logstructure{{\cal R}=\sum_{i=1}^n y_i J_i (J_b + \sum_{k=1}^i 2 k J_k).}
Here the coefficients of the $y_i$ can be viewed as logarithmic terms
of $\partial_{S_i}{\cal F}$. This leads by \expsol\ to an explicit basis of 
solutions. 

It remains to fix the $x_i$ by requiring invariance of the $N_{\vec d}$ in 
\instbase\
given the general solution \logstructure , which yields 
\eqn\xiyi{x_i=M_{i,j}y_j, \ i,j=1,\ldots, n \ ,}
where $M$ is the Cartan matrix of $A_n$. 
Using this description, we can calculate the instantons for all $A_n$. 

In the following we give some explicit numbers for $A_2.$
We arrive at these same numbers from A-model techniques, even
though this is a non-bundle case.  
$$
\vbox{\offinterlineskip\tabskip=0pt
\halign{\strut\vrule#
&\hfil~$#$
&\vrule#&~
\hfil ~$#$~
&\hfil ~$#$~
&\hfil $#$~
&\vrule#\cr
\noalign{\hrule}
&d_b=0 && d_2    & 0   &  1  &  \cr
\noalign{\hrule}
&d_1 &&          &     &     & \cr
&0   &&          &     &  -2 & \cr
&1   &&          &  -2 &  -2 & \cr
\noalign{\hrule}}}
$$
$$
\vbox{\offinterlineskip\tabskip=0pt
\halign{\strut\vrule#
&\hfil~$#$
&\vrule#&~
\hfil ~$#$~
&\hfil ~$#$~
&\hfil ~$#$~
&\hfil ~$#$~
&\hfil ~$#$~
&\hfil ~$#$~
&\hfil $#$~
&\hfil $#$~
&\vrule#\cr
\noalign{\hrule}
&d_b=1 && d_2    & 0   &  1  & 2   &     3 &     4&      5&   6&\cr
\noalign{\hrule}
&d_1 &&          &     &     &     &       &       &        &         &\cr
&1   &&          & -2  &  -2 &     &       &       &        &         &\cr
&2   &&          & -4  & -6  & -6  &-2d_3-2&   ..  &        &         &\cr
&3   &&          &  -6 & -10 & -12 & -12   &-4d_3-6&  ..    &         &\cr
&4   &&          & -8  & -14 & -18 & -20   & -20   &-6d_3-12&  ..     &\cr
&5   &&          & -10 & -18 &-24  & -28   & -30   &  -30   &-8d_3-20 &\cr
\noalign{\hrule}}}
$$
$$\eqalign{
&\vbox{\offinterlineskip\tabskip=0pt
\halign{\strut\vrule#
&\hfil~$#$
&\vrule#&~
\hfil ~$#$~
&\hfil ~$#$~
&\hfil ~$#$~
&\hfil ~$#$~
&\hfil ~$#$~
&\hfil ~$#$~
&\hfil $#$~
&\hfil $#$~
&\vrule#\cr
\noalign{\hrule}
&d_b=2 && d_2    & 0   &  1  & 2   &     3 &     4&      5&   6&\cr
\noalign{\hrule}
&d_1 &&          &     &     &     &       &       &        &          &\cr
&3   &&          &  -6 & -10 & -12 & -12   & -10   &  -4(d_3-1)-6 & .. &\cr
&4   &&          & -32 & -70 & -96 & -110  & -112  &-126    & -192     &\cr
&5   &&          & -110&-270 &-416 & -518  & -576  &  -630  &  ..      &\cr
\noalign{\hrule}}}\cr
&\vbox{\offinterlineskip\tabskip=0pt
\halign{\strut\vrule#
&\hfil~$#$
&\vrule#
&~\hfil ~$#$~
&\hfil ~$#$~
&\hfil ~$#$~
&\hfil ~$#$~
&\hfil ~$#$~
&\hfil ~$#$~
&\hfil $#$~
&\hfil $#$~
&\vrule#\cr
\noalign{\hrule}
&d_b=3 && d_2    & 0   &  1  & 2   &     3 &     4&      5&   6&\cr
\noalign{\hrule}
&d_1 &&          &     &     &     &       &       &        &          &\cr
&4   &&          & -8  & -14 & -18 & -20   & -20   &-20     & -18      &\cr
&5   &&          & -110&-270 &-416 & -518  & -576  &  -630  &  ..      &\cr
\noalign{\hrule}}}}
$$
\vskip-10pt
\noindent
{{\bf Table 4}: Gromov-Witten invariants for local $A_2$. 
For $d_3>d_2$ we have $n_{1,d_2,d_3}=-2(d_2-1)d_3-d_2(d_2-1)$. }
\vskip 10pt

For $A_3$:

$$\eqalign{
&\vbox{\offinterlineskip\tabskip=0pt
\halign{\strut\vrule#
&\hfil~$#$
&\vrule#~
&\hfil ~$#$~
&\hfil ~$#$~
&\hfil $#$~
& \hfil ~$#$~
& \hfil ~$#$~
&\vrule#\cr
\noalign{\hrule}
&d_b=0 && [d_2,d_3]    & [0,0] &  [1,0] &  [0,1] &   [1,1]  & \cr
\noalign{\hrule}
&d_1 &&                &       &        &        &          &\cr
&0   &&                &       &   -2   &  -2    &    -2    & \cr
&1   &&                &  -2   &   -2   &        &    -2    &  \cr
\noalign{\hrule}}}\cr
&\vbox{\offinterlineskip\tabskip=0pt
\halign{\strut\vrule#
&\hfil~$#$
&\vrule#~
&\hfil ~$#$~
&\hfil ~$#$~
&\hfil $#$~
&\hfil ~$#$~
&\hfil $#$~
&\hfil ~$#$~
&\hfil $#$~
&\hfil ~$#$~
&\hfil $#$~
&\hfil $#$~
&\vrule#\cr
\noalign{\hrule}
&d_b=1 && [d_2,d_4]    & [0,0] &  [1,0] &   [1,1]  
&[2,0]&[2,1]&[2,2]&[3,0]&[3,1]&[3,2]&\cr
\noalign{\hrule}
&d_1 &&                &       &        &          &     &     &     &     &     & 
    &\cr
&1   &&                &  -2   &   -2   &   -2     &     &     &     &     &     & 
    &\cr
&2   &&                & -4    &  -6    &   -6     & -6  & -8  &  -6 & -4  &  -6 & 
-6  &\cr
&3   &&                &  -6   &  -10   &   -10    & -12 &  -16& -12 & -12 & -18 & 
-   &\cr
\noalign{\hrule}}}\cr
&\vbox{\offinterlineskip\tabskip=0pt
\halign{\strut\vrule#
&\hfil~$#$
&\vrule#~
&\hfil $#$~
&\hfil $#$~
&\hfil $#$~
&\hfil $#$~
&\hfil $#$~
&\hfil $#$~
&\vrule#\cr
\noalign{\hrule}
&d_b=2 && [d_2,d_3]    & [0,0] &  [1,0] &   [1,1]  &[2,0]&[2,1]&\cr
\noalign{\hrule}
&d_1 &&                &       &        &          &     &     &\cr
&3   &&                &  -6   &   -10  &   -10    & -12 & -16 &\cr
&4   &&                & -32   &   -70  &   -70    & -96 & -   &\cr
&5   &&                & -270  &  -110  &   -      & -   & -   &\cr
\noalign{\hrule}}}}
$$
\vskip-10pt
\noindent
\centerline{{\bf Table 5}: Gromov-Witten invariants for local $A_3$.}
\vskip 10pt

Note that, as expected,  at $d_b=0$ the only Gromov-Witten
invariants occur at the degrees $\alpha^+$ with  
$N_{\alpha^+}=-2$ where $\alpha^+$ are the vector of positive roots in the
Cartan-Weyl basis.

{\baselineskip=12pt \sl
\goodbreak\midinsert
\centerline{\epsfxsize 2.5 truein\epsfbox{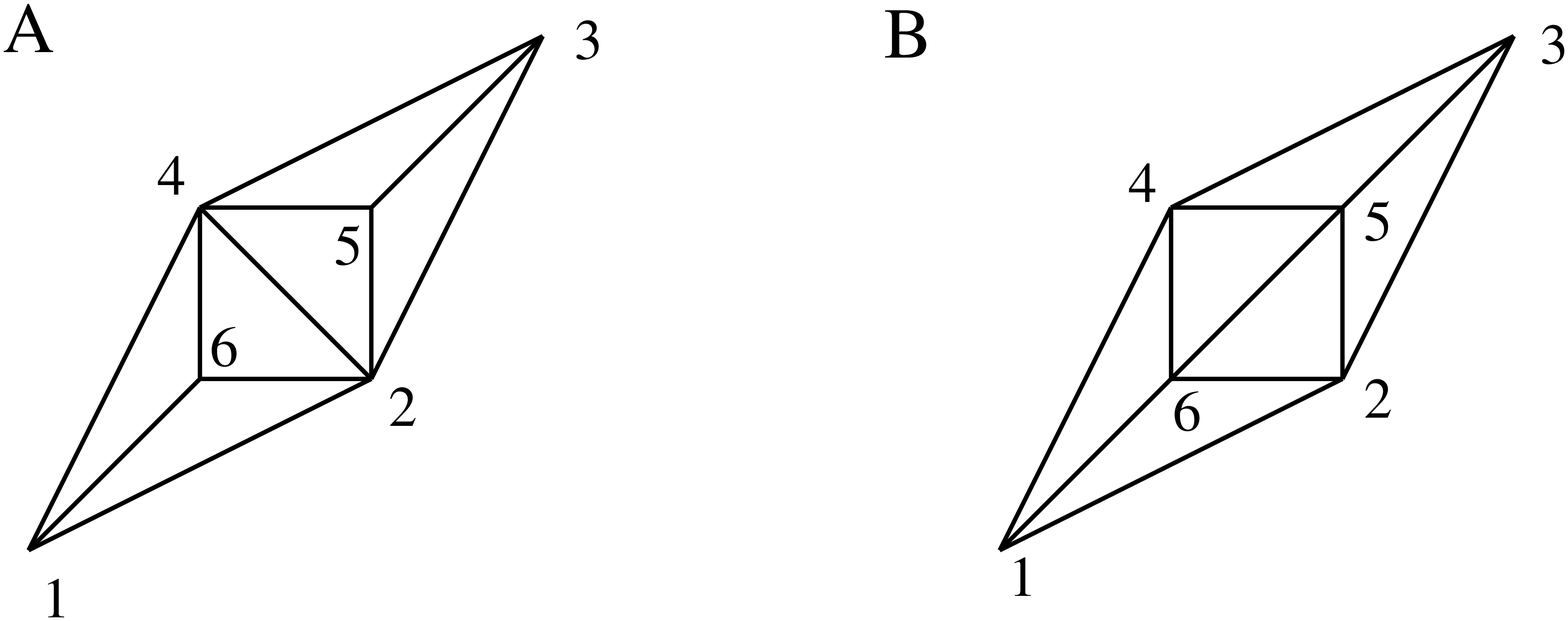}}
\leftskip 1pc\rightskip 1pc \vskip0.3cm
\noindent{\ninepoint  \baselineskip=8pt  
{{\bf Fig. 3:} Two phases of a local Calabi-Yau manifold. 
Phase $B$ is obtained by a flop transition.  
}}
\endinsert}

As a final example, we consider a toric grid diagram which admits a flop 
transition, 
describing in phase A two ${\bf P }^2$ connected by a ${\bf P}^1$.
The local geometry is defined  by the ${\bf C}^*$ operations generated by
$$\eqalign{
l_A^{(1)}=&(\tbs 0,-1,\tbs 0,-1,\tbs 1, \tbs 1) ,  
\quad\quad \qquad\qquad l_B^{(1)}=(\tbs 0,\tbs 1,\tbs 0,\tbs 1,-1,-1)\cr       
l_A^{(2)}=&(\tbs 1,\tbs 1,\tbs 0,\tbs 1,\tbs 0,-3),  
\quad\quad \qquad \qquad l_B^{(2)}=(\tbs 1,\tbs  0,\tbs 0,\tbs 0,\tbs 1,-2)\cr       
l_A^{(3)}=&(\tbs 0,\tbs 1,\tbs 1,\tbs 1,-3,\tbs 0) ,  
\quad \quad \qquad \qquad l_B^{(3)}=(\tbs 0, \tbs 0,\tbs 1,\tbs 0,-2,\tbs 1)\cr
{\cal SRI}_A=&\{x_1=x_2=x_4=0,x_5=x_6=0,  \quad  \quad {\cal 
SRI}_B=\{x_2=x_4=0,x_1=x_3=0,\cr    
\phantom{{\cal SRI}_A=\{}&x_2=x_3=x_4=0,x_1=x_3=0,  \ \ \ \   \qquad 
\qquad \qquad x_1=x_5=0,x_3=x_6=0\} \cr
\phantom{{\cal SRI}_A=\{}&x_1=x_5=0,x_3=x_6=0\} }$$ 
and the indicated Stanley-Reisner ideal. As in the $A_2$ case the mirror geometry 
is given by a genus two Riemann surface. The space of double-logarithmic solutions 
is two dimensional and can be determined from \diffident. 
Local invariants follow then for the $A$ and $B$ phase via \instbase~ from the 
following data 
$$\eqalign{
{\cal R}_A&=y_1 J_2^2+y_2(J_3^2-J_2^2),\quad x_1^A=y_1, \ \ x_2^A=-3(y_1-y_2), \ \ 
x_3^A=- 3y_2\cr 
{\cal R}_B&=y_1 (2 J_1 J_2+J_1 J_3 + 2 J_2^2+J_3^2)+ y_2 (J_2^2+J_1 J_2)\cr 
         x^B_1&=y_2,\ \  x^B_2=-3 y_1-2 y_2, \ \ x^B_3= -3 y_1 -y_2 \ .}$$

Below we list the Gromov-Witten invariants for the $B$ phase of the $({\bf 
P}^2,{\bf P}^2)$ diagram 
Fig. 3. The instantons of the $A$ phase are related to the one in the $B$ phase by 
$n^A_{k,j,i}=n^B_{i+j-k,i,j}$. The only degree for which this formula does not 
apply is 
$n^A_{1,0,0}=1$, which  counts just the flopped ${\bf P}^1$.
$$
\vbox{\offinterlineskip\tabskip=0pt
\halign{\strut\vrule#
&\hfil~$#$
&\vrule#&~
\hfil ~$#$~
&\hfil ~$#$~
&\hfil $#$~
&\vrule#\cr
\noalign{\hrule}
&d_b=0 && d_2    & 0   &  1  &  \cr
\noalign{\hrule}
&d_1 &&          &     &     & \cr
&0   &&          &     &  -2 & \cr
&1   &&          &  -2 &  -2 & \cr
\noalign{\hrule}}}
\quad
\vbox{\offinterlineskip\tabskip=0pt
\halign{\strut\vrule#
&\hfil~$#$
&\vrule#&~
\hfil ~$#$~
&\hfil ~$#$~
&\hfil ~$#$~
&\hfil ~$#$~
&\hfil ~$#$~
&\hfil ~$#$~
&\hfil $#$~
&\hfil $#$~
&\vrule#\cr
\noalign{\hrule}
&d_b=1 && d_2    & 0   &  1  & 2   &     3 &     4&      5&   6&\cr
\noalign{\hrule}
&d_1 &&          &     &     &     &       &       &        &         &\cr
&0   &&          & 1   &  3  &  5  &  7    &  9    &   11   &  13     &\cr
&1   &&          & 3   &  4  &  8  &  13   &  16   &   20   & 24      &\cr
&2   &&          & 5   & 8   & 9   &  15   &  21   &  27    &  33     &\cr
&3   &&          & 7   & 12  &  15 & 16    &  24   &  32    & 40      &\cr
&4   &&          & 9   & 16  & 21  & 24    &  25   &  35    & 45      &\cr
&5   &&          & 11  & 20  & 27  & 32    &  35   &  36    &  48     &\cr
\noalign{\hrule}}}
$$

$$
\vbox{\offinterlineskip\tabskip=0pt
\halign{\strut\vrule#
&\hfil~$#$
&\vrule#&~
\hfil ~$#$~
&\hfil ~$#$~
&\hfil ~$#$~
&\hfil ~$#$~
&\hfil ~$#$~
&\hfil ~$#$~
&\hfil $#$~
&\hfil $#$~
&\vrule#\cr
\noalign{\hrule}
&d_b=2 && d_2    & 0   &  1  & 2   &     3 &     4&      5&   6&\cr
\noalign{\hrule}
&d_1 &&          &     &     &     &       &       &        &            &\cr
&0   &&          &     &     & -6  & -32   & -110  &  -288        &-644  &\cr
&1   &&          &     &     & -10 & -70   & -270  &  -770        &-1820 &\cr
&2   &&          &  -6 & -10 & -32 & -126  & -456  & -1330        &-3264 &\cr
&3   &&          &  -32& -70 & -126& -300  & -784  &  -2052       &-4928 &\cr
&4   &&          & -110& -270& -456& -784  & - 1584& -3360  & - 7260     &\cr
&5   &&          & -288&-770 &-1330& -2052 & -3360 &  -6076 & -          &\cr
\noalign{\hrule}}}
$$

$$
\vbox{\offinterlineskip\tabskip=0pt
\halign{\strut\vrule#
&\hfil~$#$
&\vrule#
&~\hfil ~$#$~
&\hfil ~$#$~
&\hfil ~$#$~
&\hfil ~$#$~
&\hfil ~$#$~
&\hfil $#$~
&\hfil $#$~
&\vrule#\cr
\noalign{\hrule}
&d_b=3 && d_2    & 0   &  1  & 2   &     3 &     4&   5     &    \cr
\noalign{\hrule}
&d_1 &&          &     &     &     &       &       &       &            \cr
&0   &&          & 0   & 0   & 0   &  27   & 286   &1651      &            \cr
&1   &&          & 0   & 0   & 0   & 64    & 800   &5184       &            \cr
&2   &&          & 0   & 0   & 25  & 266   & 1998  &11473       &            \cr
&3   &&          &  27 & 64  &266  & 1332  & 6260  &26880       &            \cr
&4   &&          & 286 & 800 &1998 & 6260  & 21070 &70362       &            \cr
\noalign{\hrule}}}
$$
\vskip-10pt
\noindent
{{\bf Table 6}: Gromov-Witten invariants for the phase B in Fig. 3  }
\vskip 10pt

{}From the examples treated so far 
it should be clear how to proceed for a general toric grid diagram with $n$ inner 
points and $m$
boundary points. After choosing a triangulation and a corresponding basis of the 
$m+n-3$ linear 
relations $l^{(i)}$ one analyses the principal part of the differential system 
\diffident~to obtain a basis of the double-logarithmic solutions. This is the only 
additional 
information needed to specify the full set of the $2 n + m-2$ solutions from 
\expsol . 
The general structure of the solutions will be as follows. Besides the constant
solution, we get for each of the $n$ inner points of the toric diagram, whose 
total number equals 
the genus of the Riemann surface, one-single logarithmic solution and one
double-logarithmic solution coming from the period integrals around $a$- and 
$b$-type 
cycles of the Riemann surface. From additional boundary points in the toric grid 
diagram 
beyond $3$ we get additional single-logarithmic solutions which correpond to 
residues 
of the meromorphic form \lnform. Together with \mirrormap~the solutions determine 
\instbase~ and hence
the Gromov-Witten invariants, up to a choice of $x_i$ and $y_i$, which represents 
a choice of  the bases
for the double-logarithmic solutions. Requiring the Gromov-Witten invariants to be 
independent 
of this choice gives a  linear relation $x_i(y_j)$. This produces the
exact vacuum solutions and the BPS counting functions for those five dimensional theories, 
as discussed in \ln, which come from arbitrary grid diagrams \lv.  

\subsec{Cases with constraints}
The del Pezzo surfaces $B_n$ can be constructed by blowing ${\bf P}^2$
up $n$ times, $0\le n\le 8,$
in addition to ${\bf P}^1\times {\bf P}^1$. As is well known, the case $n=6$ 
can be represented as a cubic in ${\bf P}^3$ denoted by $X_{3}(1,1,1,1),$ and
$n=7,8$ are representable as degree four and six hypersurfaces in weighted 
projective
spaces:  $X_{4}(1,1,1,2)$ and $X_{6}(1,1,2,3),$
respectively. The $n=5$ case can be represented
as the degree (2,2) complete intersection in ${\bf P}^4$ $X_{2,2}(1,1,1,1,1)$. 
In addition, the quadric 
in ${\bf P}^3$, $X_{2}(1,1,1,1)$ is another representation of ${\bf P}^1\times
{\bf P}^1$. In the  representation given for the non-local Calabi-Yau geometry as the
canonical bundle over this del Pezzo surfaces below,  
the map $i^*:H^{1,1}(X)\rightarrow H^{1,1}(B)$ is not onto, as in the
previous cases. As a consequence the Gromow-Witten invariants are a 
sum over curves with degree $d_i$ in  classes in $H^{1,1}(B)$ up to degree
$d=\sum_i d_i$. These cases have been considered before \kmv\lmw\mnw\mnvw .

As all the weights in these representations are co-prime, the K\"ahler class 
associated to the hyperplane class of the ambient space is the only one 
which restricts to the surface (no exceptional divisors from the ambient space). 
We can recast the Chern classes of the surface and the of the canonical
bundle over it in terms of this class $J$. This can be stated more
generally for smooth complete intersections by the definition of
the following formal weight or charge vectors 
$l^{(k)}=(d_{1(k)},\ldots,d_{r(k)}|w_{1(k)},\ldots,w_{s(k)})$, where $d_{i(k)}$ 
are the degree of the $i$'th polynomial, $i=1,\ldots,p$, in the variables of
the 
$k$'th weighted projective space, $k=1,\ldots,s$.  
In terms of these we can express the 
total Chern class, by the adjunction formula, to obtain the
following formal expansion   
\eqn\chern{c={\prod_{k=1}^s \prod_{i=1}^{r_{(k)}} (1+w_{i,(k)} J_{(k)})\over 
\prod_{i=1}^p (1+ \sum_{k=1}^sd_{i,(k)} J_{(k)})}   {\prod_{i=1}^p 
\sum_{k=1}^s d_{j,(k)} J^{(k)}\over 
\prod_{k=1}^s \prod_{i=1}^{r_{(k)}} w_{i,{(k)}}}.}
Integrals over the top class for the non-compact case
are formally defined by multiplying with the volume form of the 
normal bundle ${\cal V}=\prod_{k=1}^t \prod_{j=1}^r d_{j,(k)} J^{(k)}$ and picking 
the coefficient of $\prod_{k=1}^t J^{(s_{(k)}-1)}$. Similarly wedge products  of
$c_2$ with $J$ and triple intersections are obtained.\foot{There are minor 
disagreements 
with the classical integrals equation (4.18) in \lmw~as well as with the 
normalization 
of the instanton numbers for  the case referred to as ``conifold'' in \lmw . 
These data should all follow from \chern .} 
We start by summarizing the weight or charge vectors for the non-compact case 
(the compact cases are obtained by deleting the last entry):
$$\eqalign{
X_{2}(1,1,1,1)\ : &\quad l=(-2|1,1,1,1,-1)\cr
X_{2,2}(1,1,1,1,1)\ : & \quad  l=(-2,-2|1,1,1,1,1,-1)\cr
X_{3}(1,1,1,1)\ : & \quad l=(-3|1,1,1,1,-1)\cr
X_{4}(1,1,1,2)\ : & \quad l=(-4|1,1,1,2,-1)\cr
X_{6}(1,1,2,3)\ : & \quad l=(-6|1,1,2,3,-1)}$$

Using \chern~we calculate for the cases in turn 
$$\eqalign{
\int J^3  =-1 ,-4 ,-3,-2,-1 \quad \int J c_2&=2 ,-4 ,-6,-8,-10\quad 
 \int c_3  =4 ,16 ,24,36,60 \ .}$$
The differential operators follow directly from \diffident\ for the five  cases 
they are
$$\eqalign{
{\cal L}^{(1)}=&\theta^3 - 4z (2\theta+1)^2\theta\ ,\quad 
{\cal L}^{(2)}=\theta^3 + 4z (2\theta+1)^2\theta\ ,\cr    
{\cal L}^{(3)}=&\theta^3 + 3z(3\theta+1)(3\theta+2)\theta\ ,\quad 
{\cal L}^{(4)}=\theta^3 + 4 z(4\theta+1)(4\theta+3)\theta\ , \cr
{\cal L}^{(5)}=&\theta^3 + 12 z (6\theta+1)(6\theta+5)\theta\ .}$$
The principal discriminant appears in front of the highest derivative 
${\cal L}=\Delta {\dd\over \dd z}^3+\ldots$, i.e.  $\Delta=(1+a z)$ with 
$a=-16,16,27,64,432,$ respectively. The following properties concerning the 
exponents at the critical loci of the discriminants are common:  the discriminat appears
with $\Delta^r$ $r=-{1\over 6},$ as for the conifold, and the $z^s$ appears with 
$s=-(12+\int Jc_2)/12$.  For these  cases, the charge vectors and the topological 
data listed above give (using INSTANTON) the following invariants.
$$
\vbox{\offinterlineskip\tabskip=0pt
\halign{\strut
\vrule#
&\hfil~$#$
&\vrule#~
&\hfil ~$#$~
&\hfil ~$#$~
&\vrule#~
&\hfil $#$~
&\hfil ~$#$~
&\vrule#~
&\hfil $#$~
&\hfil ~$#$~
&\vrule#\cr
\noalign{\hrule}
&    
&& \multispan2 $X_2(1,1,1,1)$      
&& \multispan2 $X_{2,2}(1,1,1,1,1)$     
&& \multispan2 $X_3(1,1,1,1)$  & \cr
\noalign{\hrule}
&d    
&&      {\rm rational}      & {\rm elliptic}          
&&      {\rm rational}      & {\rm elliptic}           
&&      {\rm rational}      & {\rm elliptic}     &  \cr
\noalign{\hrule} 
&1 &&  -4      &     0     && 16         &  0        &&   27     & 0&    \cr
&2 &&  -4      &     0     &&-20         &  0        &&  -54     & 0&  \cr
&3 &&  -12     &     0     && 48         &  0        &&  243     & -4& \cr
&4 &&  -48     &     9     && -192       &  5        &&  -1728   & -135 & \cr
&5 &&  -240    &   136     && 960        &  -96      &&  15255   & -3132 & \cr
&6 &&  -1356   &  1616     && -5436      &  1280     && -153576  & 62976  &    \cr
&7 &&  -8428   &  17560    && 33712      &  -14816   && 169086   &-1187892 &   \cr
&8 &&  -56000  &  183452   && -224000    &  160784   &&-20053440 &  21731112 &    
\cr
&9 &&  -392040 &  1878664  && 1588160    &  -1688800 &&249740091 &  -391298442 & 
\cr
&10&&  -2859120&  19027840 && -11436720  & -17416488 &&-3240109350&  6985791864 
&\cr
\noalign{\hrule}}}
$$
\vskip -8mm
$$
\vbox{\offinterlineskip\tabskip=0pt
\halign{\strut
\vrule#
&\hfil~$#$
&\vrule#~
&\hfil ~$#$~
&\hfil ~$#$~
&\vrule#~
&\hfil $#$~
&\hfil ~$#$~
&\vrule#\cr
\noalign{\hrule}
&    
&& \multispan2 $X_4(1,1,1,2)$      
&& \multispan2 $X_{6}(1,1,2,3)$ & \cr
\noalign{\hrule}
&d    &&      {\rm rational}      & {\rm elliptic}          
&&  {\rm rational}           & {\rm elliptic}         &  \cr
\noalign{\hrule} 
&1 &&  56      &  0      &&     252      &   -2  &    \cr
&2 &&  -272    &   3      &&    -9252     &   762  &  \cr
&3 &&  3240    &  -224      &&   848628       &  -246788 & \cr
&4 &&  -58432  & 12042       &&  -114265008       &  76413073 & \cr
&5 &&  1303840 &   -574896    &&  18958064400      & -23436186174  & \cr
&6 &&-33255216 &  26127574   &&   -3589587111852     &  7209650619780 &    \cr
&7 &&930431208 &    -1163157616   &&  744530011302420   &  -2232321201926988  &   
\cr
&8 &&-27855628544&   51336812456   &&   -165076694998001856 &  696061505044554012 
&    \cr
\noalign{\hrule}}}
$$\nobreak
{{\bf Table 7}: Gromov-Witten Invariants for local cases with constraints. }

As expected from the Segre embedding of ${\bf P}^1\times {\bf P}^1$ into
${\bf P}^3$ by the conic
constraint, this case should correspond to the diagonal part
of the local ${\bf P}^1\times {\bf P}^1$ case,
i.e.  $\sum_{i+j=r}n_{i,j}^{P^1\times P^1}=n_r^{X_{2}(1,1,1,1)}$, which is indeed 
true. The Gromov-Witten invariants for the elliptic curves are calculated using the 
holomorphic anomaly of the topological $B$-model \bcov\ . (In \klemmhg\ some of them 
are checked using localisation.)


\lref\calabi{E. Calabi,
``M\'etriques K\"ahl\`eriennes et Fibr\'es
Holomorphiques,'' Ann. Scient. \'Ec. Norm. Sup.
{\bf 12} (1979) 269-294.}
\lref\kapvas{M. Kapranov and E. Vasserot, 
``Kleinian Singularities, Derived Categories,
and Hall Algebras,'' math.AG/9812016.}

\newsec{Discussion}

We have established that mirror symmetry makes good sense
in the local setting, with the enumerative invariants
counting the effective contribution of the surface to the
Gromov-Witten invariants of a would-be Calabi-Yau
threefold which contains it.  These invariants, defined
and computed mathematically, are obtained through
analyzing solutions to differential equations, as in
the global case.  As in previous works, we see the
Seiberg-Witten curve arising from the B-model approach,
if an N=2 gauge theory is geometrically engineered.

Several interesting observations were made along the way.
We found, analyzing a reducible quadric,
that singular surfaces pose no obstacle to
defining the local invariants.
Indeed the A-model should be independent of deformations;
equivalently, calculations of Chern classes by sections
are independent of the choice of section.
Further, rather as the canonical bundle description
breaks down for singular surfaces, we find in
the fibered $A_n$ examples (in which the fibered
sphere-trees represent the singular surface)
that the bundle structure does not appear to be necessary
to proceed with the calculation.  Heuristically, one
can model not just the moduli space of maps as a
projective variety, but in fact the whole
vector bundle $U_d.$  With intersections in the Chow ring
and integration well-defined by virtue of a Thom class,
the procedure seems to yield the correct results.
This technique needs to be developed and made rigorous,
but the numbers still agree with the B-model results
in Table 4-5.

The recent work of Vafa and Gopakumar \vafagop\ 
introduces a new interpretation of these numbers and
their analogues at higher genus.  In particular, those authors
count the contributions of BPS states (D-branes) in a fixed homology
class
(but not fixed genus) to the full string partition function, which
is a sum over topological partition functions at all genera.
Their calculations tell us how to organize the partition functions
in order to extract integers, which represent BPS states
corresponding to cohomology classes on the full moduli space of
BPS states\foot{This is the standard reduction to a supersymmetric sigma
model on a moduli space.} and transforming under a
certain $SU(2)$ action in a particular way.  In genus zero,
the contribution is equivalent to the Euler characteristic.
At degree three in ${\bf P}^2,$ for example, a smooth degree
three polynomial is an elliptic curve, and the D-brane moduli
space includes the choice of a $U(1)$ bundle over the curve,
equivalently a point on the curve (if it is smooth).  The choice
of curves with points is shown in \vafagop\ to
be a ${\bf P}^8$ bundle over ${\bf P}^2,$ a space with Euler
characteristic 27 (which is indeed $n_3$ for ${\bf P}^2$).
The singular curves, however, should
be accompanied by their compactified Jacobians
as in \yz, but these can
in general no longer
be equated with the curves themselves.
Further, the compactified
Jacobians of reducible curves (e.g., the cubic $XYZ = 0$ in ${\bf P^2}$)
are particularly troublesome.  Perhaps the non-compact direction in
$K_{{\bf P}^2}$
leads to a resolution of these difficulties.  It would be very
interesting to mesh
the Gromov-Witten and D-brane explanations of these local
invariants.

Having extended traditional mirror symmetry to the non-compact case,
one naturally asks whether other viewpoints of mirror symmetry
make sense in the non-compact setting.  Is there any kind of 
special-Lagrangian fibration?  It is likely that, if so, the 
fibers would be decompactified tori, e.g. $S^1\times S^1\times {\bf R}.$
\foot{Recently, \penglu\ has found such a fibration for 
canonical bundles of projective spaces.}
This would be an interesting venue to the conjectures 
of \syz.  In particular, once the K\"ahler-Einstein metric
of the surface is
known, Calabi \calabi\ has given a method to find a Ricci-flat
metric on the total space of the canonical bundle.  In \syz\
it is argued that not only should the total D-brane moduli
space of the special-Lagrangian torus be the mirror manifold,
but also that the metric of the mirror should be computable
by an instanton expansion involving holomorphic discs bounding
the torus.  The local setting of a degenerate fiber, such as
has been studied and given an explicit metric
in \vafaoog, may prove an illustrative starting
point (though several of us have been unable to crack this
example).

What about the categorical mirror symmetry conjecture of Kontsevich?
Unfortunately, few explicit descriptions are known of
the derived categories of coherent sheaves over non-compact spaces
(or any spaces, for that matter).  In two dimensions, however,
the recent work of \kapvas\ gives a description of the derived
category over resolutions of A-D-E singularities in two dimensions.
The fibered versions of these spaces are just what we consider in
this paper.  It would be extremely interesting to calculate
Fukaya's category in these examples, especially as we
currently have
a real dearth testing grounds for Kontsevich's ideas.

We feel that the local setting may
be the best place for gleaning what's really at work
in mirror symmetry and tying together our still fragmented understanding
of this subject.

\vskip.2in

\centerline{\bf Acknowledgements}

We are indebted to B. Lian, for explaining extensions of the mirror 
principle, and for helping us at various stages of this project; and to M. Roth,
for explaining many algebro-geometric constructions and for his involvement
at the early stages of our work.  We thank them, as well as 
T. Graber, R. Pandharipande, and C. Vafa for many helpful discussions.
The work of T.-M. Chiang  and S.-T. Yau 
is supported in part by the NSF grant DMS-9709694; 
that of A. Klemm in part by 
a DFG Heisenberg fellowship and NSF Math/Phys DMS-9627351
and that of E. Zaslow in part by DE-F602-88ER-25065.

\newsec{Appendix:  Examples (A-Model)}

\subsec{${\cal O}(4) \rightarrow {\bf P}^3$}

This case is very similar, only we note that the rank of $U_d$ is
$4d + 1$ and the dimension of moduli space is $4d.$  Thus we must
take the Chern class $c_{4d}$ integrated over the moduli space.  Taking
the next-to-top Chern class has the following interpretation.
Instead of just counting the zeros of a section, $s_0,$ we take
two sections $s_0$ and $s_1$ and look at the zeros of $s_0 \wedge s_1,$
i.e. we look for points where the two sections are not linearly independent.
The number of such points also has the following interpretation.
Look at the ${\bf P}^1$ linear system generated by $s_0$ and $s_1.$
If $s_0\wedge s_1$ has a zero at a point
$(C,f)$ in moduli space, then some section
$a s_0 + b s_1$ vanishes identically on $f(C),$ i.e. $f(C)$ maps
to the zero locus of $a s_0 + b s_1.$  Therefore, the interpretation
of the next-to-top Chern class is as the number of rational curves
in any member of the linear system generated by two linearly independent
sections.

Of course, in this problem, we must account for multiple covers as above.
After doing so, the numbers $n_d$ which we get are as follows:

$$        
  \vbox{\offinterlineskip        
  \hrule        
  \halign{ &\vrule# & \strut\quad\hfil#\quad\cr     
  \noalign{\hrule}        
  height1pt&\omit&   &\omit&\cr        
  &$d$&& $n_d$ &\cr        
  \noalign{\hrule}        
  &1&& $320$ &\cr        
  &2&& $5016$ &\cr        
  &3&& $144192$ &\cr        
  &4&& $5489992$ &\cr        
  &5&& $247634752$ &\cr        
  &6&& $12472771416$ &\cr        
  &7&& $678995926336$ &\cr        
  &8&& $39156591980232$ &\cr        
  &9&& $2360737982593344$ &\cr        
 &10&& $147445147672907352$ &\cr        
  height1pt&\omit&   &\omit&\cr        
  \noalign{\hrule}}        
  \hrule}        
$$
\vskip-10pt
\noindent
{{\bf Table 8}: Gromov-Witten invariants for a $K_3$ surface inside a
Calabi-Yau threefold. }
\vskip 10pt

Practically speaking, sections are simply quartic polynomials, and the 
zero loci are quartic K3 surfaces.  Therefore, we are counting the number
of rational curves in a pencil of quartic K3 surfaces.  We may wish to
compare our results with a Calabi-Yau manifold admitting a K3 fibration.
Of course, the number of curves will depend on the nature of the fibration.
Our count pertains to a trivial total family, i.e. the zero locus of
a polynomial of bi-degree $(1,4)$ in ${\bf P}^1\times {\bf P}^3.$
The degree 8 hypersurface Calabi-Yau manifold in ${\bf P}_{2,2,2,1,1}$ 
is a pencil of quartic surfaces fibered over ${\bf P}^1$
in a different way, though the counting differs only by a factor of two.
Specifically, if we look at the Gromov-Witten invariants in
the homology class of $d$ times
the fiber (for this example $h^{11} = 2,$ one class
coming from the ${\bf P}^1$ base, one from the projective class of the K3
fiber), we get twice the numbers computed above.

\subsec{${\cal O}(3) \rightarrow {\bf P}^2$}

In this example, the rank of $U_d$ is now $3d + 1$ which is two greater
than the dimension of ${\cal M}_{00}(d,{\bf P}^2),$ so we must take
$c_{({\rm top} - 2)}(U_d).$  The interpretation is similar 
to the case of ${\cal O}(4) \rightarrow {\bf P}^3$.  We count the
number of rational curves in a two-dimensional family of cubic curves
generated by three linearly independent sections of ${\cal O}(3)$ (cubics).

The numbers $n_d$ are:  $n_1 = 21, n_2 = 21, n_3 = 18, n_4 = 21,
n_5 = 21, n_6 = 18,$ and so on, repeating these three values (as far
as we have computed them explicitly).  Some of the numbers can be easily
verified.  For example, the space of cubics on ${\bf P}^2$ (three variables)
forms (modulo scale) a ${\bf P}^9,$  while conics form a ${\bf P}^5$
and lines form a ${\bf P}^2.$  In order for a cubic curve to admit
a line, the polynomial must factor into a linear polynomial times
a conic.  To count the number of cubics in a ${\bf P}^2$ family which
do so, we must look at the intersection of ${\bf P}^2 \subset {\bf P}^2$
with the image $V$ of
$m:  {\bf P}^2 \times {\bf P}^5 \rightarrow {\bf P}^9,$ which is
just the map of multiplication of polynomials.  The Poincar\'e dual of the
${\bf P}^2$ family is just $H^7,$ where $H$ is the hyperplane class.
Therefore, we wish to compute
$\int_V H = \int_{{\bf P}^2\times {\bf P}^5}m^*(H).$
Now the map $m$ is linear in each of the coefficients (of the line and
the polynomial of the conic), so we have $m^*(H) = H_1 + H_2,$ where the $H_i$
are the hyperplane classes in ${\bf P}^2$ and ${\bf P}^5,$ respectively.
The integral just picks up the coefficient of $H_1^2 H_2^5$ in $(H_1 + H_2)^7,$
which is $21.$  Note that the same analysis applies to $n_2,$ since we
have already computed the number of cubics factoring into a conic 
(times a line).

To compute $n_3$ one needs more information about the discriminant locus
of the ${\bf P}^2$ family.  Similar calculations have been done in
\cdfkmI, where the authors consider a Calabi-Yau manifold  which is a
fibered by elliptic curves over a two-dimensional base.  The numbers differ
from the ones we have computed, since the fibration structure is different.
Nevertheless, the repeating pattern of three numbers survives.

\subsec{$K_{F_n}$}

``Local mirror symmetry'' of canonical bundle of Hirzebruch surfaces can
also be computed. The results are as follows:

{\vbox{\ninepoint{
$$
\vbox{\offinterlineskip\tabskip=0pt
\halign{\strut\vrule#
&\hfil~$#$
&\vrule#&~
\hfil ~$#$~
&\hfil ~$#$~
&\hfil $#$~
&\hfil $#$~
&\hfil $#$~
&\hfil $#$~
&\hfil $#$~
&\hfil $#$~
&\vrule#\cr
\noalign{\hrule}
& && d_F& 0 &  1  &  2  &  3 & 4 & 5 & 6&\cr
\noalign{\hrule}
&d_B &&  & &     &            &       &        &         &        &\cr
&0   &&  & & -2&   0&        0&     0&         0&     0&\cr
&1   &&  &-2&-4&  -6&       -8&     -10&     -12& -14&\cr
&2   &&  & 0&-6& -32&     -110&    -288&    -644& -1280&\cr
&3   &&  & 0&-8&-110&     -756&   -3556&  -13072& -40338&\cr
&4   &&  & 0&-10&-288&   -3556&  -27264& -153324& -690400&\cr
&5   &&  & 0&-12&-644&  -13072& -153324&-1252040& -7877210 &\cr
& 6  &&  & 0&-14&-1280& -40338& -690400&-7877210& -67008672&\cr
\noalign{\hrule}}
\hrule}$$
\vskip-7pt
\noindent
{\bf Table 9}: Invariants of $K_{{\bf F}_0}$ ($B$ and $F$ denote the ${\bf 
P}^1$'s)
\vskip7pt}}

{\vbox{\ninepoint{
$$
\vbox{\offinterlineskip\tabskip=0pt
\halign{\strut\vrule#
&\hfil~$#$
&\vrule#&~
\hfil ~$#$~
&\hfil ~$#$~
&\hfil $#$~
&\hfil $#$~
&\hfil $#$~
&\hfil $#$~
&\hfil $#$~
&\hfil $#$~
&\vrule#\cr
\noalign{\hrule}
& && d_F& 0 &  1  &  2  &  3 & 4 & 5 & 6&\cr
\noalign{\hrule}
&d_B &&  & &     &      &       &        &         &        &\cr
&0   &&  & &  -2&   0&    0&     0&      0&     0&\cr
&1   &&  &1& 3& 5& 7& 9& 11& 13& \cr
&2   &&  &0 & 0 & -6 & -32 & -110 & -288 & -644& \cr
&3   &&  &0 & 0 & 0 & 27 & 286 & 1651 & 6885 &  \cr
&4   &&  &0 & 0 & 0 & 0 & -192 & -3038 & -25216 & \cr
&5   &&  &0 & 0 & 0 & 0 & 0 & 1695 & 35870 & \cr
& 6  &&  &0 & 0 & 0 & 0& 0 & 0 & -17064 & \cr
\noalign{\hrule}}
\hrule}$$
\vskip-7pt
\noindent
{\bf Table 10}: Invariants of $K_{{\bf F}_1}$ ($B$ and $F$ denote 
the base and fiber class respectively)
\vskip7pt}}

The numbers for $d_B = d_F$ in the above table are the same as
that for $K_{\bP^2}$. As ${{\bf F}_1}$ is the blowup of $\bP^2$
at a point and the homology class of a line in $\bP^2$ pulled
back to ${{\bf F}_1}$ is $B + F$, this is what we expect.

{\vbox{\ninepoint{
$$
\vbox{\offinterlineskip\tabskip=0pt
\halign{\strut\vrule#
&\hfil~$#$
&\vrule#&~
\hfil ~$#$~
&\hfil ~$#$~
&\hfil $#$~
&\hfil $#$~
&\hfil $#$~
&\hfil $#$~
&\hfil $#$~
&\hfil $#$~
&\vrule#\cr
\noalign{\hrule}
& && d_F& 0&   1  &  2  &  3 & 4 & 5 & 6&\cr
\noalign{\hrule}
&d_B &&  &&      &      &       &        &         &        &\cr
&0   &&  &&   -2&   0&    0&     0&      0&     0&\cr
&1   &&  &-1/2&  -2& -4& -6& -8& -10& -12& \cr
&2   &&  &0&  0 & 0& -6 & -32 & -110 & -288& \cr
&3   &&  &0&  0 & 0 & 0 & -8 & -110 & -756 &  \cr
&4   &&  &0&  0 & 0 & 0 & 0 & -10& -288 & \cr
&5   &&  &0&  0 & 0 & 0 & 0 & 0 & -12 & \cr
& 6  &&  &0&  0 & 0 & 0& 0 & 0 & 0& \cr
\noalign{\hrule}}
\hrule}$$
\vskip-7pt
\noindent
{\bf Table 11}: Invariants of $K_{{\bf F}_2}$ ($B$ and $F$ denote the
base and fiber class respectively)
\vskip7pt}}

	We do not understand the result of $-1/2$ above, but 
it reflects the fact that the moduli space of stable maps into the
base, which is a curve of negative self-intersection,  is not convex.
Therefore the $A$-model calculation is suspect and we consider 
the $B$-model result of $0$ for this invariant to be the right answer. 
(For higher degree maps into the base,  the $A$- and $B$-model 
results agree again.)  
\listrefs

\end